\documentclass[printer]{aa}
\usepackage{multirow}
\usepackage[varg]{txfonts}
\usepackage{natbib}
\usepackage{graphicx}
\usepackage{txfonts}
\usepackage{color} 
\usepackage{gensymb} 
\bibpunct{(}{)}{;}{a}{}{,}
\usepackage{tabularx}
\begin{document}  
\title{A CO survey on a sample of \textit{Herschel} cold clumps}
   \author{O. Feh\'er\inst{1,2},
          M. Juvela\inst{3,4}, 
		  T. Lunttila\inst{5},
          J. Montillaud\inst{4},
          I. Ristorcelli\inst{6,7},
          S. Zahorecz\inst{8,9},
          L. V. T\'oth\inst{2}
          }
   \institute{Konkoly Observatory, Research Centre for Astronomy and Earth Sciences, Hungarian Academy of Sciences, H-1121 Budapest, Konkoly Thege Mikl\'os \'ut 15-17, Hungary
   \and
   E\"otv\"os Lor\'and University, Department of Astronomy, P\'azm\'any P\'eter s\'et\'any 1/A, 1117 Budapest, Hungary
         \and
		Department of Physics, PO Box 64, 00014 University of Helsinki, Finland
                \and
Institut UTINAM - UMR 6213 - CNRS - Univ Bourgogne Franche Comt\'e, 41 bis avenue de l’Observatoire, 25000 Besançon, France
        \and
        Chalmers University of Technology, Department of Earth and Space Sciences, Onsala Space Observatory, 439 92 Onsala, Sweden
        \and
        Universit\'e de Toulouse, UPS-OMP, IRAP, Toulouse, France
        \and
        CNRS, IRAP, 9 Av. colonel Roche, BP 44346, F-31028 Toulouse cedex 4, France
        \and
        Department of Physical Science, Graduate School of Science, Osaka
Prefecture University, 1-1 Gakuen-cho, Naka-ku, Sakai, Osaka 599-8531, Japan
        \and
        Chile Observatory, National Astronomical Observatory of Japan, National
Institutes of Natural Science, 2-21-1 Osawa, Mitaka, Tokyo 181-8588, Japan
        }
\authorrunning{Feh\'er et al.}

  \abstract
   {The physical state of cold cloud clumps has a great impact on the process and efficiency of star formation and the masses of the forming stars inside these objects. The sub-millimetre survey of the \textit{Planck} space observatory and the far-infrared follow-up mapping of the \textit{Herschel} space telescope provide an unbiased, large sample of these cold objects.}
   {We have observed $^{12}$CO(1$-$0) and $^{13}$CO(1$-$0) emission in 35 high-density clumps in 26 \textit{Herschel} fields sampling different environments in the Galaxy. We derive the physical properties of the objects and estimate their gravitational stability.}
   {The densities and temperatures of the clumps were calculated from both the dust continuum and the molecular line data. Kinematic distances were derived using $^{13}$CO(1$-$0) line velocities to verify previous distance estimates and the sizes and masses of the objects were calculated by fitting 2D Gaussian functions to their optical depth distribution maps on 250\,$\mu$m. The masses and virial masses were estimated assuming an upper and lower limit on the kinetic temperatures and considering uncertainties due to distance limitations.}
   {The derived excitation temperatures are between 8.5$-$19.5\,K, and for most clumps between 10$-$15\,K, while the \textit{Herschel}-derived dust colour temperatures are more uniform, between 12$-$16\,K. The sizes (0.1$-$3\,pc), $^{13}$CO column densities (0.5$-$44\,$\times$\,10$^{15}$\,cm$^{-2}$) and masses (from less than 0.1\,$M_{\odot}$ to more than 1500\,$M_{\odot}$) of the objects all span broad ranges. We provide new kinematic distance estimates, identify gravitationally bound or unbound structures and discuss their nature.}
  {The sample contains objects on a wide scale of temperatures, densities and sizes. Eleven gravitationally unbound clumps were found, many of them smaller than 0.3\,pc, but large, parsec-scale clouds with a few hundred solar masses appear as well. Colder clumps have generally high column densities but warmer objects appear at both low and higher column densities. The clump column densities derived from the line and dust observations correlate well, but are heavily affected by uncertainties of the dust properties, varying molecular abundances and optical depth effects.}
   \maketitle

\section{Introduction}
\label{intro}

The properties of star formation and the parameters of the forming young stars depend on the initial conditions within the molecular cloud they are located in. A large sample of different star forming environments have to be examined to investigate the connection between the physical parameters (density and temperature structure, turbulence, kinematics) of a star-forming clump, its star formation efficiency and the properties of the forming stars.

The \textit{Planck} telescope \citep{tauber2010} mapped the whole sky in nine frequency bands covering 30-857\,GHz with high sensitivity and angular resolution (below 5\,$\arcmin$ at the highest frequencies) and found many cold and dense regions in the Galaxy. The Early Cold Cores catalogue \citep[ECC;][]{planck2011a} contains the 915 most reliable detections of these cold structures. A statistical analysis of the properties of the whole catalogue of more than 10\,000 sources (C3PO: Cold Clump Catalogue of \textit{Planck} Objects) was performed by \citet{planck2011b}. Most of these objects proved to be parsec-scale so-called clumps that are likely to contain one or several cores in the early stages of pre-stellar or protostellar evolution. However, we can find a broad range of other objects among the C3PO sample, from low-mass, individual cores to large cloud complexes. Based on the $Planck$ observations, the H$_2$ column densities of the objects are between 10$^{20}-$10$^{23}$\,cm$^{-2}$, the dust temperatures vary between 7 and 19\,K and most objects are located closer than 2\,kpc. The catalogue was further improved using the full \textit{Planck} survey and the \textit{Planck} Catalogue of Galactic Cold Clumps \citep[PGCC;][]{planck2016} was released, containing 13\,188 sources.

\begin{table}[t]
	\footnotesize
	\centering
		\caption{The observed clumps.}
		\begin{tabular}{l c l l}
		\hline
		\multicolumn{1}{c}{GCC ID} & \multicolumn{1}{c}{clump} & \multicolumn{1}{c}{RA (J2000)} & \multicolumn{1}{c}{DEC (J2000)} \\
		\multicolumn{1}{c}{\ } & \multicolumn{1}{c}{\ } & \multicolumn{1}{c}{[hh:mm:ss]} & \multicolumn{1}{c}{[dd:mm:ss]} \\
		\hline
		\hline
G26.34+8.65 & A & 18:08:35.8 & -01:49:55.4 \\
G37.49+3.03 & A & 18:48:55.8 & +05:27:01.8 \\
 & B & 18:49:09.4 & +05:37:33.2 \\
G37.91+2.18 & A & 18:52:53.4 & +05:25:10.8 \\
 & B & 18:53:31.5 & +05:25:29.8 \\
G39.65+1.75-1 & A & 18:57:02.5 & +06:58:22.4 \\
 & B & 18:57:30.7 & +06:48:23.8 \\
G62.16-2.92 & A & 19:59:49.8 & +24:13:59.7 \\
G69.57-1.74-1 & A & 20:13:32.6 & +31:21:52.4 \\
 & B & 20:13:05.8 & +31:18:23.4 \\
G70.10-1.69-1 & A & 20:14:35.4 & +31:57:26.6 \\
 & B & 20:14:07.7 & +31:38:29.2 \\
G71.27-11.32 & A & 20:53:20.6 & +26:52:00.9 \\
G91.09-39.46 & A & 23:10:30.3 & +17:05:32.4 \\
G95.76+8.17-1 & A & 20:56:49.9 & +58:01:35.1 \\
 & B & 20:57:52.6 & +58:06:48.2 \\
G109.18-37.59 & A & 00:03:48.9 & +23:58:47.0 \\
G110.62-12.49-1 & A & 23:38:02.1 & +48:35:15.8 \\
G115.93+9.47 & A & 23:23:52.6 & +71:08:45.5 \\
 & B & 23:24:25.2 & +71:11:03.8 \\
G116.08-2.40-1 & A & 23:56:45.0 & +59:42:20.0 \\
G126.24-5.52 & A & 01:15:46.8 & +57:12:45.3 \\
G139.60-3.06-1 & A & 02:50:21.6 & +55:50:53.6 \\
G141.25+34.37 & A & 08:48:35.8 & +72:43:10.2 \\
G159.12-14.30 & A & 03:50:31.9 & +35:40:50.9 \\
G159.23-34.51-1 & A & 02:56:00.5 & +19:26:17.7 \\
 & C & 02:57:49.6 & +19:23:25.2 \\
G171.35-38.28 & A & 03:18:15.4 & +10:17:58.2 \\
G174.22+2.58 & A & 05:41:34.5 & +35:10:25.6 \\
 & B & 05:40:28.0 & +35:04:56.4 \\
G188.24-12.97-1 & A & 05:16:28.3 & +15:08:52.9 \\
G189.51-10.41-1 & A & 05:29:07.3 & +15:30:53.6 \\
G195.74-2.29 & A & 06:10:58.5 & +14:09:26.2 \\
G203.42-8.29-1 & A & 06:04:19.1 & +04:11:43.4 \\
G205.06-6.04-1 & A & 06:16:04.3 & +04:00:47.3 \\
\hline
		\end{tabular}
        \tablefoot{The columns are: (1) ID of the \textit{Herschel} field; (2) ID of the clumps (column density peaks) on the \textit{Herschel} field classified from the highest column density (A) to the lowest (C); (3,4) equatorial coordinates of the clump. }
	\label{sources}
\end{table}

The \textit{Herschel} Open Time Key Programme \textit{Galactic Cold Cores} (GCC) selected \textit{Planck} C3PO objects to observe with the \textit{Herschel} \textit{PACS} and \textit{SPIRE} instruments in the 60$-$500\,$\mu$m wavelength range \citep{poglitsch2010, griffin2010}. During the survey 116 fields (390 individual \textit{Planck} cold clumps) were mapped, providing a representative cross-section of the \textit{Planck} clump population regarding their Galactic position, dust colour temperature and mass. The higher spatial resolution (down to 7$\arcsec$ at 100\,$\mu$m) of \textit{Herschel} \citep{pilbratt2010} revealed the structures of the cold sources, and often individual cores were resolved as well. The shorter wavelength observations help to determine the physical characteristics of the objects. \citet{juvela2012} investigated the morphology and parameters of the clumps in 71 fields observed by \textit{Herschel}, deriving column densities and dust temperatures and found that about half of the fields are associated with active star formation. Most of the examined cloud structures were filamentary, but they occasionally show cometary and compressed shapes. A catalogue of sub-millimetre sources on all 116 fields in the programme was built by \citet{montillaud2015}. They derived the general properties of the fields, separated starless sources from those containing protostellar objects and provided distance estimates for the observed clouds with uncertainties and reliability flags. \citet{juvela2015b} investigated the variations and systematic errors of sub-millimetre dust opacity estimates relative to near-infrared optical depths. They found that the ratio of the two values correlates with Galactic location and star formation activity and the sub-millimeter opacity increases in the densest and coldest regions. \citet{juvela2015a} determined that the dust opacity spectral index anti-correlates with temperature, correlates with column density and is lower closer to the Galactic plane. Although the results are affected by various error sources and the used data sets, they are robust concerning the observing wavelength and the detected spatial variations. A statistical survey of filaments was performed by \citet{rivera2016} where they extracted, fitted and characterized filaments on the GCC fields. They found that their linear mass density is connected to the local environment which supports the accretion-based filament evolution theory \citep{arzoumanian2011, palmeirim2013}. According to a survey by \citet{toth2016} about 30\% of the Taurus-Auriga-Perseus PGCC clumps have associated YSOs. The $Planck$ clumps also form massive clusters as shown by \citet{toth2017}. \citet{zahorecz2016} tested a number of $Planck$ clumps massive enough to form high-mass stars and star clusters since they exceed the empirical threshold for massive star formation. Seven of those clumps are without associated YSOs.

Studies of molecular line emission in these cold clumps and cores provide mass and stability estimates and information about their kinematics. \citet{wu2012} carried out a $^{12}$CO(1$-$0), $^{13}$CO(1$-$0) and C$^{18}$O(1$-$0) single pointing survey towards 674 ECCs, estimating kinematic distances and column densities. Additionally they mapped 10 fields and found 22 cores of which 7 are gravitationally bound. \citet{meng2013} mapped 71 ECCs from this previous sample and derived excitation temperatures, column densities and velocity dispersions. They identified 38 cores: 90\% of them are starless and the majority are gravitationally unbound. Their dust temperatures were usually higher than the gas temperatures. \citet{parikka2015} observed $^{12}$CO(1$-$0), C$^{18}$O(1$-$0) and N$_2$H$^+$(1$-$0) lines in 21 cold clumps in 20 \textit{Herschel} fields, calculated clump masses and densities from both the dust continuum and molecular line data and found them to be in reasonable agreement. They also derived $^{13}$CO and C$^{18}$O relative abundances with radiative transfer modelling of two clumps. They conclude that most cold clumps are not necessarily pre-stellar. There was also a CO mapping survey of 96 ECCs in the second quadrant of the Galaxy by \citet{zhang2016} to derive temperatures, densities and velocity dispersions. The two PGCCs in the well-known dense filament TMC-1 were observed in NH$_3$(1,1) and (2,2) to investigate the structure of the cloud \citep{feher2016} and extended ammonia surveys on several \textit{Herschel} fields were also carried out to constrain their temperature, density and velocity structure (Tóth et al. 2017 in prep.).

\begin{table*}[h]
	\centering
	\footnotesize
	\caption{Parameters of the $^{12}$CO(1$-$0) and $^{13}$CO(1$-$0) lines at the centre of each clump.}
		\begin{tabular}{l c r r r r r r r r}
		\hline
\multicolumn{1}{c}{clump ID} & \multicolumn{1}{c}{L} & \multicolumn{1}{c}{$W\mathrm{^{^{12}CO}}$}& \multicolumn{1}{c}{$v\mathrm{_{LSR}^{^{12}CO}}$} & \multicolumn{1}{c}{$\Delta v\mathrm{^{^{12}CO}}$} & \multicolumn{1}{c}{$T\mathrm{_{MB}^{^{12}CO}}$} & \multicolumn{1}{c}{$W\mathrm{^{^{13}CO}}$}& \multicolumn{1}{c}{$v\mathrm{_{LSR}^{^{13}CO}}$} & \multicolumn{1}{c}{$\Delta v\mathrm{^{^{13}CO}}$} & \multicolumn{1}{c}{$T\mathrm{_{MB}^{^{13}CO}}$}  \\
	\multicolumn{1}{c}{\ } & \multicolumn{1}{c}{\ } & \multicolumn{1}{c}{[Kkms$^{-1}$]} & \multicolumn{1}{c}{[kms$^{-1}$]} & \multicolumn{1}{c}{[kms$^{-1}$]} & \multicolumn{1}{c}{[K]} & \multicolumn{1}{c}{[Kkms$^{-1}$]} & \multicolumn{1}{c}{[kms$^{-1}$]} & \multicolumn{1}{c}{[kms$^{-1}$]} & \multicolumn{1}{c}{[K]} \\
		\hline
		\hline
G26.34-A & 1 &   9.9 $\pm$ 1.2 &  11.02 $\pm$ 0.10 & 1.7 $\pm$ 0.3 &  5.3 $\pm$  2.2 &  2.5 $\pm$ 0.2 &  11.28 $\pm$ 0.02 & 0.49 $\pm$ 0.04 &  4.8 $\pm$ 0.8 \\
G37.49-A & 1 &  27.7 $\pm$ 1.5 &  15.49 $\pm$ 0.06 & 2.0 $\pm$ 0.1 & 12.7 $\pm$  2.9 &  5.7 $\pm$ 0.3 &  15.42 $\pm$ 0.03 & 1.08 $\pm$ 0.06 &  4.9 $\pm$ 0.7 \\
G37.49-B & 1 &   ... &   ... & ... &  ... &  3.1 $\pm$ 0.3 &  15.82 $\pm$ 0.04 & 0.79 $\pm$ 0.10 &  3.7 $\pm$ 0.9 \\
G37.91-A & 1 &  36.4 $\pm$ 1.3 &  34.37 $\pm$ 0.07 & 4.0 $\pm$ 0.2 &  8.6 $\pm$  1.9 &  8.4 $\pm$ 0.4 &  34.23 $\pm$ 0.04 & 1.93 $\pm$ 0.10 &  4.1 $\pm$ 0.7 \\
G37.91-B & 1 &   ... &   ... & ... &  ... &  5.1 $\pm$ 0.3 &  35.16 $\pm$ 0.03 & 1.05 $\pm$ 0.10 &  4.5 $\pm$ 0.8 \\
G39.65-A & 1 &   ... &   ... & ... &  ... & 24.3 $\pm$ 0.5 &  29.96 $\pm$ 0.03 & 2.79 $\pm$ 0.06 &  8.2 $\pm$ 0.8 \\
G39.65-B & 1 &   ... &   ... & ... &  ... & 14.0 $\pm$ 0.2 &  28.56 $\pm$ 0.07 & 2.71 $\pm$ 0.07 &  4.8 $\pm$ 0.7 \\
 & 2 &   ... &   ... & ... &  ... & 11.6 $\pm$ 0.2 &  31.61 $\pm$ 0.07 & 2.04 $\pm$ 0.07 &  5.4 $\pm$ 0.7 \\
G62.16-A & 1 &   ... &   ... & ... &  ... &  2.2 $\pm$ 0.2 &  13.08 $\pm$ 0.04 & 0.83 $\pm$ 0.13 &  2.5 $\pm$ 0.6 \\
G69.57-A & 1 &  44.1 $\pm$ 1.4 &  12.50 $\pm$ 0.08 & 5.1 $\pm$ 0.2 &  8.1 $\pm$  1.7 & 16.1 $\pm$ 0.5 &  12.54 $\pm$ 0.05 & 3.16 $\pm$ 0.10 &  4.8 $\pm$ 0.7 \\
G69.57-B & 1 &  15.3 $\pm$ 2.1 &  10.03 $\pm$ 0.03 & 1.3 $\pm$ 0.1 & 11.0 $\pm$  1.5 & 13.8 $\pm$ 0.3 &  11.15 $\pm$ 0.03 & 2.29 $\pm$ 0.07 &  5.7 $\pm$ 0.5 \\
 & 2 &  43.3 $\pm$ 2.9 &  12.16 $\pm$ 0.14 & 4.2 $\pm$ 0.3 &  9.6 $\pm$  1.5 &  3.5 $\pm$ 0.2 &  13.49 $\pm$ 0.02 & 0.87 $\pm$ 0.06 &  3.8 $\pm$ 0.5 \\
G70.10-A & 1 &  96.3 $\pm$ 1.5 &  11.05 $\pm$ 0.05 & 5.6 $\pm$ 0.1 & 16.1 $\pm$  1.7 & 26.1 $\pm$ 0.5 &  11.35 $\pm$ 0.03 & 3.31 $\pm$ 0.07 &  7.4 $\pm$ 0.7 \\
G70.10-B & 1 &   ... &   ... & ... &  ... & 16.7 $\pm$ 0.3 &  14.75 $\pm$ 0.03 & 3.09 $\pm$ 0.07 &  5.1 $\pm$ 0.5 \\
G71.27-A & 1 &   3.6 $\pm$ 0.5 &   6.01 $\pm$ 0.03 & 0.4 $\pm$ 0.1 &  8.9 $\pm$  2.2 &  0.7 $\pm$ 0.1 &   5.94 $\pm$ 0.02 & 0.28 $\pm$ 0.09 &  2.4 $\pm$ 0.5 \\
G91.09-A & 1 &  40.0 $\pm$ 1.9 &  -4.82 $\pm$ 0.11 & 4.7 $\pm$ 0.3 &  8.1 $\pm$  2.2 &  2.9 $\pm$ 0.2 &  -4.66 $\pm$ 0.04 & 0.95 $\pm$ 0.09 &  2.9 $\pm$ 0.6 \\
G95.76-A & 1 &  18.1 $\pm$ 0.9 &  -0.12 $\pm$ 0.07 & 2.6 $\pm$ 0.1 &  6.4 $\pm$  1.5 &  6.8 $\pm$ 0.5 &   0.20 $\pm$ 0.07 & 1.74 $\pm$ 0.14 &  3.7 $\pm$ 1.0 \\
G95.76-B & 1 &  29.9 $\pm$ 0.9 &  -0.28 $\pm$ 0.05 & 3.2 $\pm$ 0.1 &  8.9 $\pm$  1.4 &  7.7 $\pm$ 0.5 &  -0.39 $\pm$ 0.06 & 1.87 $\pm$ 0.12 &  3.9 $\pm$ 0.9 \\
G109.18-A & 1 &   9.5 $\pm$ 0.8 &  -4.09 $\pm$ 0.05 & 1.2 $\pm$ 0.1 &  7.5 $\pm$  1.9 &  1.2 $\pm$ 0.2 &  -4.09 $\pm$ 0.04 & 0.51 $\pm$ 0.08 &  2.3 $\pm$ 0.7 \\
G110.62-A & 1 &  10.7 $\pm$ 0.4 &  -7.86 $\pm$ 0.02 & 1.0 $\pm$ 0.0 & 10.4 $\pm$  1.2 &  5.0 $\pm$ 0.1 &  -7.89 $\pm$ 0.01 & 0.66 $\pm$ 0.02 &  7.2 $\pm$ 0.4 \\
G115.93-A & 1 &  41.6 $\pm$ 2.0 &  -2.56 $\pm$ 0.07 & 3.1 $\pm$ 0.2 & 12.5 $\pm$  3.0 &  9.7 $\pm$ 0.4 &  -3.45 $\pm$ 0.03 & 1.48 $\pm$ 0.07 &  6.2 $\pm$ 0.8 \\
G115.93-B & 1 &   ... &   ... & ... &  ... &  4.0 $\pm$ 0.2 &  -4.02 $\pm$ 0.02 & 0.58 $\pm$ 0.04 &  6.4 $\pm$ 0.8 \\
G116.08-A & 1 &  16.5 $\pm$ 0.7 &  -0.96 $\pm$ 0.05 & 2.3 $\pm$ 0.1 &  6.8 $\pm$  1.3 &  5.2 $\pm$ 0.4 &  -0.60 $\pm$ 0.05 & 1.28 $\pm$ 0.12 &  3.8 $\pm$ 0.9 \\
G126.24-A & 1 &  18.7 $\pm$ 2.6 & -16.50 $\pm$ 0.21 & 2.8 $\pm$ 0.3 &  6.2 $\pm$  4.4 &  0.9 $\pm$ 0.2 & -16.98 $\pm$ 0.04 & 0.37 $\pm$ 0.08 &  2.2 $\pm$ 0.7 \\
G139.60-A & 1 & 101 $\pm$ 1.2 & -32.44 $\pm$ 0.02 & 3.4 $\pm$ 0.1 & 28.3 $\pm$  1.8 & 27.1 $\pm$ 0.7 & -32.62 $\pm$ 0.04 & 2.65 $\pm$ 0.08 &  9.6 $\pm$ 1.1 \\
G141.25-A & 1 &  10.1 $\pm$ 0.6 &   0.86 $\pm$ 0.04 & 1.5 $\pm$ 0.1 &  6.5 $\pm$  1.2 &  1.1 $\pm$ 0.2 &   0.72 $\pm$ 0.03 & 0.47 $\pm$ 0.10 &  2.1 $\pm$ 0.6 \\
G159.12-A & 1 &  14.3 $\pm$ 1.1 &  15.84 $\pm$ 0.04 & 1.0 $\pm$ 0.1 & 14.0 $\pm$  2.8 &  2.7 $\pm$ 0.1 &  15.92 $\pm$ 0.01 & 0.45 $\pm$ 0.02 &  5.5 $\pm$ 0.6 \\
G159.23-A & 1 &  46.1 $\pm$ 1.7 &  -4.43 $\pm$ 0.10 & 5.3 $\pm$ 0.2 &  8.2 $\pm$  2.0 &  7.7 $\pm$ 0.5 &  -5.54 $\pm$ 0.03 & 0.90 $\pm$ 0.07 &  8.0 $\pm$ 1.4 \\
G159.23-C & 1 &  18.2 $\pm$ 0.8 &  -1.28 $\pm$ 0.04 & 1.8 $\pm$ 0.1 &  9.7 $\pm$  1.7 &  5.8 $\pm$ 0.3 &  -1.50 $\pm$ 0.02 & 0.87 $\pm$ 0.06 &  6.3 $\pm$ 0.9 \\
G171.35-A & 1 &  10.1 $\pm$ 0.7 &   6.76 $\pm$ 0.03 & 0.9 $\pm$ 0.1 & 10.0 $\pm$  1.7 &  0.9 $\pm$ 0.2 &   6.59 $\pm$ 0.06 & 0.55 $\pm$ 0.15 &  1.6 $\pm$ 0.6 \\
G174.22-A & 1 &  35.0 $\pm$ 1.7 & -21.87 $\pm$ 0.07 & 3.3 $\pm$ 0.2 & 10.0 $\pm$  2.0 & 10.8 $\pm$ 0.2 & -22.09 $\pm$ 0.02 & 1.64 $\pm$ 0.04 &  6.2 $\pm$ 0.5 \\
G174.22-B & 1 &  14.8 $\pm$ 0.8 & -12.70 $\pm$ 0.07 & 2.3 $\pm$ 0.1 &  5.9 $\pm$  1.4 &  7.4 $\pm$ 0.3 & -12.91 $\pm$ 0.03 & 1.57 $\pm$ 0.07 &  4.4 $\pm$ 0.6 \\
G188.24-A & 1 &  13.7 $\pm$ 0.7 &   1.47 $\pm$ 0.03 & 1.2 $\pm$ 0.1 & 10.8 $\pm$  1.7 &  1.0 $\pm$ 0.2 &   1.50 $\pm$ 0.04 & 0.57 $\pm$ 0.09 &  1.7 $\pm$ 0.5 \\
 & 2 &   7.6 $\pm$ 0.6 &   6.44 $\pm$ 0.03 & 0.8 $\pm$ 0.1 &  9.4 $\pm$  1.7 &  1.7 $\pm$ 0.1 &   6.44 $\pm$ 0.02 & 0.42 $\pm$ 0.04 &  3.7 $\pm$ 0.5 \\
G189.51-A & 1 &  17.0 $\pm$ 1.0 &   9.06 $\pm$ 0.07 & 2.4 $\pm$ 0.1 &  6.7 $\pm$  1.8 &  3.6 $\pm$ 0.3 &   9.36 $\pm$ 0.03 & 0.84 $\pm$ 0.08 &  4.1 $\pm$ 0.8 \\
G195.74-A & 1 &  57.8 $\pm$ 1.0 &   4.33 $\pm$ 0.02 & 2.7 $\pm$ 0.1 & 19.9 $\pm$  1.6 & 21.6 $\pm$ 0.5 &   4.17 $\pm$ 0.02 & 1.98 $\pm$ 0.04 & 10.3 $\pm$ 0.9 \\
G203.42-A & 1 &  17.5 $\pm$ 1.3 &  11.72 $\pm$ 0.06 & 1.7 $\pm$ 0.1 &  9.8 $\pm$  2.5 &  6.1 $\pm$ 0.3 &  11.60 $\pm$ 0.02 & 0.83 $\pm$ 0.04 &  6.9 $\pm$ 0.9 \\
G205.06-A & 1 &  19.9 $\pm$ 1.4 &  11.13 $\pm$ 0.07 & 1.9 $\pm$ 0.2 &  9.6 $\pm$  3.8 &  6.8 $\pm$ 0.4 &  11.33 $\pm$ 0.03 & 0.98 $\pm$ 0.06 &  6.5 $\pm$ 1.1 \\
\hline
		\end{tabular}
        \tablefoot{The columns are: (1) ID of the clump; (2) number of the velocity component; (3) integrated intensity of the $^{12}$CO(1$-$0) line; (4) central velocity of the $^{12}$CO(1$-$0) line; (5) $^{12}$CO(1$-$0) linewidth; (6) $^{12}$CO(1$-$0) peak main beam brightness temperature; (7,8,9,10) the same line parameters for $^{13}$CO(1$-$0).}
		\label{lineparam1}
\end{table*}

We selected 26 fields from the GCC sample, where no high spectral resolution observations of $^{12}$CO(1$-$0) and $^{13}$CO(1$-$0) emission were made before. The molecular emission was observed at the positions of the \textit{Herschel}-based column density maxima using a five point cross and a single pointing, respectively. The goal of this paper is to extend the statistical characterization of the Herschel-detected clumps and cores using this new molecular line survey. We analyse the correlation between dust and molecular emission, determine the kinematic distances of individual clumps, estimate excitation temperatures, column densities and masses, and assess the stability of the clumps.

\section{Observations and data analysis}
\label{observ}

\subsection{The selected clumps}
\label{clumpselect}

We have selected 35 clumps in 26 \textit{Herschel} GCC fields where no high spectral resolution molecular line observations were performed before. Each of our fields includes one or two PGCCs that often break up into several clumps embedded in cometary or filamentary structures in the far-infrared. Our clumps were defined as areas in the $Herschel$-based column density maps with a peak H$_2$ column density of $N$(H$_2$)$_{\rm dust}$\,>\,10$^{21}$\,cm$^{-2}$ (A$_{\rm V}$\,>\,1.4). The fields are located between -30$\degree$ and 10$\degree$ Galactic latitudes, but not closer than approximately 2$\degree$ to the Galactic plane. Seven ECCs coinciding with one or two of our clumps were observed during earlier surveys \citep{wu2012,zhang2016} but the observations discussed in this paper have clear advantages. The previous measurements were based on $Planck$ maps and might have missed the actual clumps or cores, while this sample is based on $Herschel$ observations that have higher spatial resolution that ensures accurate pointing at the column density maxima. Our selection is not restricted to certain star-forming regions, we sample many different environments throughout the Galaxy. Due to higher spectral and spatial resolution we can compare the derived parameters with observations of the dust continuum by $Herschel$. Together with the work by \citet{parikka2015} the survey of the main clumps in all GCC fields observable from the Onsala Space Observatory is completed.

We note that many of our selected objects may be in fact much larger than traditional gravity-bound cores, since depending on the distance our resolution will correspond to larger structures. \citet{bergin2007} defined clumps as dense condensations with masses of 50$-$500\,M$_{\odot}$, sizes of 0.3$-$3\,pc, and densities of 10$^3-$10$^4$\,cm$^{-3}$. Cores have masses of 0.5$-$5\,M$_{\odot}$, sizes of 0.03$-$0.2\,pc and an order of magnitude larger densities. Some of our objects might be individual cores, others larger clumps containing unresolved cores. In this paper all of our sources are referred to as clumps.

\subsection{Molecular line observations}
\label{codata}

The $^{12}$CO(1$-$0) and $^{13}$CO(1$-$0) observations were carried out in January 14-16, 2014, using the 20-m telescope of the Onsala Space Observatory (OSO) in Sweden. The observations were completed using a SIS-mixer and two correlators simultaneously: the RCC (Radio Camera Correlator) and the FFTS (Fast Fourier Transform Spectrometer). The RCC was used with a bandwidth of 40\,MHz and a spectral resolution of 25\,kHz and the FFTS with a bandwidth of 100\,MHz and a spectral resolution of 12.2\,kHz. Both instruments were centred on 115.271\,GHz to observe $^{12}$CO(1$-$0) and on 110.201\,GHz to observe $^{13}$CO(1$-$0). The FFTS also measured two polarisation directions simultaneously.

The centre position of each clump in our sample was defined as the position of the local peak on the \textit{Herschel}-based column density map (see Sect. \ref{herscheldata}) with an accuracy of 10\,$\arcsec$. We first observed a single $^{12}$CO(1$-$0) spectrum at the centre of the clumps, then completed a 5-point cross through them with a spacing of 33\,$\arcsec$ (the half power beamwidth on this frequency). The $^{12}$CO(1$-$0) observations were made with position switching mode (PSW) and their reference positions were chosen by selecting low flux areas on \textit{IRAS} 100\,$\mu$m maps inside the 1$\degree$ radius of the centre of the clumps. The resulting spectrum from the PSW observations was an on-off spectrum, which is the reference position spectrum subtracted from the spectrum observed on target. After this a single $^{13}$CO(1$-$0) spectrum towards the clump centre was observed with frequency switching mode (FSW) with a frequency throw of 7.5\,MHz. The integration time was chosen to result in an antenna temperature rms noise of 0.3\,K on the $^{12}$CO(1$-$0) spectra and 0.1\,K on the $^{13}$CO(1$-$0) spectra. The typical integration time was 8 minutes/position. The telescope has a pointing accuracy of 3$\arcsec$. The pointing and focus was regularly checked during the observation with bright SiO masers (R Cas, TX Cam, S UMi).

Before measuring the PSW spectra on the targets, each reference position was observed by taking a single FSW spectrum to check for contamination. If a line was found in the reference spectrum, it was then added to the observed spectrum at the target. This correction occasionally resulted in other lines appearing at separate velocities on the final spectra, originating from CO emission on the reference positions (e.g. emission around 26.3\,kms$^{-1}$ in G37.49-A). We exclude these lines from our analysis. Furthermore, both the PSW and FSW spectra were checked for telluric lines but no contamination was found.

The data were reduced with the software package GILDAS CLASS\footnote{http://www.iram.fr/IRAMFR/GILDAS} (version: aug15b). The calibration was performed with the chopper wheel method, and the conversion of antenna temperature to main beam brightness temperature ($T\mathrm{_{MB}}$) was made dividing the spectra with the main beam efficiency at 115\,GHz measured at the time of each observation (this value was in the range of 0.33-0.47, depending on the observing elevation). The baselines of all $^{12}$CO(1$-$0) PSW spectra were modelled with second or third order polynomials and subtracted. The two polarisation directions measured with the FFTS were averaged together weighted by noise, then the spectra measured on each of the 5 positions of the 5-point cross were averaged together weighted by noise. The $^{13}$CO(1$-$0) FSW spectra were first folded, the baselines were modelled with second or third order polynomials and subtracted. The measurements on each position were averaged together weighted by noise. Due to its higher spectral resolution and problems with the response of the RCC during some measurements, only the data from the FFTS instrument are presented in this paper. 

We observed 35 clumps in 26 \textit{Herschel} GCC fields in total; we name the clumps on a field with the abbreviated ID of the GCC and an A, B or C affix, from the one with the highest peak $N$(H$_2$)$_{\rm dust}$ to the lowest. $^{13}$CO(1$-$0) emission was measured in 35 clumps and $^{12}$CO emission was measured in only 28 clumps due to time constraints. In 16 fields all the clumps with peak $N$(H$_2$)$_{\rm dust}$ above the selected 10$^{21}$\,cm$^{-2}$ column density threshold were observed in one or both transitions and in 10 fields only the clump with the highest peak $N$(H$_2$)$_{\rm dust}$ has measurements. Two clumps were dropped because of emission found on the reference position and a missing FSW observation on that position. Table \ref{sources} lists the observed clumps.

\subsection{Analysis of the molecular line data}

\subsubsection{Temperature and density calculations}
\label{cocalc}

The $^{12}$CO(1$-$0) and $^{13}$CO(1$-$0) lines at each observed position were fitted with a Gaussian line profile to obtain the peak main beam brightness temperature $T\mathrm{_{MB}}$, the line centre velocity in the Local Standard of Rest frame $v\mathrm{_{LSR}}$, and the linewidth $\Delta v$. The calculation of the excitation temperature $T\mathrm{_{ex}}$ and the $^{13}$CO column density $N$($^{13}$CO) at the centre of each clump were performed according to the method described by \citet{rohlfs1996}. 

\begin{table*}[t]
	\centering
	\footnotesize
		\caption{The calculated physical parameters of the clumps.}
		\begin{tabular}{l l r r r r r}
		\hline
		\multicolumn{1}{c}{clump ID} & \multicolumn{1}{c}{L} & \multicolumn{1}{c}{$T\mathrm{_{ex}}$} & \multicolumn{1}{c}{$N$($\mathrm{^{13}CO})\mathrm{_{l}}$} & \multicolumn{1}{c}{$N$($\mathrm{^{13}CO})\mathrm{_{u}}$} & \multicolumn{1}{c}{$T\mathrm{_{dust}}$} & \multicolumn{1}{c}{$N$(H$_2$)$\mathrm{_{dust}}$} \\
	\multicolumn{1}{c}{\ } & \multicolumn{1}{c}{\ } &  \multicolumn{1}{c}{[K]} & \multicolumn{2}{c}{[10$^{15}$ cm$^{-2}$]} & \multicolumn{1}{c}{[K]} & \multicolumn{1}{c}{[10$^{21}$ cm$^{-2}$]} \\
		\hline
		\hline
G26.34-A & 1 &  8.6 $\pm$ 1.4 &  2.5 $\pm$ 0.1 &  3.3 $\pm$ 0.2 & 12.8 $\pm$ 0.6 &  8.8 $\pm$  1.5 \\
G37.49-A & 1 & 16.1 $\pm$ 1.0 &  5.9 $\pm$ 0.2 &  7.7 $\pm$ 0.3 & 15.1 $\pm$ 0.9 &  7.7 $\pm$  1.5 \\
G37.49-B & 1 &  ... &  2.6 $\pm$ 0.2 &  4.0 $\pm$ 0.4 & 15.3 $\pm$ 0.9 &  5.5 $\pm$  1.0 \\
G37.91-A & 1 & 12.0 $\pm$ 0.9 &  7.5 $\pm$ 0.2 & 11.0 $\pm$ 0.5 & 14.0 $\pm$ 0.8 & 14.9 $\pm$  2.7 \\
G37.91-B & 1 &  ... &  4.9 $\pm$ 0.2 &  6.8 $\pm$ 0.4 & 14.7 $\pm$ 0.9 & 10.8 $\pm$  2.0 \\
G39.65-A & 1 &  ... &  ... & 36.9 $\pm$ 0.6 & 14.1 $\pm$ 0.8 & 22.8 $\pm$  4.2 \\
G39.65-B & 1 &  ... & 14.3 $\pm$ 0.1 & 18.9 $\pm$ 0.2 & \multirow{2}{*}{14.0 $\pm$ 0.8} & \multirow{2}{*}{21.8 $\pm$  3.9} \\
G39.65-B & 2 &  ... & 13.2 $\pm$ 0.1 & 16.0 $\pm$ 0.2 & &  \\
G62.16-A & 1 &  ... &  1.6 $\pm$ 0.2 &  2.7 $\pm$ 0.3 & 15.8 $\pm$ 1.0 &  3.0 $\pm$  0.6 \\
G69.57-A & 1 & 11.4 $\pm$ 0.8 & 16.3 $\pm$ 0.3 & 21.7 $\pm$ 0.6 & 13.1 $\pm$ 0.7 & 33.1 $\pm$  6.0 \\
G69.57-B & 1 & 14.4 $\pm$ 0.6 & 17.0 $\pm$ 0.2 & 19.2 $\pm$ 0.4 & \multirow{2}{*}{13.9 $\pm$ 0.8} & \multirow{2}{*}{22.3 $\pm$  4.1} \\
G69.57-B & 2 & 13.0 $\pm$ 0.6 &  3.0 $\pm$ 0.1 &  4.6 $\pm$ 0.3 &  &  \\
G70.10-A & 1 & 19.5 $\pm$ 0.5 &  ... & 38.4 $\pm$ 0.6 & 12.8 $\pm$ 0.6 & 29.3 $\pm$  5.0 \\
G70.10-B & 1 &  ... & 17.8 $\pm$ 0.2 & 22.6 $\pm$ 0.4 & 13.0 $\pm$ 0.7 & 23.3 $\pm$  4.1 \\
G71.27-A & 1 & 12.2 $\pm$ 1.0 &  0.5 $\pm$ 0.1 &  0.9 $\pm$ 0.2 & 14.9 $\pm$ 0.9 &  0.8 $\pm$  0.2 \\
G91.09-A & 1 & 11.4 $\pm$ 1.1 &  2.2 $\pm$ 0.2 &  3.7 $\pm$ 0.3 & 14.2 $\pm$ 0.8 &  1.3 $\pm$  0.2 \\
G95.76-A & 1 &  9.8 $\pm$ 0.9 &  5.8 $\pm$ 0.3 &  8.8 $\pm$ 0.6 & 12.1 $\pm$ 0.5 & 13.1 $\pm$  2.2 \\
G95.76-B & 1 & 12.2 $\pm$ 0.7 &  6.8 $\pm$ 0.3 & 10.1 $\pm$ 0.6 & 13.4 $\pm$ 0.7 &  5.6 $\pm$  1.0 \\
G109.18-A & 1 & 10.8 $\pm$ 1.0 &  0.9 $\pm$ 0.1 &  1.5 $\pm$ 0.2 & 15.4 $\pm$ 1.0 &  0.6 $\pm$  0.1 \\
G110.62-A & 1 & 13.8 $\pm$ 0.5 &  ... &  7.4 $\pm$ 0.2 & 13.4 $\pm$ 0.7 &  5.9 $\pm$  1.1 \\
G115.93-A & 1 & 15.9 $\pm$ 1.0 & 14.0 $\pm$ 0.2 & 13.7 $\pm$ 0.5 & 14.1 $\pm$ 0.8 &  4.7 $\pm$  0.8 \\
G115.93-B & 1 &  ... &  6.4 $\pm$ 0.2 &  5.7 $\pm$ 0.3 & 13.4 $\pm$ 0.7 &  4.2 $\pm$  0.8 \\
G116.08-A & 1 & 10.1 $\pm$ 0.7 &  4.6 $\pm$ 0.3 &  6.8 $\pm$ 0.5 & 13.3 $\pm$ 0.7 &  9.3 $\pm$  1.7 \\
G126.24-A & 1 &  9.5 $\pm$ 2.7 &  0.6 $\pm$ 0.1 &  1.1 $\pm$ 0.2 & 15.1 $\pm$ 0.9 &  1.5 $\pm$  0.3 \\
G139.60-A & 1 & 31.8 $\pm$ 0.3 &  ... & 43.9 $\pm$ 0.9 & 13.3 $\pm$ 0.7 & 19.6 $\pm$  3.6 \\
G141.25-A & 1 &  9.8 $\pm$ 0.7 &  0.8 $\pm$ 0.1 &  1.3 $\pm$ 0.2 & 14.9 $\pm$ 0.9 &  0.8 $\pm$  0.1 \\
G159.12-A & 1 & 17.4 $\pm$ 0.9 &  3.1 $\pm$ 0.1 &  3.7 $\pm$ 0.1 & 14.5 $\pm$ 0.8 &  2.5 $\pm$  0.5 \\
G159.23-A & 1 & 11.5 $\pm$ 1.0 &  ...  & 11.7 $\pm$ 0.6 & 12.5 $\pm$ 0.6 & 14.5 $\pm$  2.6 \\
G159.23-C & 1 & 13.1 $\pm$ 0.7 &  8.8 $\pm$ 0.2 &  8.3 $\pm$ 0.3 & 14.0 $\pm$ 0.8 &  4.4 $\pm$  0.8 \\
G171.35-A & 1 & 13.4 $\pm$ 0.7 &  0.6 $\pm$ 0.1 &  1.1 $\pm$ 0.2 & 15.3 $\pm$ 1.0 &  1.1 $\pm$  0.2 \\
G174.22-A & 1 & 13.4 $\pm$ 0.8 & 15.6 $\pm$ 0.2 & 15.2 $\pm$ 0.3 & 13.1 $\pm$ 0.7 & 15.2 $\pm$  2.8 \\
G174.22-B & 1 &  9.2 $\pm$ 0.9 &  7.0 $\pm$ 0.2 &  9.8 $\pm$ 0.4 & 12.4 $\pm$ 0.6 & 11.0 $\pm$  1.9 \\
G188.24-A & 1 & 14.2 $\pm$ 0.7 &  0.7 $\pm$ 0.1 &  1.3 $\pm$ 0.2 & \multirow{2}{*}{15.2 $\pm$ 0.9} &  \multirow{2}{*}{2.3 $\pm$  0.4} \\
G188.24-A & 2 & 12.7 $\pm$ 0.8 &  1.4 $\pm$ 0.1 &  2.2 $\pm$ 0.2 &  &   \\
G189.51-A & 1 & 10.0 $\pm$ 1.0 &  3.3 $\pm$ 0.2 &  4.8 $\pm$ 0.4 & 13.5 $\pm$ 0.7 &  4.2 $\pm$  0.7 \\
G195.74-A & 1 & 23.4 $\pm$ 0.4 &  ... & 36.1 $\pm$ 0.6 & 14.3 $\pm$ 0.8 & 40.2 $\pm$  7.3 \\
G203.42-A & 1 & 13.2 $\pm$ 1.1 & 13.3 $\pm$ 0.2 &  8.8 $\pm$ 0.4 & 12.1 $\pm$ 0.6 &  7.9 $\pm$  1.4 \\
G205.06-A & 1 & 13.0 $\pm$ 1.6 & 11.3 $\pm$ 0.3 &  9.7 $\pm$ 0.5 & 12.9 $\pm$ 0.6 &  6.6 $\pm$  1.2 \\
  \hline
		\end{tabular}
        \tablefoot{The columns are: (1) ID of the clump; (2) number of the velocity component; (3) excitation temperature; (4,5) $N$($^{13}$CO) lower and upper limits; (6,7) \textit{Herschel}-based dust colour temperature and H$_2$ column density. }
	\label{nt}
\end{table*}

Expressing the radiative transfer equation with the measured main beam brightness temperature, we obtain
\begin{equation}
T\mathrm{_{MB}}(\nu) = T\mathrm{_0}\left(\frac{1}{e^{T\mathrm{_0}/T\mathrm{_{ex}}}-1}-\frac{1}{e^{T\mathrm{_0}/2.7}-1}\right)(1-e^{-\tau_{\nu}}),
\end{equation}
where $T_0$\,=\,$h\nu$/$k\mathrm{_B}$ and we assume a beam filling factor of 1 since the sources are extended clouds. Assuming that the $^{12}$CO(1$-$0) emission is optically thick, the excitation temperature can be calculated with
\begin{equation}
T\mathrm{_{ex}}=\frac{5.5\,\mathrm{K}}{\ln \left[1+\left(\frac{5.5\,\mathrm{K}}{T\mathrm{_{MB}}^{^{12}\mathrm{CO}}+0.82\,\mathrm{K}}\right)\right]}
\label{co_tkin}
\end{equation}
where  5.5\,K\,=\,$h\nu$($^{12}$CO)/$k\mathrm{_B}$. In the case of a local thermodynamic equilibrium (LTE) and an isothermal medium (assuming that $T\mathrm{_{ex}}$ is uniform for all molecules along the line of sight in the J\,=\,1$-$0 transition and for all different isotopic species, and that the $^{12}$CO(1$-$0) and $^{13}$CO(1$-$0) lines are emitted from the same volume) and if the $^{13}$CO(1$-$0) emission is optically thin, the $\tau_{13}$ optical depth of $^{13}$CO(1$-$0) can be derived using
\begin{equation}
\tau_{13}=-\ln\left[1-\frac{T\mathrm{_{MB}}^{^{13}\mathrm{CO}}/5.3\,\mathrm{K}}{1/(e^{5.3\,\mathrm{K}/T\mathrm{_{ex}}}-1)-0.16}\right].
\end{equation}
The total column density of $^{13}$CO, $N$($^{13}$CO) can be calculated as
\begin{equation}
N({^{13}\mathrm{CO}})=\left[\frac{\tau_{13}}{1-e^{-\tau_{13}}}\right]\,3\times10^{14}\frac{W^{^{13}\mathrm{CO}}}{1-e^{-5.3/T\mathrm{_{ex}}}}.
\label{cocoldens}
\end{equation}
where $W^{\rm ^{13}CO}$ is the integrated intensity of the $^{13}$CO(1$-$0) line. Both $T\mathrm{_{ex}}$ and $N$($^{13}$CO) were calculated from the fitted line parameters at the centre of each clump for all observed line components. Since the excitation temperatures of most of our clumps vary between 8.5\,K and 19.5\,K (see Sect. \ref{res:physpar}), we use these values to calculate the lower and upper limits of the $^{13}$CO column density even for those clumps where we did not have $^{12}$CO observations thus could not calculate $T_{\rm ex}$. We adopt a [$^{13}$CO]/[H$_2$] relative abundance of 10$^{-6}$ found by \citet{parikka2015} and calculate the line-based hydrogen column densities $N$(H$_2$)$_{\rm gas}$ at the centre of the clumps.

\begin{figure}[t]
	\centering
		\includegraphics[width=\linewidth]{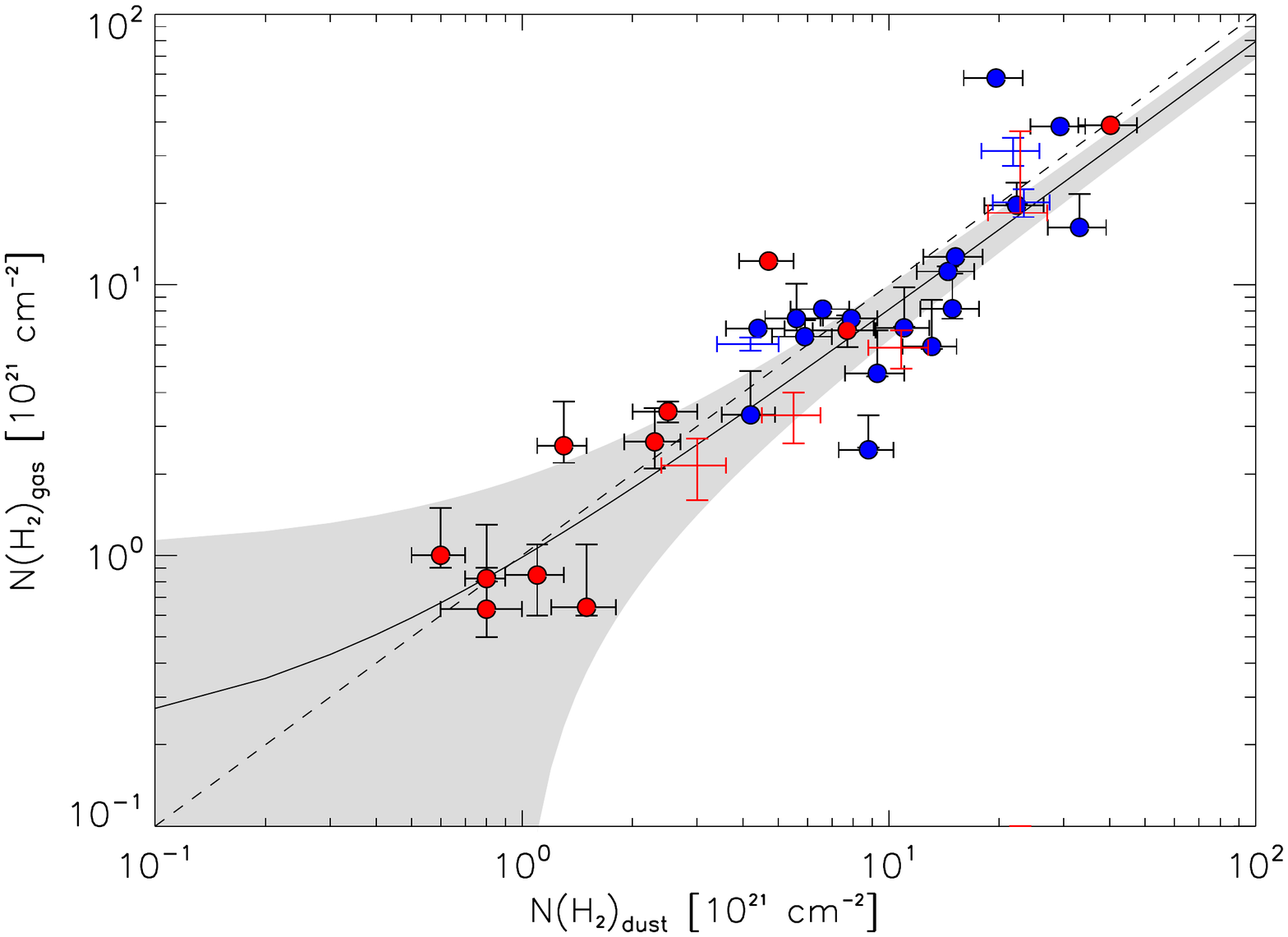}
	\caption{Correlation of $N$(H$_2$) calculated from dust continuum and from $^{13}$CO(1$-$0) using a $^{13}$CO abundance of 10$^{-6}$. The circles mark the column densities derived from the excitation temperature estimates (where $^{12}$CO measurements were available). The vertical error bars show $N$(H$_2$)$_{\rm gas}$ lower and upper limits calculated using 8.5 and 19.5\,K as excitation temperature and the horizontal error bars are from the uncertainty of the $\tau_{250}$ values in the centre of the clumps. The colours indicate the dust colour temperature: blue is below 14\,K and red is above 14\,K. Where only lower and upper limits of $N$(H$_2$)$_{\rm gas}$ could be calculated only error bars are shown with the colour code. The dashed line indicates $N$(H$_2$)$\mathrm{_{dust}}$\,=\,$N$(H$_2$)$\mathrm{_{gas}}$, the solid line is the linear fit to the plotted values and the shaded area shows the 1$\sigma$uncertainty range of this fit.}
	\label{NvsN}
\end{figure}

Although we assume optically thin $^{13}$CO emission, the optical depth of the lines can reach and even exceed unity in some clumps. This results in systematically lower $^{13}$CO column density estimates. We investigated this with radiative transfer simulations using a series of spherically symmetric Bonnor-Ebert models. Assuming a kinetic temperature of 12\,K, we derive $^{12}$CO and $^{13}$CO line profiles for model clouds that have $^{13}$CO column densities and turbulent linewidths similar to the observed clumps. We estimate column densities from the synthetic observations and compared those with the actual values in the models. We find that when $\tau_{13}$ is close to unity, we underestimate the true column density by 20-40\%, but the error is not likely to exceed 50\% even when the optical depth is around 3. See the details of this modeling in Appendix \ref{model11}.

\subsubsection{Virial mass calculation}

The virial mass of the clumps $M_{\rm vir}$ was determined using the equation from \citet{maclaren1988}
\begin{equation}
M\mathrm{_{vir}} [\mathrm{M_{\odot}}]=k_2\,R [\mathrm{pc}]\,\Delta v_{\rm H_2}^2 [\mathrm{kms^{-1}}],
\end{equation}
where $R$ is the effective radius of the clump (the root mean square of the semi-major and semi-minor axes of the 2D Gaussian fitted to it, see Sect. \ref{herscheldata}), $\Delta v_{\rm H_2}$ is the FWHM linewidth calculated from the total velocity dispersion of H$_2$ and $k_2$\,=\,168 assuming Gaussian velocity distribution and that the density varies with the radius as $\rho$ $\propto$ $R^{-1.5}$. 

The total velocity dispersion of H$_2$ was calculated by adding the turbulent velocity dispersion of the $^{13}$CO lines to the thermal velocity dispersion of the hydrogen molecules as
\begin{equation}
\sigma_{\rm H_2}=\sqrt{\frac{k_\mathrm{B} T_{\rm kin}}{m_\mathrm{H_2}}+\left(\frac{\Delta v_\mathrm{{^{13}CO}}^2}{8\ln2}-\frac{k_\mathrm{B} T_{\rm kin}}{m_\mathrm{^{13}CO}}\right)},
\label{h2disp}
\end{equation}
where $T_{\rm kin}$ is the kinetic temperature in the clump and $m_{\mathrm{H_2}}$ and $m_{\rm ^{13}CO}$ are the masses of a H$_2$ and a $^{13}$CO molecule. The velocity dispersion of H$_2$ is then converted to FWHM linewidth by multiplying it with $\sqrt{8\ln2}$.

The temperature in Eq. \ref{h2disp} is the kinetic temperature which does not necessarily equal to $T_{\rm ex}$ in our objects. The two temperatures are equal if the densities are high enough so that LTE conditions hold. However, the two temperatures should generally vary within a similar interval, thus we adopt the same limiting temperatures as before (8.5\,K and 19.5\,K) and calculate $M_{\rm vir}$ lower and upper limits. We discuss the relationship between kinetic and excitation temperatures further in Sect. \ref{disc:tempdens}.

The linewidth uncertainty results in virial mass errors of less than 1\%, but optical depth effects of $^{13}$CO can lead to overestimating the velocity dispersion and thus the virial masses. From our radiative transfer modeling (see Appendix \ref{modeling1}) we determined that when $\tau_{13}$ is around 3 or less, the difference between the observed and true FWHM values is 30\% or less. Thus the effect on virial masses is no more than a factor of two. The uncertainty in the distance of the clump also affects the calculated virial masses linearly (see Sect. \ref{clumpstab}).

\subsection{\textit{Herschel} observations}
\label{herscheldata}

The selected \textit{Herschel} GCC fields were mapped with the \textit{Herschel SPIRE} instrument at 250, 350 and 500 $\mu$m in November-December 2009 and May 2011. The observations were reduced with the \textit{Herschel} Interactive Processing Environment\footnote{http://herschel.esac.esa.int/hipe} (HIPE) v.12.0 using the official pipeline and the absolute zero point of the intensity scale was determined using \textit{Planck} maps complemented by the IRIS version of the \textit{IRAS} 100\,$\mu$m data, as described by \citet{juvela2012}.
The resolution of the \textit{SPIRE} maps is 18\,$\arcsec$, 25\,$\arcsec$ and 37\,$\arcsec$ at 250, 350 and 500\,$\mu$m, respectively. The calibration accuracy of the \textit{Herschel SPIRE} data is expected to be better than 7\%. For our cold sources, spectral energy distribution (SED) fitting to the SPIRE bands 250-500\,$\mu$m is sufficient to determine the colour temperature to an accuracy better than 1\,K \citep{juvela2012}, corresponding to $\sim$20\% column density
uncertainty at 15\,K. For warmer sources, the shorter wavelength data would be necessary but in the case of cold sources the addition of $PACS$ data may bias the column density estimates \citep{shetty2009a, shetty2009b, malinen2011,juvela2013}. Our method works well up to 20\,K dust temperatures.

The \textit{Herschel SPIRE} 250 and 350\,$\mu$m intensity maps were first convolved to the resolution of the 500\,$\mu$m map (37$\arcsec$). To obtain $T\mathrm{_{dust}}$ we fitted the SED at these three wavelengths in each pixel with a modified blackbody function
\begin{equation}
I_{\nu} \propto B_{\nu}(T\mathrm{_{dust}})\nu^{\beta},
\end{equation}
where $I_{\nu}$ is the intensity at a frequency, $B$($T\mathrm{_{dust}}$) is the Planck function and the dust spectral index $\beta$ had a value of 2, which is consistent with observations of dense clumps \citep{juvela2015b, juvela2015a}. The dust optical depth $\tau_{\nu_0}$ was calculated using the formula
\begin{equation}
I_{\nu_0}=B_{\nu_0}(T\mathrm{_{dust}})(1-e^{-\tau_{\nu_0}}) \approx B(T\mathrm{_{dust}})\times\tau_{\nu_0},
\end{equation}
where $I_{\nu_0}$ is the observed intensity at $\nu_0$\,=\,1\,200\,GHz or $\lambda_0$\,= 250\,$\mu$m. The equation assumes that the emission is optically thin in the far infrared. For our sources, the estimated 250\,$\mu$m optical depth is at most of the order of 10$^{-2}$. We note that analysis of thermal dust emission can lead to biased column density estimates as studied and quantified for example by \citet{malinen2011, juvela2013, juvela2015a, pagani2015} and \citet{steinacker2016}. Assuming an error of a factor of two, the 250\,$\mu$m optical depth could be as high as 0.01, which is still optically thin.

When converting optical depth to column density the sub-millimetre dust opacity is likely to be a large source of uncertainty, because the values may increase by a factor of a few from diffuse clouds to dense cores. Therefore, to make the conversion between submillimetre opacity and total gas mass, we make use of the new empirical result by \citet{juvela2015b}, where the ratio $\tau_{250}$/$\tau\mathrm{_J}$, where $\tau_{\rm J}$ is the optical depth in the J-band, was found to be on average 1.6\,$\times$\,10$^{-3}$ in dense clumps. The ratio between $\tau\mathrm{_J}$ and total gas mass is a well-characterised quantity that (unlike sub-millimetre opacity) does not significantly vary from region to region. 
Thus the $N$(H$_2$) hydrogen column density is calculated using the ratio $\tau_{\rm J}$/$N$(H)\,=\,1.994\,$\times$\,10$^{-22}$ for an extinction curve with $R_{\rm V}$\,=\,5.5 \citep{draine2003}. The difference in the column density values when using $R_V$\,=\,3.1 is less than 30\%.

For the mass calculation we first fitted 2D Gaussian functions to the clumps on the $\tau_{250}$ maps to determine their size and orientation. The resulting parameters were the maximum $\tau_{250}$ value, the standard deviations of the fitted 2D Gaussian $\sigma\mathrm{_x}$ and $\sigma\mathrm{_y}$, and the position angle PA (the angle between the celestial Equator and the major axis of the clump, counted counter-clockwise). The total number of H$_2$ molecules, $c_{\rm H_2}$ in the clumps was calculated as
\begin{equation}
c_{H_2}=2\pi\frac{\mathrm{FWHM_x}\mathrm{FWHM_y}}{\sqrt{8\ln2}}N(H_2)\mathrm{_{max}},
\end{equation}
where $N$(H$_2$)$_{\rm max}$ is the peak column density of the clump from the 2D Gaussian fitting and $FWHM_{\rm x}$ and $FWHM_{\rm y}$ are the full-width at half-maximum sizes of the clump and can be derived from the standard deviation with $FWHM$\,=\,$\sqrt{8\ln2}\sigma$. The FWHM sizes were converted from arcminutes to centimeters, thus accounting for the distances of the clumps. 
The mass of each clump $M_{\rm clump}$ was then calculated as
\begin{equation}
M\mathrm{_{clump}}=c_{H_2}m\mathrm{_H}\mu,
\end{equation}
where $m\mathrm{_H}\mu$ with $\mu$\,=\,2.8 is the mean molecular mass per H$_2$ molecule. As we later discuss the dominant error in $M_{\rm clump}$ originates from the uncertainties of the clump distances and not the uncertainty of $\tau_{250}$.

\section{Results}

\begin{table}[t]
	\centering
	\caption{The results of the 2D Gaussian fit to the continuum maps of the clumps.}
	\resizebox{.45\textwidth}{!}{
		\begin{tabular}{l r r r r r r r r}
		\hline
		\multicolumn{1}{c}{clump ID} & \multicolumn{1}{c}{FWHM$\mathrm{_{x}}$} & \multicolumn{1}{c}{FWHM$\mathrm{_{y}}$} & \multicolumn{1}{c}{PA} & \multicolumn{1}{c}{$N$(H$_2$)$\mathrm{_{max}}$} \\
	\multicolumn{1}{c}{\ } & \multicolumn{1}{c}{[arcmin]} & \multicolumn{1}{c}{[arcmin]} & \multicolumn{1}{c}{[deg]} & \multicolumn{1}{c}{[10$^{21}$\,cm$^{-2}$]} \\
		\hline
		\hline
G26.34-A & 1.79 $\pm$ 0.05 & 1.70 $\pm$ 0.04 &  44.6 $\pm$  20.3 &  6.46 $\pm$  0.21 \\
G37.49-A & 2.88 $\pm$ 0.27 & 2.01 $\pm$ 0.12 & 114.9 $\pm$   4.9 &  3.69 $\pm$  0.16 \\
G37.49-B & 1.54 $\pm$ 0.18 & 1.12 $\pm$ 0.12 &  83.9 $\pm$  10.6 &  2.31 $\pm$  0.27 \\
G37.91-A & 1.42 $\pm$ 0.29 & 0.89 $\pm$ 0.06 & 124.8 $\pm$   6.6 &  9.04 $\pm$  1.12 \\
G37.91-B & 1.95 $\pm$ 0.18 & 0.91 $\pm$ 0.06 & 112.8 $\pm$   3.1 &  5.22 $\pm$  0.42 \\
G39.65-A & 2.43 $\pm$ 0.12 & 1.03 $\pm$ 0.05 & 101.4 $\pm$   1.7 & 13.86 $\pm$  0.79 \\
G39.65-B & 1.52 $\pm$ 0.11 & 1.34 $\pm$ 0.19 &  77.3 $\pm$  26.4 & 10.95 $\pm$  0.81 \\
G39.65-B & 1.52 $\pm$ 0.11 & 1.34 $\pm$ 0.19 &  77.3 $\pm$  26.4 & 10.95 $\pm$  0.81 \\
G62.16-A & 2.30 $\pm$ 2.81 & 1.72 $\pm$ 1.20 &  20.1 $\pm$  31.6 &  0.59 $\pm$  0.49 \\
G69.57-A & 2.31 $\pm$ 0.10 & 1.25 $\pm$ 0.04 &  87.4 $\pm$   1.5 & 19.21 $\pm$  0.83 \\
G69.57-B & 1.38 $\pm$ 0.07 & 1.12 $\pm$ 0.05 & 129.5 $\pm$   7.6 & 14.88 $\pm$  0.90 \\
G69.57-B & 1.38 $\pm$ 0.07 & 1.12 $\pm$ 0.05 & 129.5 $\pm$   7.6 & 14.88 $\pm$  0.90 \\
G70.10-A & 4.52 $\pm$ 0.29 & 1.16 $\pm$ 0.04 & 168.8 $\pm$   0.8 & 17.00 $\pm$  0.70 \\
G70.10-B & 3.28 $\pm$ 0.21 & 1.12 $\pm$ 0.05 &  15.1 $\pm$   1.4 & 14.52 $\pm$  0.69 \\
G71.27-A & 4.55 $\pm$ 0.23 & 2.44 $\pm$ 0.08 & 147.2 $\pm$   1.5 &  0.57 $\pm$  0.01 \\
G91.09-A & 2.71 $\pm$ 0.15 & 1.56 $\pm$ 0.06 &  27.4 $\pm$   1.7 &  0.87 $\pm$  0.03 \\
G95.76-A & 2.16 $\pm$ 0.09 & 1.08 $\pm$ 0.03 &  83.0 $\pm$   1.6 &  9.64 $\pm$  0.38 \\
G95.76-B & 2.47 $\pm$ 0.12 & 1.35 $\pm$ 0.06 & 125.1 $\pm$   2.5 &  3.28 $\pm$  0.15 \\
G109.18-A & 3.22 $\pm$ 0.16 & 1.63 $\pm$ 0.10 & 179.3 $\pm$   2.3 &  0.29 $\pm$  0.01 \\
G110.62-A & 2.20 $\pm$ 0.07 & 1.37 $\pm$ 0.04 & 128.6 $\pm$   1.5 &  4.74 $\pm$  0.16 \\
G115.93-A & 4.42 $\pm$ 0.30 & 1.36 $\pm$ 0.06 &  42.5 $\pm$   0.9 &  2.73 $\pm$  0.10 \\
G115.93-B & 2.73 $\pm$ 0.09 & 1.90 $\pm$ 0.10 &  15.2 $\pm$   4.3 &  2.58 $\pm$  0.08 \\
G116.08-A & 1.91 $\pm$ 0.08 & 1.68 $\pm$ 0.06 &  91.3 $\pm$  10.0 &  5.73 $\pm$  0.21 \\
G126.24-A & 4.73 $\pm$ 9.75 & 1.60 $\pm$ 0.30 &  23.9 $\pm$   4.1 &  0.51 $\pm$  0.06 \\
G139.60-A & 1.77 $\pm$ 0.05 & 1.29 $\pm$ 0.04 &  93.3 $\pm$   2.7 & 15.80 $\pm$  0.58 \\
G141.25-A & 2.88 $\pm$ 0.28 & 0.84 $\pm$ 0.05 & 105.9 $\pm$   2.0 &  0.38 $\pm$  0.03 \\
G159.12-A & 2.48 $\pm$ 0.08 & 1.02 $\pm$ 0.03 &  50.4 $\pm$   0.9 &  1.86 $\pm$  0.07 \\
G159.23-A & 2.43 $\pm$ 0.18 & 1.24 $\pm$ 0.07 &  53.7 $\pm$   2.7 &  8.70 $\pm$  0.38 \\
G159.23-C & 2.65 $\pm$ 0.14 & 2.30 $\pm$ 0.11 & 131.3 $\pm$  10.4 &  2.52 $\pm$  0.08 \\
G171.35-A & 2.94 $\pm$ 2.45 & 1.61 $\pm$ 1.44 &  23.7 $\pm$  16.5 &  0.28 $\pm$  2.07 \\
G174.22-A & 3.21 $\pm$ 0.13 & 1.15 $\pm$ 0.04 & 106.7 $\pm$   0.8 & 10.86 $\pm$  0.42 \\
G174.22-B & 2.40 $\pm$ 0.10 & 1.12 $\pm$ 0.05 & 135.1 $\pm$   1.3 &  7.35 $\pm$  0.29 \\
G188.24-A & 1.87 $\pm$ 0.08 & 1.46 $\pm$ 0.05 &  37.5 $\pm$   6.3 &  1.47 $\pm$  0.07 \\
G188.24-A & 1.87 $\pm$ 0.08 & 1.46 $\pm$ 0.05 &  37.5 $\pm$   6.3 &  1.47 $\pm$  0.07 \\
G189.51-A & 1.36 $\pm$ 0.08 & 0.72 $\pm$ 0.04 & 174.1 $\pm$   2.9 &  3.01 $\pm$  0.22 \\
G195.74-A & 1.67 $\pm$ 0.06 & 1.15 $\pm$ 0.04 &  30.0 $\pm$   2.3 & 28.93 $\pm$  1.20 \\
G203.42-A & 1.52 $\pm$ 0.11 & 1.14 $\pm$ 0.17 & 153.3 $\pm$  12.7 &  5.59 $\pm$  0.42 \\
G205.06-A & 2.92 $\pm$ 0.12 & 1.48 $\pm$ 0.07 & 155.4 $\pm$   1.6 &  3.73 $\pm$  0.13 \\
  \hline
			\end{tabular}}
    \tablefoot{The columns are: (1) ID of the clump; (2,3) FWHM of the 2D Gaussian; (4) position angle of the 2D Gaussian; (5) maximum H$_2$ column density of the 2D Gaussian.}
	\label{gauss}
\end{table}

\subsection{Velocities and linewidths}

\label{res:vellinewidth}

Lines with two velocity components were found in 2 clumps, G69.57-B and G188.24-A, where both the $^{12}$CO(1$-$0) and $^{13}$CO(1$-$0) spectra showed two peaks. Additionally, G39.65-B which only has $^{13}$CO(1$-$0) measurements, shows a double peak as well. In other cases, such as G70.10-A or G110.62-A, the multiple peaks might be caused by the self-absorption of the emission. All the lines were detected with S/N\,>\,3 (at 0.07\,kms$^{-1}$ velocity resolution) in the centre of the clumps, except for G26.34-A, where S/N\,=\,1.8 for $^{12}$CO(1$-$0) and S/N\,=\,1.4 for $^{13}$CO(1$-$0), for G126.24-A, where S/N\,=\,2.3 for $^{13}$CO(1$-$0) and for G171.35-A, where S/N\,=\,2.5 for $^{13}$CO(1$-$0). The parameters of the Gaussian functions fitted to the lines in the centre of each clump are given in Table \ref{lineparam1} and the spectra of both CO species in the centre of each observed clump are plotted in Fig. \ref{spec1}--\ref{spec5} with their fitted Gaussian profiles. 

The $^{12}$CO(1$-$0) linewidths are in the 0.5$-$5.6\,kms$^{-1}$ range and the $^{13}$CO(1$-$0) linewidths in the 0.28$-$3.3\,kms$^{-1}$ range, similar to the values found by \citet{wu2012}. In most cases we observed the $^{12}$CO(1$-$0) and $^{13}$CO(1$-$0) lines in our clumps at similar $v_{\rm LSR}$ velocities as previous existing measurements detected them \citep{dame2001, wu2012}. For other clumps, the differences in the detected central velocity could be explained by the lower spectral resolution (2\,kms$^{-1}$) and sparser sampling (8.5$\arcmin$) of the previous measurements. For example, we observed two lines at $\approx$\,1.5 and 6.5\,kms$^{-1}$ towards G188.24-A, while \citet{dame2001} only detected one line around 0.7\,kms$^{-1}$. We did not detect the $-$16 and $-$9\,kms$^{-1}$ components observed by them towards G139.60-A only the velocity component around 35\,kms$^{-1}$. Towards the GCC field containing the clumps G37.91-A and G37.91-B they detected two $^{12}$CO(1$-$0) velocity components around 11\,kms$^{-1}$ and 30\,kms$^{-1}$. The higher velocity line component dominates their spectrum and they assume that it corresponds to background emission, while the other line corresponds to the cloud. However, we did not detect the 11\,kms$^{-1}$ line in our data, only the higher velocity component.

The five-point $^{12}$CO(1$-$0) maps were searched for systematic changes in the central velocities of the lines and in 13 clumps signs of a velocity gradient higher than 3 times our channel width was found across the size of the five-point map (66\,$\arcsec$). The real gradient can be calculated using the distances of the clumps (see Sect. \ref{clumpdist}). The highest velocity gradient, 4.25\,kms$^{-1}$pc$^{-1}$ in an east-west direction was measured in the clump G141.25-A. In G159.23-C a 2\,kms$^{-1}$pc$^{-1}$ velocity gradient can be observed in the east-west direction and the other clump in the same GCC field, G159.23-A also shows gradients towards the east (0.87\,kms$^{-1}$pc$^{-1}$) and north (0.27\,kms$^{-1}$pc$^{-1}$). The more distant G126.24-A has a gradient of 1\,kms$^{-1}$pc$^{-1}$ to the south.

\subsection{Physical parameters in the clumps}
\label{res:physpar}

Table \ref{nt} lists the physical properties of the gas at the centre of the clumps calculated with the equations in Sect. \ref{cocalc} and \ref{herscheldata}. The error bars of the line-based parameters were calculated by propagating the uncertainties of the Gaussian function fits of the lines and the uncertainties of the continuum-based parameters originate mainly from the calibration error of the \textit{SPIRE} maps and the modified blackbody function fit. The excitation temperatures of the clumps are between 8.5 and 19.5\,K, most clumps have temperatures between 10 and 15\,K and the coldest clumps are G26.34-A and G174.22-B with $\approx$\,9\,K. Two clumps, G139.60-A and G195.74-A have excitation temperatures of 23 and 32\,K but both clearly host a YSO as seen on the maps by \citet{montillaud2015}.

\begin{table*}[t]
	\centering
	\footnotesize
    	\caption{The kinematic distances of the clumps.}
		\begin{tabular}{l l l r r r r r r r r}
		\hline
\multicolumn{1}{c}{ID} & \multicolumn{1}{c}{L} & \multicolumn{1}{c}{F} & \multicolumn{1}{c}{D} & \multicolumn{1}{c}{D$\mathrm{_{kin}}$($v_{\rm s}$=0\,kms$^{-1}$)} & \multicolumn{1}{c}{D$\mathrm{_{kin}}$($v_{\rm s}$=-15\,kms$^{-1}$)} & \multicolumn{1}{c}{D$\mathrm{_{adopted}}$} \\
 & & & \multicolumn{1}{c}{[kpc]} & \multicolumn{1}{c}{[kpc]} & \multicolumn{1}{c}{[kpc]} & \multicolumn{1}{c}{[kpc]}\\
		\hline
		\hline
G26.34-A & 1 & 1 & 1.000 $\pm$ 0.300 & 0.52 $\pm$ 0.07 & 1.05 $\pm$ 0.07 & 1.000 $\pm$ 0.300 \\
G37.49-A & 1 & 0 & 0.800 $\pm$ 0.600 & 0.60 $\pm$ 0.06 & 1.22 $\pm$ 0.06 & 0.800 $\pm$ 0.600 \\
G37.49-B & 1 & 0 & 0.800 $\pm$ 0.600 & 0.62 $\pm$ 0.06 & 1.25 $\pm$ 0.06 & 0.800 $\pm$ 0.600 \\
G37.91-A & 1 & 1 & 1.060 $\pm$ 0.790 & 1.71 $\pm$ 0.06 & 2.34 $\pm$ 0.06 & 1.700 $\pm$ 0.700 \\
G37.91-B & 1 & 1 & 1.060 $\pm$ 0.790 & 1.76 $\pm$ 0.06 & 2.40 $\pm$ 0.06 & 1.700 $\pm$ 0.700 \\
G39.65-A & 1 & 1 & 1.500 $\pm$ 0.500 & 1.45 $\pm$ 0.06 & 2.11 $\pm$ 0.06 & 1.500 $\pm$ 0.500 \\
G39.65-B & 1 & 1 & 1.500 $\pm$ 0.500 & 1.37 $\pm$ 0.06 & 2.02 $\pm$ 0.06 & 1.500 $\pm$ 0.500 \\
G39.65-B & 2 & 1 & 1.500 $\pm$ 0.500 & 1.54 $\pm$ 0.06 & 2.20 $\pm$ 0.06 & 1.500 $\pm$ 0.500 \\
G62.16-A & 1 & 1 & 1.110 $\pm$ 0.350 & 0.33 $\pm$ 0.08 & 1.67 $\pm$ 0.12 & 1.110 $\pm$ 0.350 \\
G69.57-A & 1 & 1 & 1.780 $\pm$ 0.810 & 0.32 $\pm$ 0.11 & 2.90 $\pm$ 0.72 & 1.780 $\pm$ 0.810 \\
G69.57-B & 1 & 1 & 1.780 $\pm$ 0.810 & 0.17 $\pm$ 0.11 & 2.91 $\pm$ 0.72 & 1.780 $\pm$ 0.810 \\
G69.57-B & 2 & 1 & 1.780 $\pm$ 0.810 & 0.43 $\pm$ 0.12 & 2.91 $\pm$ 0.72 & 1.780 $\pm$ 0.810 \\
G70.10-A & 1 & 1 & 2.090 $\pm$ 0.830 & 0.19 $\pm$ 0.11 & 2.82 $\pm$ 0.73 & 2.090 $\pm$ 0.830 \\
G70.10-B & 1 & 1 & 2.090 $\pm$ 0.830 & 0.58 $\pm$ 0.12 & 2.86 $\pm$ 0.73 & 2.090 $\pm$ 0.830 \\
G71.27-A & 1 & - & ... & 5.74 $\pm$ 0.10* & 1.63 $\pm$ 0.30* & ... \\
G91.09-A & 1 & - & ... & 2.53 $\pm$ 0.12* & 0.40 $\pm$ 0.39* & ... \\
G95.76-A & 1 & 0 & 0.800 $\pm$ 0.100 & 1.71 $\pm$ 0.12 & ... & 0.800 $\pm$ 0.100 \\
G95.76-B & 1 & 0 & 0.800 $\pm$ 0.100 & 1.76 $\pm$ 0.12 & ... & 0.800 $\pm$ 0.100 \\
G109.18-A & 1 & 0 & 0.160 $\pm$ 0.160 & 1.08 $\pm$ 0.09 & ... & 0.620 $\pm$ 0.460 \\
G110.62-A & 1 & 1 & 0.440 $\pm$ 0.100 & 1.51 $\pm$ 0.08 & 0.33 $\pm$ 0.10 & 0.385 $\pm$ 0.160 \\
G115.93-A & 1 & 0 & 0.650 $\pm$ 0.500 & 1.00 $\pm$ 0.08 & ... & 0.650 $\pm$ 0.500 \\
G115.93-B & 1 & 0 & 0.650 $\pm$ 0.500 & 1.04 $\pm$ 0.09 & ... & 0.650 $\pm$ 0.500 \\
G116.08-A & 1 & 1 & 0.250 $\pm$ 0.050 & 0.77 $\pm$ 0.08 & ... & 0.250 $\pm$ 0.050 \\
G126.24-A & 1 & 1 & 1.000 $\pm$ 0.200 & 1.78 $\pm$ 0.08 & 0.96 $\pm$ 0.07 & 1.000 $\pm$ 0.200 \\
G139.60-A & 1 & 2 & 2.500 $\pm$ 0.500 & 3.14 $\pm$ 0.10 & 2.39 $\pm$ 0.10 & 2.500 $\pm$ 0.500 \\
G141.25-A & 1 & 1 & 0.110 $\pm$ 0.010 & 0.32 $\pm$ 0.07 & .... & 0.215 $\pm$ 0.115 \\
G159.12-A & 1 & 0 & 0.800 $\pm$ 0.800 & ... & ... & 0.800 $\pm$ 0.800 \\
G159.23-A & 1 & 2 & 0.325 $\pm$ 0.050 & 0.86 $\pm$ 0.12 & 0.39 $\pm$ 0.11 & 0.325 $\pm$ 0.050 \\
G159.23-C & 1 & 2 & 0.325 $\pm$ 0.050 & 0.41 $\pm$ 0.11 & ... & 0.325 $\pm$ 0.050 \\
G171.35-A & 1 & - & ... & ... & ... & ... \\
G174.22-A & 1 & 2 & 2.000 $\pm$ 0.400 & 72.6 $\pm$ 49.0* &	67.8 $\pm$ 46.3* & 2.000 $\pm$ 0.400 \\
G174.22-B & 1 & 2 & 2.000 $\pm$ 0.400 & 9.60 $\pm$ 1.60* & 8.53 $\pm$ 1.54* & 2.000 $\pm$ 0.400 \\
G188.24-A & 1 & 2 & 0.445 $\pm$ 0.050 & 0.86 $\pm$ 0.30* & 0.32 $\pm$ 0.28* & 0.445 $\pm$ 0.050 \\
G188.24-A & 2 & 2 & 0.445 $\pm$ 0.050 & 2.54 $\pm$ 0.42* & 1.90 $\pm$ 0.40* & 0.445 $\pm$ 0.050 \\
G189.51-A & 1 & 2 & 0.445 $\pm$ 0.050 & 3.19 $\pm$ 0.40* & 2.51 $\pm$ 0.38* & 0.445 $\pm$ 0.050 \\
G195.74-A & 1 & 1 & 1.000 $\pm$ 0.500 & 1.01 $\pm$ 0.16 & 0.44 $\pm$ 0.15 & 1.000 $\pm$ 0.500 \\
G203.42-A & 1 & 2 & 0.400 $\pm$ 0.100 & 1.67 $\pm$ 0.12 & 1.05 $\pm$ 0.12 & 0.400 $\pm$ 0.100 \\
G205.06-A & 1 & 2 & 0.400 $\pm$ 0.100 & 1.57 $\pm$ 0.12 & 0.95 $\pm$ 0.11 & 0.400 $\pm$ 0.100 \\
  \hline
		\end{tabular}
    \tablefoot{The columns are: (1) ID of the clump; (2) number of the velocity component; (3) distance reliability flag by \citet{montillaud2015}; (4) distance estimate by \citet{montillaud2015}; (5,6) kinematic distance calculated from our $^{13}$CO(1$-$0) data with source peculiar velocity $v_{\rm s}$=0 and -15\,kms$^{-1}$; (7) adopted distance value. Asterisk marks the kinematic distance results that we rejected according to Sect. \ref{clumpdist}.}
	\label{dist}
\end{table*}

The \textit{Herschel}-based dust colour temperatures are more uniform: all the clumps show $T_{\rm dust}$ values between 12 and 16\,K and the histogram peaks at 14\,K. The $T_{\rm dust}$ values do not correlate well with the derived excitation temperatures. The reason for this is that dust and gas in the interstellar matter are only coupled at volume densities above $\approx$\,10$^{5}$\,cm$^{-3}$ \citep{goldsmith2001}, which are typically not traced by our observed molecules. The peak $N$(H$_2$)$_{\rm dust}$ in the clumps have values from 0.6$-$40\,$\times$\,10$^{21}$\,cm$^{-2}$ but more than half of the clumps have column densities under 10$^{22}$\,cm$^{-2}$. The densest clumps are G195.74-A, G69.57-A and G70.10-A with peak $N$(H$_2$)$_{\rm dust}$\,$\geq$\,3\,$\times$\,10$^{22}$\,cm$^{-2}$. The \textit{Herschel}-derived $\tau_{250}$ and $T\mathrm{_{dust}}$ maps of our GCC fields are shown in Fig. \ref{herschel1}-\ref{herschel2}, where the 2D Gaussians fitted to the clumps are overplotted and the clump names are shown as well. The parameters of the fitted 2D Gaussians of the clumps are in Table \ref{gauss}. The sizes of the clumps on the plane of the sky have a median value of 10 square arcminutes with G71.27-A being the most extended with 34.8 arcmin$^2$ and G189.51-A the smallest with 3.1 arcmin$^2$. The centres of the fitted 2D Gaussian functions are usually very close to the peak position on the \textit{Herschel} $N$(H$_2$)$_{\rm dust}$ map. The difference between them is generally under 1.5\,$\arcsec$, except in the case of G37.49-A and G116.08-A where it is around 15$\arcsec$. The most elongated clumps are G70.10-A, G115.93-A and G141.25-A, while G26.34-A, G39.65-B and G116.08-A are nearly circular. Almost all our clumps are well resolved by the $Herschel$ observations or at least resolved in one direction and marginally resolved in the other direction (G141.25-A, G189.51-A). The peaks of the \textit{Herschel}-derived $N$(H$_2$)$_{\rm dust}$ maps coincide with the measured highest integrated intensity values on the $^{12}$CO(1$-$0) five-point maps only in 5 clumps (G37.49-A, G91.09-A, G95.76-A, G116.08-A, G139.60-A) but a more detailed analysis of the differences of the peaks cannot be made without larger spectral mapping.

The lower and upper limits of the $^{13}$CO column densities, $N$($^{13}$CO)$_{\rm l}$ and $N$($^{13}$CO)$_{\rm u}$ are between 0.5$-$44\,$\times$10$^{15}$\,cm$^{-2}$, resulting in N(H$_2$)$_{\rm gas}$ line-based column densities of 0.5$-$44\,$\times$10$^{21}$\,cm$^{-2}$ using a $^{13}$CO abundance of 10$^{-6}$. Fig. \ref{NvsN} shows the correlation of $N$(H$_2$)$_{\rm gas}$ with the \textit{Herschel}-based peak H$_2$ column densities. The error bars on $N$(H$_2$)$_{\rm gas}$ represent the interval between the calculated lower and upper limits of the value, while the circles mark the values calculated with the derived excitation temperatures, where possible. The uncertainties of $N$(H$_2$)$_{\rm dust}$ again come from the error of the $\tau_{250}$ map at the centre position of the clump. The figure shows a similar correlation to the one found by \citet{parikka2015}. The two values have a linear correlation coefficient of $r$\,=\,0.85 with a 95\% confidence interval of 0.72$-$0.92. We find a linear least-squares fit corresponding to a $^{13}$CO abundance of 0.8\,$\pm$\,0.1\,$\times$\,10$^{-6}$. Considering the uncertainties of the linear fit parameters this is close to the 1$-$1 correlation. However, as discussed before, optical depth effects of the $^{13}$CO emission and the variation of the $^{13}$CO relative abundances can shift the fitted line. The real scatter also might be smaller or larger than in the figure. The factors contributing to the uncertainty of gas- and dust-based H$_2$ column densities are summarized in Sect. \ref{disc:tempdens}. The correlation with $T\mathrm{_{dust}}$ is also shown in the figure with colour code. Clumps with $T\mathrm{_{dust}}$\,<\,14\,K (corresponding to the average dust temperature in our sample) appear only above $N$(H$_2$)$\mathrm{_{dust}}$\,=\,0.4\,$\times$\,10$^{22}$\,cm$^{-2}$, but some warmer clumps with $T\mathrm{_{dust}}$\,$\geq$\,14\,K have high column densities e.g. G195.74-A with $T\mathrm{_{dust}}$\,=\,14.3\,K has the highest peak $N$(H$_2$)$_{\rm dust}$ in the sample. 

\subsection{Clump distances}
\label{clumpdist}

The distances of more than one hundred GCC fields were provided by \citet{montillaud2015} using extinction method, kinematic distance calculation (based on the surveys of \citet{dame2001} and \citet{wu2012}) and association with molecular cloud complexes or star clusters. They included reliability flags with their results that indicate the level of confidence of the adopted distances: a flag equal to 2 indicates a good level of confidence due to the agreement of the used methods, a value of 1 indicates a reasonable estimate which needs to be confirmed independently and a value of 0 indicates unreliable estimates. They have distance estimates for all but 3 out of our 35 clumps: 9 clumps have distances with a reliability flag of 2, 15 clumps have distances with a flag of 1 and 8 clumps only have an estimate with a reliability flag of 0. Since the spatial and spectral resolution of our survey is higher than what was used by them, we also performed kinematic distance calculations to assess the confidence of the distance estimation of our clumps, to verify the previous results or provide more reliable numbers. We used the revised method of \citet{reid2009} and similarly to the work of \citet{montillaud2015} the kinematic distances were calculated both assuming a source peculiar velocity of 0\,kms$^{-1}$ and $-$15\,kms$^{-1}$ compared to the rotation of the Galaxy. The error in the $^{13}$CO $v_{\rm LSR}$ velocities was assumed to be a conservative value of 1\,kms$^{-1}$.

We chose the near solution versus the far solution where the two were not equal, since these galactic cold objects are expected to be mostly under 2\,kpc distances. This is supported by the fact that most of our clumps (except G91.09-A, G109.18-A and G141.25-A) can be seen in optical extinction maps. The three exceptions have a higher Galactic latitude (around 30$\degree$) therefore are likely to be close-by. The far solution can be also rejected in most cases since it generally would imply unrealistic Galactic altitudes, while the vertical scale height of the molecular material in the Galaxy is around 70$-$80\,pc \citep{bronfman1988, clemens1988}. Thus even the near kinematic distance estimates of G71.27-A and G91.09-A will result in very high Galactic altitudes. These distance results were rejected and not used further. The clumps G174.22-A, G174.22-B, G188.24-A and G189.51-A are located close to a Galactic longitude of 180$\degree$ where the velocity component due to the rotation of the Galaxy is the tangential component which cannot be measured with spectroscopy. Indeed the calculated kinematic distances of these clumps proved to be unrealistic and were not used further. We also note that for nearby clumps distance calculations based on galactic rotational properties are unreliable because peculiar velocities dominate their measured $v_{\rm LSR}$ values.

The calculated kinematic distances of the clumps are listed in Table \ref{dist} along with the previously determined distances by \citet{montillaud2015}. The rejected kinematic distance results are marked with an asterisk. In case of clumps with reliability flags of 2 the previous distance estimates were mostly based on association with cloud complexes. We adopt these estimates for these clumps since either at least one of our kinematic distances is consistent with them or we do not have a reliable result for them from the kinematic method. For most of the clumps with reliability flags of 1 we also accept the previous estimates because at least one of our kinematic distance results is close to the value determined by extinction methods, making it more reliable. In the case of 4 clumps (G37.91-A, G37.91-B, G110.62-A and G141.25-A) our kinematic distances were consistent with but somewhat different from previous extinction-based values. Here we adopt a mean value with a formal error that covers both results. It was already noted in Sect. \ref{res:vellinewidth} that we only observed the higher velocity $^{13}$CO line component towards the two clumps G37.91-A and G37.91-B. Thus we adopted a higher distance value for them with a smaller error bar that is still consistent with the result of the extinction method but does not cover the kinematic distance calculated from the low velocity component. For G110.62-A our estimate with a $v_s$\,=\,$-$15\,kms$^{-1}$ coincides well with the spectroscopic distances of associated stars derived by \citet{aveni1969}, thus we adopted 0.385\,$\pm$\,0.16\,kpc which covers both values. The clump G141.25-A has a distance estimate from spectroscopic analysis of background stars for the Ursa Major complex MBM29$-$31 that is associated with the clump \citep{penprase1993}. Our kinematic distance estimate is somewhat higher and a value of 0.215\,$\pm$\,0.115\,kpc was adopted that covers both. The clump G126.24-A had only an extinction-based distance estimate so far. Here our kinematic distance result with $v_{\rm s}$\,=\,$-$15\,kms$^{-1}$ is consistent with that value and it was accepted. Finally, for the clumps with reliability flags of 0, we were not able to significantly improve the previously determined distances, except in the case of G109.18-A where no kinematic distance calculation existed before. The extinction-based distance cited by \citet{montillaud2015} is 160\,pc \citep{juvela2012} but due to the lack of complementary data the value is highly uncertain and received a flag of 0. Our result is much higher than this (around 1\,kpc), thus we adopted the mean of the extinction-based and our kinematic result with an appropriately large error bar that covers both values. The clumps G71.27-A, G91.09-A and G171.35-A remain without reliable distance estimates. The final, adopted distances are included in Table \ref{dist}.

\begin{table*}[t]
	\centering
\footnotesize
	\caption{The calculated sizes and masses and adopted distances of the clumps.}
		\begin{tabular}{l r r r r c c c}
		\hline
		\multicolumn{1}{c}{clump ID} & \multicolumn{1}{c}{FWHM$\mathrm{_{x}}$} & \multicolumn{1}{c}{FWHM$\mathrm{_{y}}$} & \multicolumn{1}{c}{PA} & \multicolumn{1}{c}{D} & \multicolumn{1}{c}{$M\mathrm{_{clump}}$} & \multicolumn{1}{c}{$M\mathrm{_{vir,8.5K}}$}  & \multicolumn{1}{c}{$M\mathrm{_{vir,19.5K}}$}\\
	\multicolumn{1}{c}{\ } & \multicolumn{1}{c}{[pc]} & \multicolumn{1}{c}{[pc]} & \multicolumn{1}{c}{[deg]} & \multicolumn{1}{c}{[kpc]} & \multicolumn{1}{c}{[$M_{\odot}$]} & \multicolumn{1}{c}{[$M_{\odot}$]} & \multicolumn{1}{c}{[$M_{\odot}$]}\\
		\hline
		\hline
G26.34-A & 0.52 $\pm$ 0.16 & 0.49 $\pm$ 0.15 &  44.6 $\pm$  20.3 & 1.00 $\pm$ 0.30 &   21 --   72 &   13 --   23 &   20 --   36 \\
G37.91-A & 0.70 $\pm$ 0.29 & 0.44 $\pm$ 0.18 & 124.8 $\pm$   6.6 & 1.70 $\pm$ 0.70 &   25 --  142 &  114 --  273 &  120 --  289 \\
G37.91-B & 0.96 $\pm$ 0.40 & 0.45 $\pm$ 0.19 & 112.8 $\pm$   3.1 & 1.70 $\pm$ 0.70 &   20 --  115 &   48 --  115 &   57 --  136 \\
G39.65-A & 1.06 $\pm$ 0.35 & 0.45 $\pm$ 0.15 & 101.4 $\pm$   1.7 & 1.50 $\pm$ 0.50 &   75 --  300 &  365 --  730 &  376 --  752 \\
G62.16-A & 0.74 $\pm$ 0.23 & 0.56 $\pm$ 0.18 &  20.1 $\pm$  31.6 & 1.11 $\pm$ 0.35 &    3 --   11 &   33 --   63 &   42 --   80 \\
G69.57-A & 1.20 $\pm$ 0.54 & 0.65 $\pm$ 0.29 &  87.4 $\pm$   1.5 & 1.78 $\pm$ 0.81 &  113 --  804 &  450 -- 1201 &  460 -- 1229 \\
G70.10-A & 2.75 $\pm$ 1.09 & 0.71 $\pm$ 0.28 & 168.8 $\pm$   0.8 & 2.09 $\pm$ 0.83 &  306 -- 1642 & 1137 -- 2636 & 1161 -- 2691 \\
G70.10-B & 1.99 $\pm$ 0.79 & 0.68 $\pm$ 0.27 &  15.1 $\pm$   1.4 & 2.09 $\pm$ 0.83 &  183 --  982 &  738 -- 1710 &  756 -- 1751 \\
G109.18-A & 0.58 $\pm$ 0.43 & 0.29 $\pm$ 0.22 & 179.3 $\pm$   2.3 & 0.62 $\pm$ 0.46 &    0.085 --    4 &    4 --   30 &    7 --   46 \\
G110.62-A & 0.25 $\pm$ 0.10 & 0.15 $\pm$ 0.06 & 128.6 $\pm$   1.5 & 0.38 $\pm$ 0.16 &    2 --    9 &    6 --   15 &    9 --   21 \\
G116.08-A & 0.14 $\pm$ 0.03 & 0.12 $\pm$ 0.02 &  91.3 $\pm$  10.0 & 0.25 $\pm$ 0.05 &    2 --    4 &   16 --   24 &   18 --   27 \\
G126.24-A & 1.38 $\pm$ 0.28 & 0.47 $\pm$ 0.09 &  23.9 $\pm$   4.1 & 1.00 $\pm$ 0.20 &    5 --   12 &   22 --   33 &   38 --   57 \\
G139.60-A & 1.29 $\pm$ 0.26 & 0.94 $\pm$ 0.19 &  93.3 $\pm$   2.7 & 2.50 $\pm$ 0.50 &  312 --  701 &  548 --  822 &  566 --  849 \\
G141.25-A & 0.18 $\pm$ 0.10 & 0.05 $\pm$ 0.03 & 105.9 $\pm$   2.0 & 0.22 $\pm$ 0.12 &    0.02 --    0.2 &    2 --    7 &    3 --   11 \\
G159.23-A & 0.23 $\pm$ 0.04 & 0.12 $\pm$ 0.02 &  53.7 $\pm$   2.7 & 0.32 $\pm$ 0.05 &    4 --    8 &   13 --   18 &   16 --   22 \\
G159.23-C & 0.25 $\pm$ 0.04 & 0.22 $\pm$ 0.03 & 131.3 $\pm$  10.4 & 0.32 $\pm$ 0.05 &    3 --    5 &   16 --   21 &   20 --   27 \\
G174.22-A & 1.87 $\pm$ 0.37 & 0.67 $\pm$ 0.13 & 106.7 $\pm$   0.8 & 2.00 $\pm$ 0.40 &  222 --  499 &  272 --  408 &  294 --  441 \\
G174.22-B & 1.40 $\pm$ 0.28 & 0.65 $\pm$ 0.13 & 135.1 $\pm$   1.3 & 2.00 $\pm$ 0.40 &  109 --  246 &  195 --  292 &  212 --  318 \\
G189.51-A & 0.18 $\pm$ 0.02 & 0.09 $\pm$ 0.01 & 174.1 $\pm$   2.9 & 0.44 $\pm$ 0.05 &    1 --    2 &    9 --   12 &   12 --   15 \\
G195.74-A & 0.49 $\pm$ 0.24 & 0.33 $\pm$ 0.17 &  30.0 $\pm$   2.3 & 1.00 $\pm$ 0.50 &   30 --  270 &   72 --  217 &   76 --  229 \\
G203.42-A & 0.18 $\pm$ 0.04 & 0.13 $\pm$ 0.03 & 153.3 $\pm$  12.7 & 0.40 $\pm$ 0.10 &    2 --    5 &    9 --   14 &   11 --   18 \\
G205.06-A & 0.34 $\pm$ 0.08 & 0.17 $\pm$ 0.04 & 155.4 $\pm$   1.6 & 0.40 $\pm$ 0.10 &    3 --    9 &   19 --   32 &   23 --   39 \\
  \hline
			\end{tabular}
    \tablefoot{The columns are: (1) ID of the clump; (2,3) FWHM size of the clump; (4) position angle of the clump; (5) adopted distance of the clump; (6) mass of the clump; (7,8) virial mass of the clump calculated with $T\mathrm{_{kin}}$\,=\,8.5\,K and 19.5\,K.}
	\label{stability}
\end{table*}

\subsection{Clump stability}
\label{clumpstab}

We use the adopted distances in Table \ref{dist} to perform the virial analysis of the clumps. The uncertainties in clump mass are dominated by the error of the distance, since the errors originating from the uncertainties of $\tau_{250}$ and the conversion to $N$(H$_2$)$_{\rm dust}$ (due to different $R_V$ values) are both less than 30\%. Through the calculated sizes of the clumps the virial masses are also affected by the distance uncertainty. For this reason we only considered the stability of clumps where the distance values are reasonably reliable: the clumps with a flag of 1 and 2 and G109.18-A, where we revised the previously calculated value. The clumps with two velocity components were excluded from this analysis since the dimensions and masses of the two structures corresponding to the velocity components cannot be separated using a column density map. The sizes, masses and virial masses of these clumps are in Table \ref{stability}, including the lower and upper limits of $M_{\rm clump}$ clump mass and $M_{\rm vir}$ virial mass due to the distance uncertainty. The virial mass calculations were done using the two limiting temperatures (8.5 and 19.5\,K). The errors of the FWHM sizes were also calculated from the distance uncertainty.

The FWHM sizes of the clumps vary in the range 0.1$-$5\,pc, where the smallest is G141.25-A and the most extended are G70.10-A, G70.10-B and G174.22-A. Considering the error in the distance estimates, clump masses vary from 0.02\,$M_{\odot}$ to more than 1600\,$M_{\odot}$, covering a very wide range of values. From the 22 clumps with virial mass estimates eleven have masses lower than either of the calculated virial masses, making these clumps gravitationally unbound. Except G39.65-A and G62.16-A they all have masses less than around 10\,$M_{\odot}$, are mostly rather small (from 0.18 to 0.5\,pc) and are located closer to us (under 0.6\,kpc). The rest of the clumps tend to have larger distances and show several hundred solar masses. Their stability cannot be decided with a good reliability because of the large errors caused by the distance uncertainty in both clump masses and virial masses.

\section{Discussion}
\label{discussion}

\subsection{The temperatures and column densities of the clumps}

\label{disc:tempdens}

According to Fig. \ref{NvsN} colder clumps have a peak $N$(H$_2$)$_{\rm dust}$\,>\,4\,$\times$\,10$^{21}$\,cm$^{-2}$ and $N$($^{13}$CO)\,>\,5\,$\times$\,10$^{15}$\,cm$^{-2}$ but while most of the clumps with $T_{\rm dust}$ above 14\,K have lower densities, some of them show high values. The correlation of the peak $N$(H$_2$)$_{\rm dust}$ and $N$(H$_2$)$_{\rm gas}$ seems to be good but there are many factors that can affect their relation. As discussed before, the line-based column densities are underestimated due to optical depth effects. The radiative transfer models used to investigate the optical depth and its effect on column density in model clumps with similar densities as in our sample are described in Appendix \ref{modeling1} and \ref{modeling2}. With modelling (adopting density-dependent abundances with a maximum
value of [$^{13}$CO]/[H$_2$]\,=\,1.5\,$\times$\,10$^{-6}$) we found that the gas-derived column densities of our model clouds can have values by a factor of two lower than the ones derived from dust emission. Differences between $N$(H$_2$)$_{\rm dust}$ and $N$(H$_2$)$_{\rm gas}$ may also originate from the uncertainties of $N$(H$_2$)$\mathrm{_{dust}}$, the lack of background subtraction and the use of the incorrect dust spectral index.

The agreement of the two values is also affected by the assumed relative abundance of $^{13}$CO. Most studies find values close to 2\,$\times$\,10$^{-6}$, like 1.5\,$\times$\,10$^{-6}$ in Taurus and 2.9\,$\times$\,10$^{-6}$ in Ophiuchus \citep{frerking1982}, 2.3\,$\times$\,10$^{-6}$ in IC5146 \citep{lada1994} or 2.5\,$\times$\,10$^{-6}$ in Perseus \citep{pineda2008}. \citet{harjunpaa2004} found values between 0.9$-$3.5\,$\times$\,10$^{-6}$ in their survey of the three globules B\,133, B\,335 and L\,466. We use an abundance of 10$^{-6}$ in our calculations of $N$(H$_2$)$_{\rm gas}$ that results in a good agreement seen on Fig. \ref{NvsN}. 
We note that the abundance of $^{13}$CO can vary inside a clump as well, especially towards starless clumps due to freeze-out.

We note that although we usually expect $T_{\rm ex}$ to be somewhat lower than $T_{\rm kin}$, this may not be always the case. Since CO is not the best in probing the temperatures inside the densest clumps and we get less radiation from the clump centres, the observed $T_{\rm ex}$ could sometimes be higher than $T_{\rm kin}$ in the centre, even if the true $T_{\rm ex}$ were always locally lower than $T_{\rm kin}$. The expected $T_{\rm ex}$ values inside the clumps were checked during the radiative transfer modelling and according to the results we expect the values of $T_{\rm ex}$ and $T_{\rm kin}$ to be similar (thus the observed $T_{\rm ex}$ limits 8.5\,K and 19.5\,K were used for the stability calculations as well).

Association with possibly embedded young star objects (YSOs) could explain the high $T_{\rm dust}$ or $T_{\rm ex}$ derived for some of the clumps. According to the young star object candidate catalogue by \citet{marton2016} 26 of our clumps from the 35 have no associated Class I/II or Class III YSO candidates inside a radius that equals to the major axis of their fitted 2D Gaussian function. The clumps G37.49-A, G37.91-A and B, G69.57-A and B and G139.60-A are associated with 1$-$1 Class I/II YSO candidates, G39.65-A and G70.10-A have 2 associated Class I/II YSO candidates. The clump G195.74-A is associated with 6 Class I/II YSO candidates. Additionally, 2 Class III YSO candidates were found associated with G70.10-A and 4-4 Class III sources are associated with G37.91-A and G39.65-A \citep{marton2016c}. As mentioned previously the two clumps with high $T_{\rm ex}$ are clearly associated with YSOs both by \citet{marton2016} and \citet{montillaud2015}. 

\subsection{The nature of the clumps}

The ratio of gravitationally bound clumps was 5 out of 21 clumps by \citet{parikka2015} in their analysis of a similar GCC sample and most core-sized clumps in our sample also seem sub-critical. However, many of the more distant clumps have large distance uncertainties that makes it difficult to assess their state. There is a group of objects that have physical sizes close to 1\,pc or more, masses of more than a few hundred $M_{\odot}$ and are located at high distances (1$-$3\,kpc).
This suggests that they are large-scale clouds or clumps that may contain many smaller clumps or cores which remain unresolved in our observations. Some of our objects have sizes of less than 0.3\,pc and are at smaller distances, e.g. G110.62-A, G116.08-A, G141.25-A or G189.51-A. These objects are closer to the traditional definition of individual cores, seem to be gravitationally unbound and none of them are associated with young stars. We also find gravitationally unbound objects that are larger than cores, like G39.65-A and G126.24-A. The large-scale clouds or clumps are mostly dense with peak $N$(H$_2$)$_{\rm dust}$\,$\gtrsim$\,10$^{22}$\,cm$^{-2}$ but they appear with both low and high $T_{\rm dust}$ and $T_{\rm ex}$ values (e.g. G195.74-A with $T_{\rm dust}$\,=\,14.3\,K and $T_{\rm ex}$\,=\,23.4\,K is dense and warm, while G174.22-B with $T_{\rm dust}$\,=\,12.4 and $T_{\rm ex}$\,=\,9.5\,K is dense but cold). The large ratio of gravitationally unbound, core-sized objects might be explained by the $^{13}$CO emission overestimating the velocity dispersion inside the clumps since the line broadening due to optical depth can cause an overestimate of the virial mass by a factor of up to two, according to our models.

We compared the $T_{90\%}$ values calculated on the respective GCC fields of our clumps by \citet{montillaud2015} to the virial state of the clumps. The $T_{90\%}$ value gives the temperature below which 90\% of the pixels are on the \textit{Herschel}-derived dust colour temperature maps of the GCC fields and is used as a proxy of the intensity of the local interstellar radiation field. \citet{montillaud2015} found a correlation between $T_{90\%}$ and Galactic radius which is consistent with the radiation field being stronger in the inner Galaxy and examined the relation between this temperature and the cumulative fractions of their sources. A correlation between the boundedness of clumps and the surrounding temperature would mean that clumps are more bound in colder environments. This would support the implication that the heating by the surrounding visible$-$UV radiation field may prevent the collapse of cores. We find that the average value of $T_{90\%}$ is 17.1\,K for the gravitationally unbound clumps and 16.4\,K for the larger clumps with higher mass and distance where the stability cannot be assessed. Due to the large error in $M_{\rm clump}$ and $M_{\rm vir}$ and the usual error of around 1\,K in $SPIRE$-derived dust temperature maps in cold environments, the observed difference is not significant enough.

\section{Conclusions}

We obtained good S/N, high spectral and spatial resolution $^{12}$CO(1$-$0) and $^{13}$CO(1$-$0) spectra in the direction of 35 clumps on 26 fields selected from the \textit{Herschel} GCC sample. Based on the molecular line data, $T_{\rm ex}$ and $N$($^{13}$CO) values inside the clumps were calculated and the far-infrared dust continuum maps of \textit{Herschel} were used to derive $T_{\rm dust}$ and $N$(H$_2$)$_{\rm dust}$ distribution. 

The clumps in our sample have excitation temperatures generally between 8.5 and 19.5\,K, with G195.74-A and G139.60-A showing even higher values, suggesting the presence of embedded YSOs. The derived dust colour temperatures are between 12$-$16\,K. Clumps with temperatures above the average 14\,K appear both at low and high column densities but colder clumps have peak $N$(H$_2$)$_{\rm dust}$ above 4\,$\times$\,10$^{21}$\,cm$^{-2}$. The gas- and dust-based column densities of the clumps correlate well but can be affected by many uncertainties from CO optical depth effects to the assumed dust properties. We performed radiative transfer modeling to predict CO emission, optical depths and column densities to verify our results and estimate the uncertainties.

Kinematic distance calculations were performed using $^{13}$CO line velocities in an attempt to obtain more reliable values. The previous distance estimates of 5 clumps were refined by our new results. We identified 11 gravitationally unbound clumps in the sample with a few solar masses that are core-sized and are located close-by. Large, massive objects were also found, which are located at larger distances, suggesting that these clouds may contain further cores and clumps. Some of these large objects are associated with YSOs, while no young stars were found around most of the unbound, core-sized sources. The dominant uncertainty in assessing the state of our clumps originates from the large errors in their distance estimates. No correlation was found between the temperature of the surrounding environment and the level of boundedness of the clumps.

\begin{acknowledgements}
This research was partly supported by the OTKA grants K101393 and NN-111016 and it was supported by the Momentum grant of the MTA CSFK Lend\"ulet Disk Research Group. M.J. acknowledges the support of the Academy of Finland Grant No. 285769. T.L. acknowledges the support from the Swedish National Space Board (SNSB). This work was supported by NAOJ ALMA Scientific Research Grant Number 2016-03B.
The Onsala Space Observatory (OSO), the Swedish National Facility for Radio Astronomy, is hosted by Department of Earth and Space Sciences at Chalmers University of Technology, and is operated on behalf of the Swedish Research Council. We thank the staff of OSO for their assistance with the observations.
Herschel was an ESA space observatory with science instruments provided by European-led Principal Investigator consortia and with participation from NASA. SPIRE was developed by a consortium of institutes led by Cardiff Univ. (UK) and including Univ. Lethbridge (Canada); NAOC (China); CEA, LAM (France); IFSI, Univ. Padua (Italy); IAC (Spain); Stockholm Observatory (Sweden); Imperial College London, RAL, UCL-MSSL, UKATC, Univ. Sussex (UK); Caltech, JPL, NHSC, Univ. Colorado (USA). This development was supported by national funding agencies: CSA (Canada); NAOC (China); CEA, CNES, CNRS (France); ASI (Italy); MCINN (Spain); SNSB (Sweden); STFC (UK); and NASA (USA). PACS was developed by a consortium of institutes led by MPE (Germany) and including UVIE (Austria); KUL, CSL, IMEC (Belgium); CEA, OAMP (France); MPIA (Germany); IFSI, OAP/AOT, OAA/CAISMI, LENS, SISSA (Italy); IAC (Spain). This development has been supported by the funding agencies BMVIT (Austria), ESA-PRODEX (Belgium), CEA/CNES (France), DLR (Germany), ASI (Italy), and CICT/MCT (Spain).\\
\end{acknowledgements}

\bibliography{feher_sample_herschel.bib}
\bibliographystyle{aa}


\begin{appendix}

\section{Tests with radiative transfer models}
\label{radtransfermodel} 

\subsection{Modelling the line emission of Bonnor$-$Ebert spheres}

\label{modeling1}

To further quantify the possible bias of the column density estimates
derived from $^{13}$CO spectra, we carried out radiative transfer
calculations for a series of spherically symmetric models. The density
distributions follow the Bonnor$-$Ebert solutions for a kinetic temperature of 12\,K, a stability
parameter $\xi=6.7$, and with 0.3, 1.0, or 3.0 solar masses. The
$^{12}$CO abundance is fixed to $5\times 10^{-4}$. The $^{13}$CO
abundance is varied in a range of 0.3-3.0 times the default value of
$10^{-6}$. The turbulent linewidth is $\sigma_{\rm v}$=1\,km\,s$^{-1}$
and there is additionally a linear radial velocity gradient that
increases from zero at the centre to 1\,km\,s$^{-1}$ at the cloud
boundary (infall).

Line emission depends on the volume density, column density, velocity
field, and kinetic temperature of the clouds. A Bonnor$-$Ebert model may
not accurately describe the density profiles of our clumps. However,
here it is only necessary that the models cover the relevant parameter
space. The selected kinetic temperature and line widths are typical of
the values listed in Table \ref{lineparam1} and indicated by the $T_{\rm ex}$ values of
Table \ref{nt}. Already for the default abundance of 10$^{-6}$, the $^{13}$CO
column densities range from $10^{15}$\,cm$^{-2}$ to
$10^{17}$\,cm$^{-2}$, covering the entire range of column density
estimates in Table \ref{nt}. The mass of the Bonnor$-$Ebert spheres is a parameter that scales the cloud column density. The model clouds have mean volume densities of 5.5\,$\times$\,10$^{3}$, 5\,$\times$\,10$^{4}$ and 5.5\,$\times$\,10$^{5}$\,cm$^{-3}$  while the real clumps have peak volume densities between 4.5$\times$\,10$^{3}$\,cm$^{-3}$ and 3.2\,$\times$\,10$^{4}$\,cm$^{-3}$. Since all used densities are above the CO critical density of 10$^3$\,cm$^{-3}$, the model clumps are applicable for calculating excitation.

\begin{figure}[tp]
\includegraphics[width=8.8cm]{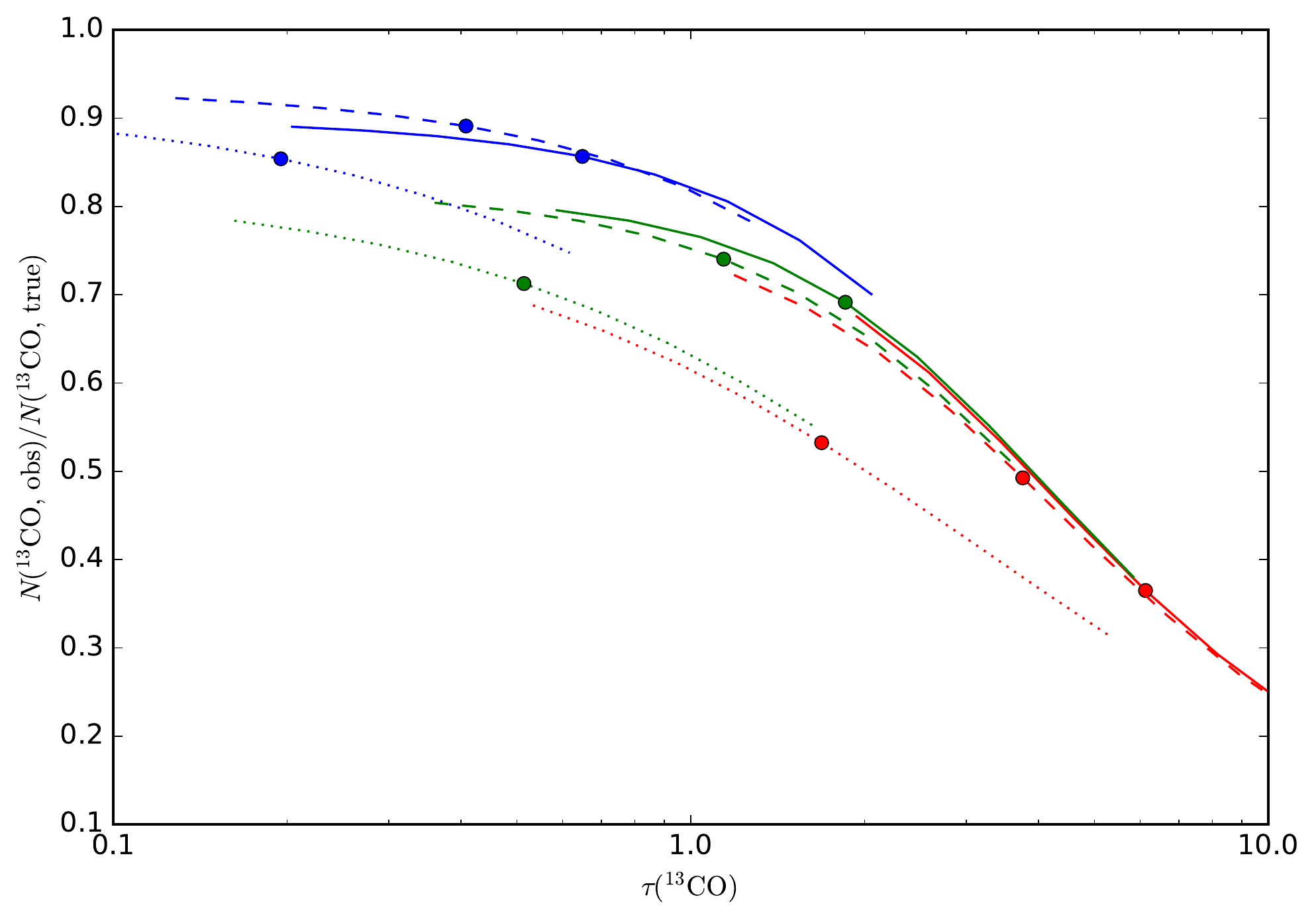}
\caption{
Ratio of the observed and the true $^{13}$CO column densities for
synthetic observations of Bonnor-Ebert spheres. There are three sets
of curves that corresponds to cloud masses of 0.3, 1.0, and 3.0 solar
masses (from right to left, colours red, green, and blue,
respectively). The solid, dashed, and dotted lines refer to averaging
of the data (observed intensity and true values) within 10\%, 50\%,
and 100\% of the cloud outer radius. The length of each line
corresponds to a variation of the $^{13}$CO abundance between 0.3 and
3 times the default value of 10$^{-6}$.
}
\label{model11}
\end{figure}

The radiative transfer calculations give $^{12}$CO and $^{13}$CO line
profiles as a function of the distance from the cloud centre. We
calculate the average spectra for emission within 10\%, 50\%, and
100\% of the outer radius of the Bonnor-Ebert spheres. Using the
equations of Sect. 2.3.1, the peak antenna temperatures, and the
integrated area of the averaged $^{13}$CO spectra, we estimate the
``observed'' $^{13}$CO column densities. For comparison, we extract
directly from the cloud models the area-averaged values of the peak
$^{13}$CO optical depth and the true $^{13}$CO column density. These
are averaged over the same regions as the synthetic observations, to
allow direct comparison.

Figure~\ref{model11} shows the ratio of the observed and the true
column densities as a function of the true $^{13}$CO optical depth.
The ratio is mainly a function of the optical depth. The averaging of
the emission over the whole cloud results in some 20\% lower values
(relative to the true values) than when the spectra represent the
inner 10\% of the model clouds. The main parameter affecting the ratio of observed and true column densities is the line optical depth. When the average optical depth of the $^{13}$CO line is close to one, the observations
underestimate the true column density by 20-40\%. In these models, the
error is likely to remain below 50\% up to optical depths of about
$\tau=3$.

\subsection{Modelling the continuum and line emissions of two fields}

\label{modeling2}

To quantify the potential systematic errors in the column density estimates, we carried
out radiative transfer modelling of the continuum \citep[][Juvela in prep.]{juvela2003} and the line emission \citep{juvela1997} for the fields
G26.34+8.65 and G195.74-2.29, where the column densities are representative of the
column density range of our clump sample. The models correspond to a projected area of
30\,$\times$\,30$\arcmin$, which is much larger than the size of the clumps, to avoid
any border effect (e.g., strong temperature gradients near the model boundaries).

In the first stage, we create 3D cloud models that reproduce the $Herschel$ dust
continuum observations at 250\,$\mu$m, 350\,$\mu$m, and 500\,$\mu$m. The models consist
of 200$^3$ cells with a cell size equal to 9\,$\arcsec$ on the sky. The dust properties
are taken from \citet{ossenkopf1994} and correspond to thin ice mantles (after 10$^5$
years at density 10$^6$\,cm$^{-3}$). The model is appropriate mainly for the densest
regions. However, in this context one of the most relevant parameters is the assumed
ratio of hydrogen column density to sub-millimetre opacity, which affects the
density field used in the following line modelling. In our basic models this is
$\tau_{250}/N({\rm
H}_2)=1.52\times 10^{-24}\,{\rm cm}^2$. However, the assumptions in Sect. \ref{herscheldata} correspond to a value that
is almost twice as large and would thus result in model clouds with higher $N({\rm
H}_2)_{\rm dust}$ values and potentially larger problems for the column density estimation. Thus,
in the line modelling, we will also test a case where the density values obtained from
the continuum modelling are scaled by a factor of 2. The density distribution in the
plane of the sky is constrained by the continuum observations. Along the line of sight,
the density is assumed to follow a profile
\begin{equation}
\rho(x) \propto   \left[1+(x/R)^2 \right]^{-p/2},
\end{equation}
where $x$ is the line-of-sight distance from the model mid-plane, $R$ is set to a 
distance that corresponds to one arcmin on the sky, and $p$\,=\,2. The uncertainty of the
density distribution along this axis is a significant source of uncertainty. However,
for the adopted parameters the clumps have roughly similar extent along the line of sight
and in the plane of the sky.

\begin{figure}[tp]
\includegraphics[width=8.8cm]{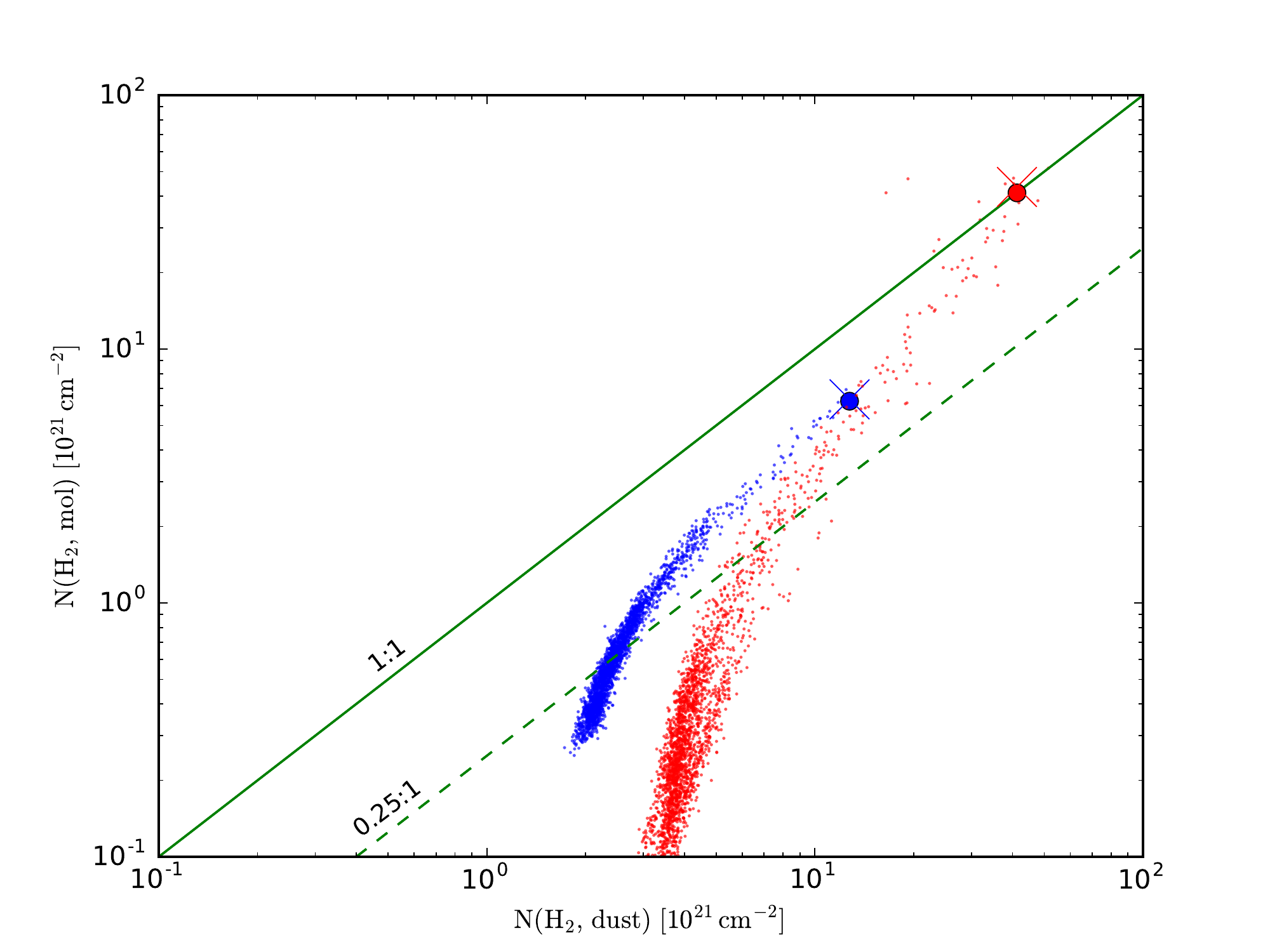}
\caption{
Comparison of column densities derived from radiative transfer models of continuum and line emission. The dots correspond to all the pixels within the central
$15\arcmin\times15\arcmin$ areas of the G26.34+8.65 (blue dots) and G195.74-2.29 (red dots) models. The coloured circles correspond to the positions of the clumps listed in Table 1. The crosses denote the column density estimates obtained assuming
$T_{\rm ex}=14$\,K.
}
\label{model1}
\end{figure}

\begin{figure}[tp]
\includegraphics[width=8.8cm]{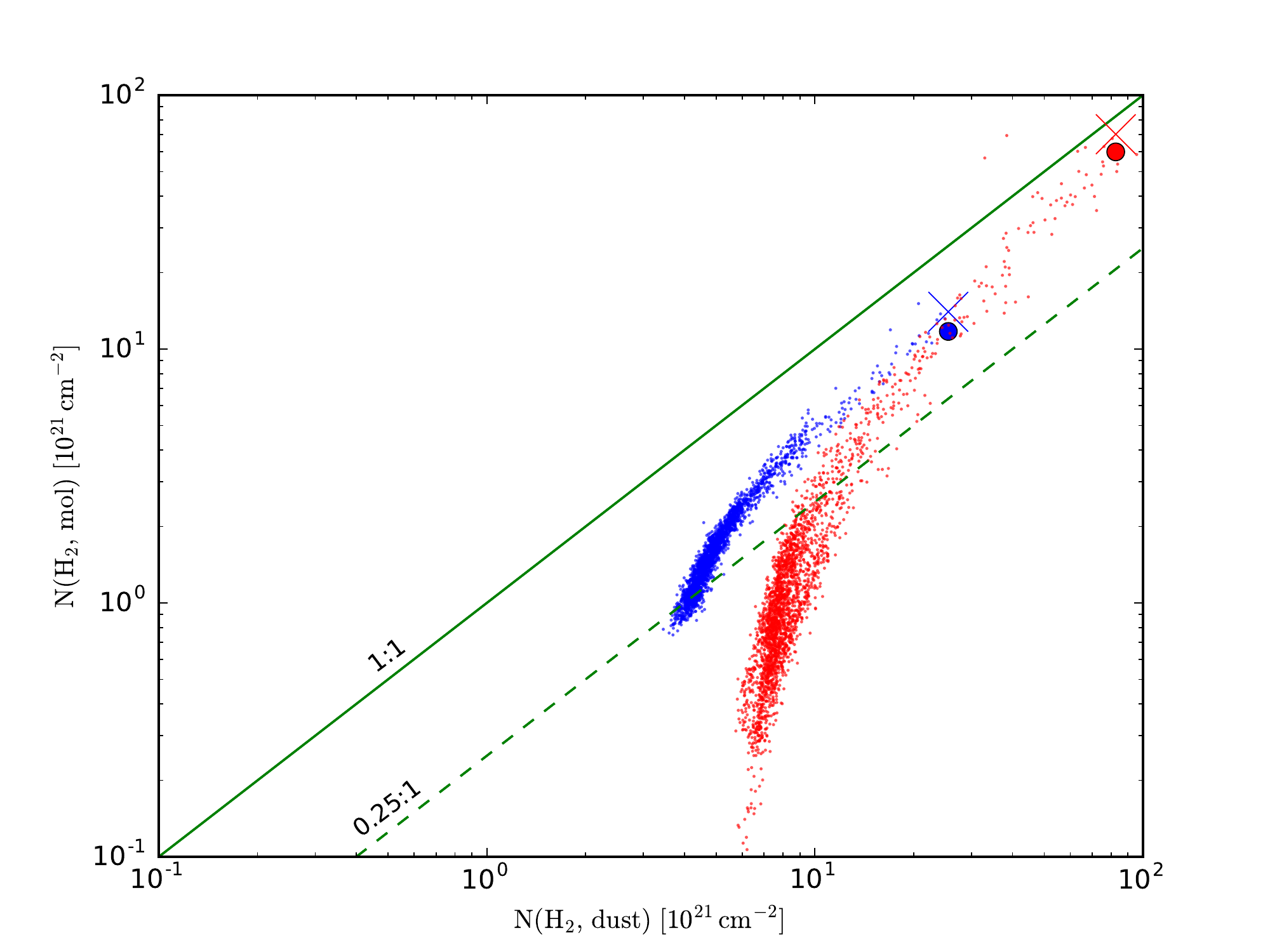}
\caption{
Comparison of column densities derived from radiative transfer models where the
densities have been increased by a factor of two.
}
\label{model2}
\end{figure}

The continuum observations are fitted in an iterative manner. The strength of the
external radiation field is scaled based on the observed and modelled ratios between the
250\,$\mu$m and 500\,$\mu$m bands. The ratios are estimated as averages over the central
area of 15$\times$\,15$\arcmin$. The 350\,$\mu$m data are used, pixel by pixel, to
adjust the model column densities. The procedure leads to a model that reproduces the
350\,$\mu$m surface brightness (final relative errors below $\sim$1\%) and gives the
correct average shape of the dust emission spectrum. In practice, the relative model errors at 250\,$\mu$m
and 500\,$\mu$m are also typically no more than $\sim$3\%. The procedure results in a 3D
model of the density field and also provides synthetic surface brightness maps at the
three $Herschel$ wavelengths. The surface brightness maps are analysed in a way similar
to the actual observations, resulting in 200\,$\times$\,200 pixel maps of the dust optical
depth. Because the 250\,$\mu$m-500\,$\mu$m spectral index of the employed dust model is
$\beta$\,=\,1.92 instead of $\beta$\,=\,2.0, the dust-derived column densities are
systematically overestimated but only by $\sim10\%$.

In the second stage, we use the derived density cube for the radiative transfer
modelling of the $^{12}$CO and $^{13}$CO lines. We assume a maximum fractional abundance of 1.5\,$\times$\,10$^{-6}$ for $^{13}$CO \citep{frerking1982,harjunpaa2004}. The abundances
are further scaled by a factor
\begin{equation}
k =  \frac{16 n^2}{5\times 10^{5}+16 n^2},
\end{equation}
which depends on the local H$_2$ density, $n$ (in units cm$^{-3}$). With this scaling,
the maximum abundance is reached only at densities above $n$\,=\,10$^3$\,cm$^{-3}$. The
scaling still results in a faster rising of abundance (as a function of density) than
most of the models discussed in \citet{glover2010}. The gas kinetic temperature is set to
a constant value of 15\,K. The model includes two components of the velocity field, one
below the model resolution (microturbulence within individual cells) and one as random
velocities between the cells. We assume that the two components are equal and together
correspond to the observed $^{13}$CO line widths, under the assumption of optically thin
line emission.

The line radiative transfer modelling results in maps of 200\,$\times$\,200 $^{12}$CO and
$^{13}$CO spectra, each with a velocity resolution of 0.07\,km\,s$^{-1}$. We process
these in a way similar as the actual observations. The spectra are fitted with Gaussian
functions and the analysis results in maps of the estimated $^{13}$CO column density, 
also at 9\,$\arcsec$ resolution. However, when $T_{\rm ex}$ is derived from the $^{12}$CO,
we use the actual peak temperature instead of the peak of the fitted Gaussian. Unlike in
the real observations, the modelled $^{12}$CO spectra are strongly self-absorbed and in
the densest regions therefore clearly double-peaked. 

Figure~\ref{model1} shows the result of the modelling exercise with the default densities.
On the x-axis are the dust-derived column densities where the scaling between dust
optical depth and column density is done using a conversion factor that is correct for
the used dust model. On the y-axis are the estimates derived from the synthetic line
spectra, using the adopted maximum $^{13}$CO abundance used in the modelling. At low
column densities the gas-derived column density is more than four times below the dust-derived
values, because of the true abundance is lower than assumed in the conversion to
N(H$_2$). For the lines of sight towards the clumps, where the abundance should
already have reached its maximum value, the difference to dust column density is
smaller. However, the results are also different for the positions of the two clumps,
which are marked in the figure as coloured circles. The estimated $^{13}$CO optical
depths are mostly below one but do rise to 1$-$1.5 in the two clumps. 

Figure~\ref{model2} shows the results when the densities are set higher by a factor
of two. The dust-derived column densities are simply multiplied by two \footnote{In
reality, the higher column density would also increase the amount by which continuum
analysis underestimates the true column density. Thus, the adopted simple rescaling 
gives slightly higher and more correct estimates of $N$(H$_2$).} but the line
calculations were repeated, starting with the simulations of the radiative transfer.

The modelling suggests that, as long as the dust opacity and the $^{13}$CO abundance
used in the analysis are correct, there should exist a fairly good correlation between
the two column density estimates. This is true only for the densest regions while in the more
diffuse regions the line-derived estimates decrease rapidly because of the decreasing
fractional abundances. In the higher density models (more consistent with the
assumptions of Sect. \ref{herscheldata}) the line analysis would be expected to lead to column
densities that are by a factor of two lower than the values derived from dust emission.

\section{\textit{Herschel} $\tau_{250}$ and $T_{\rm dust}$ maps}

Fig. \ref{herschel1}-\ref{herschel2} show the optical depths of the clumps on 250\,$\mu$m and their dust colour temperature maps based on \textit{Herschel} measurements. Black ellipses mark the 2D Gaussians fitted to the clumps and they are marked with the letter ID of the respective clumps.

 \begin{figure*}[th]
 	\centering		
 		\includegraphics[width=.45\linewidth, trim=0 2cm 0 0]{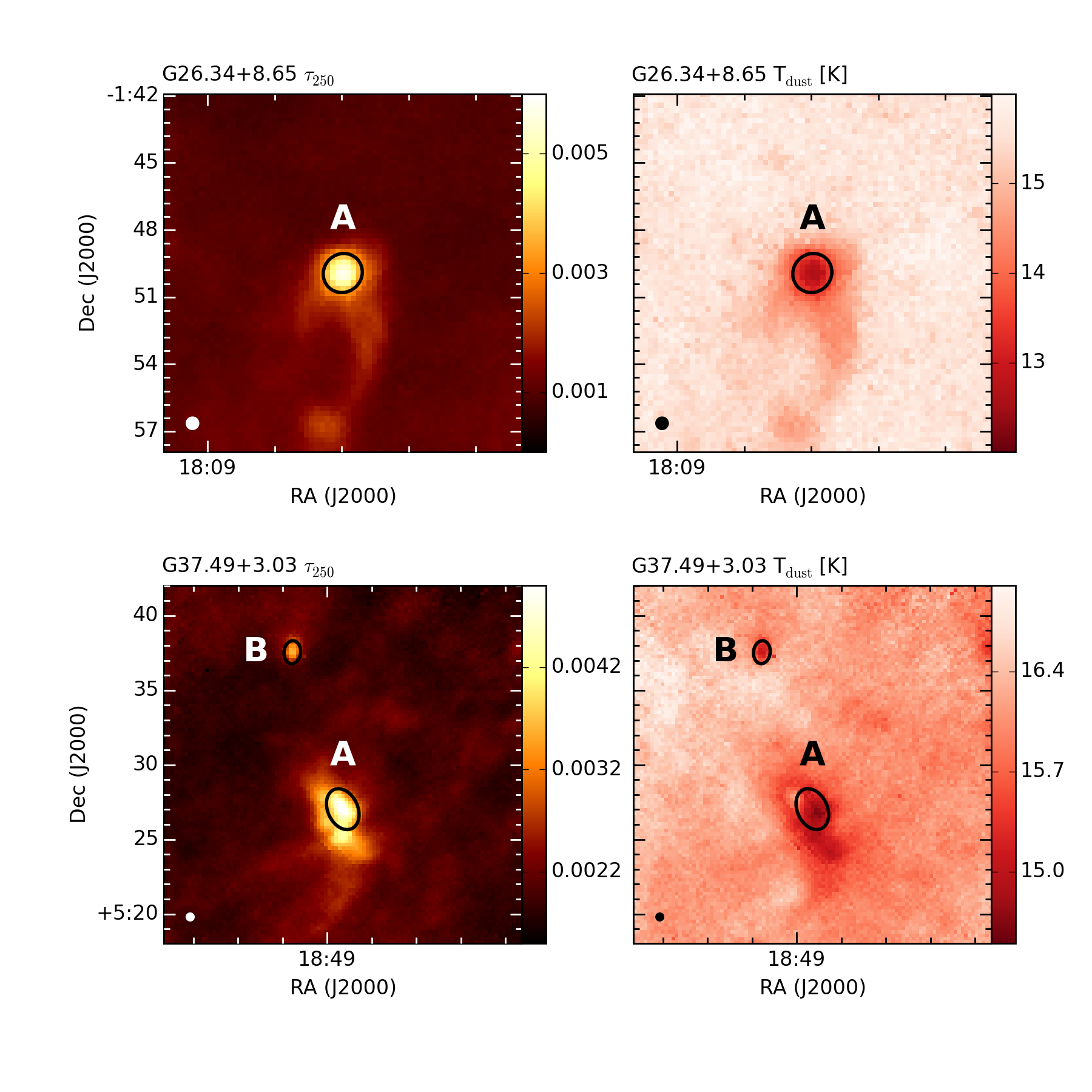}
 		\includegraphics[width=.45\linewidth, trim=0 2cm 0 0]{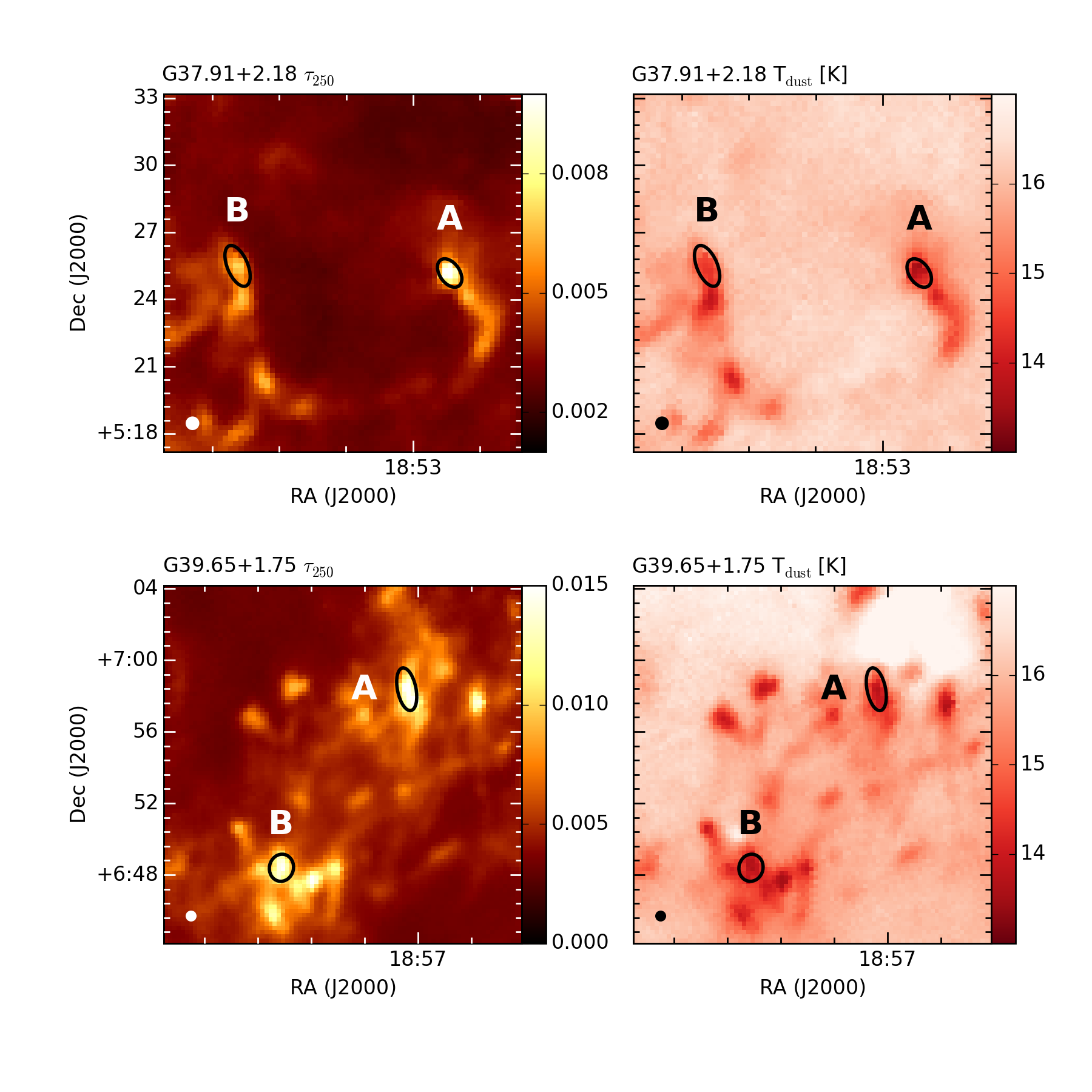}
 		\includegraphics[width=.45\linewidth, trim=0 2cm 0 0]{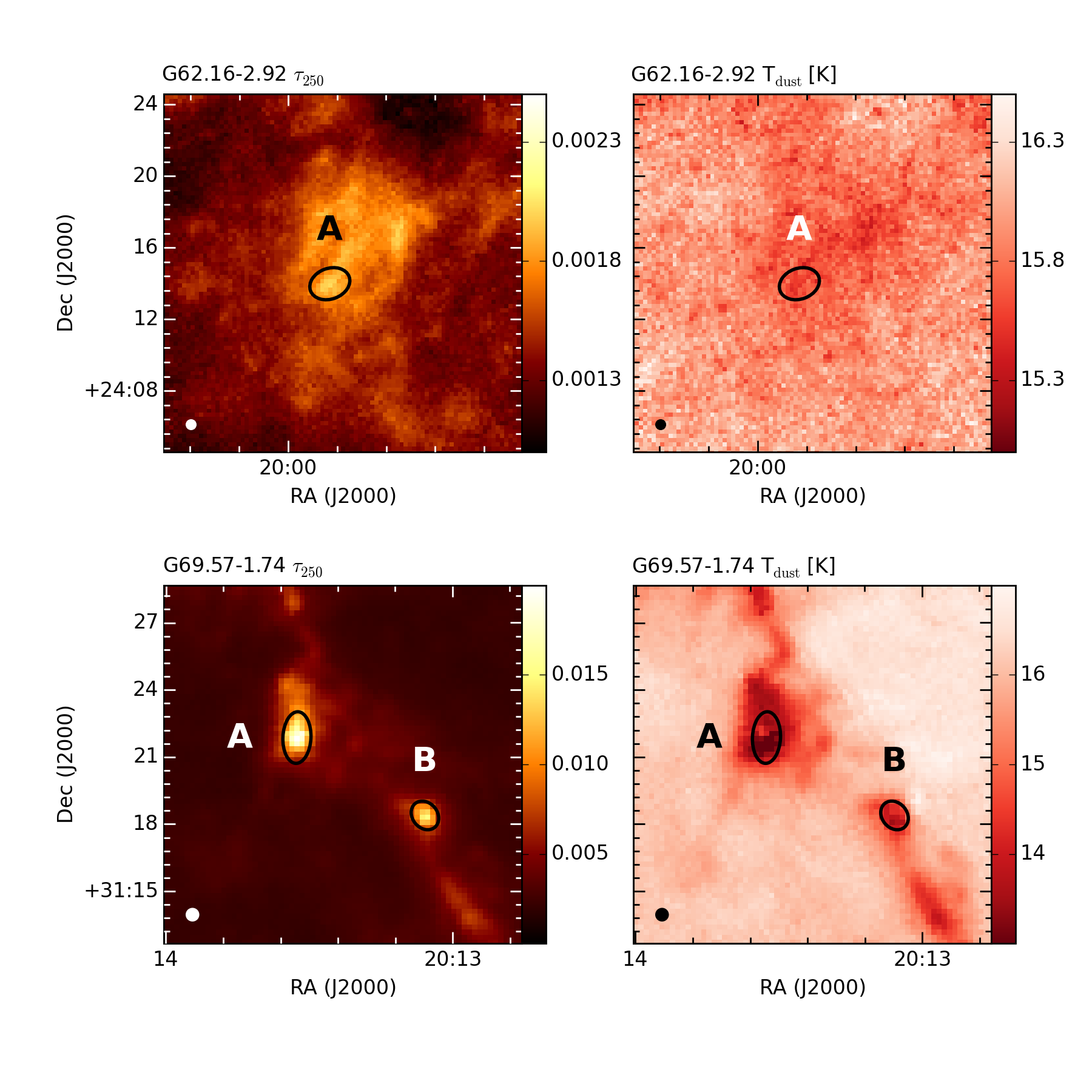}
 		\includegraphics[width=.45\linewidth, trim=0 2cm 0 0]{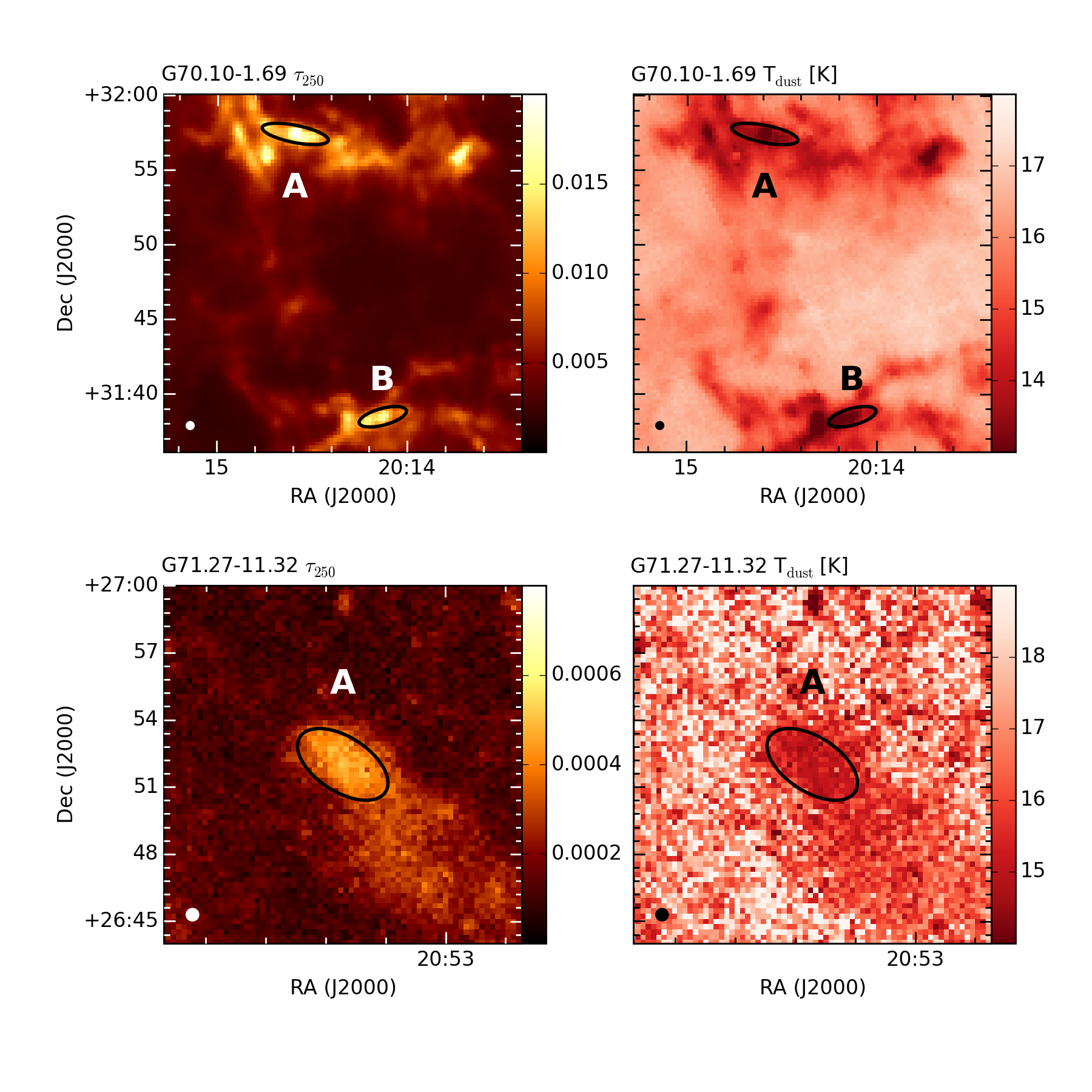}
 		\includegraphics[width=.45\linewidth,  trim=0 2cm 0 0]{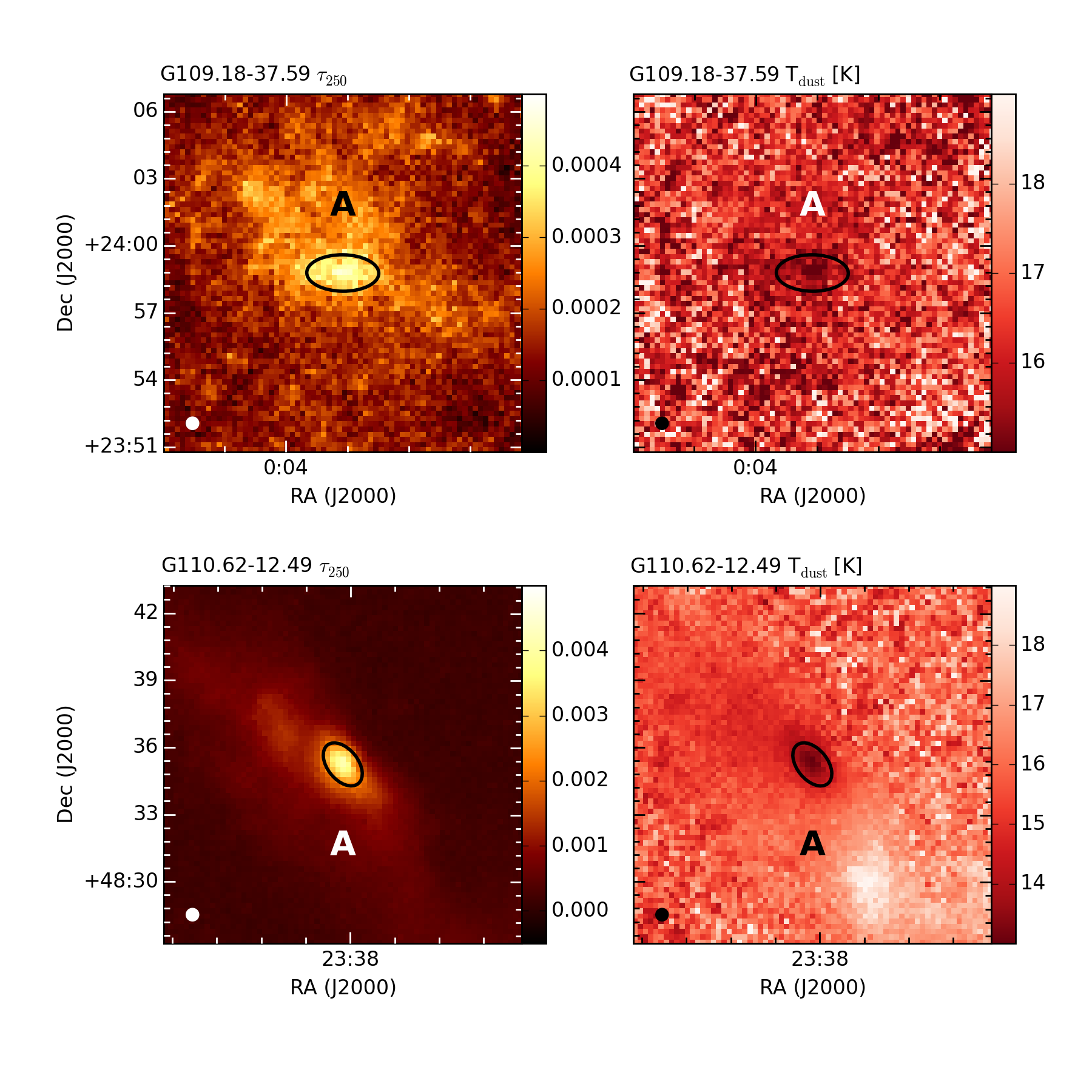}  
        \includegraphics[width=.45\linewidth, trim=0 2cm 0 0]{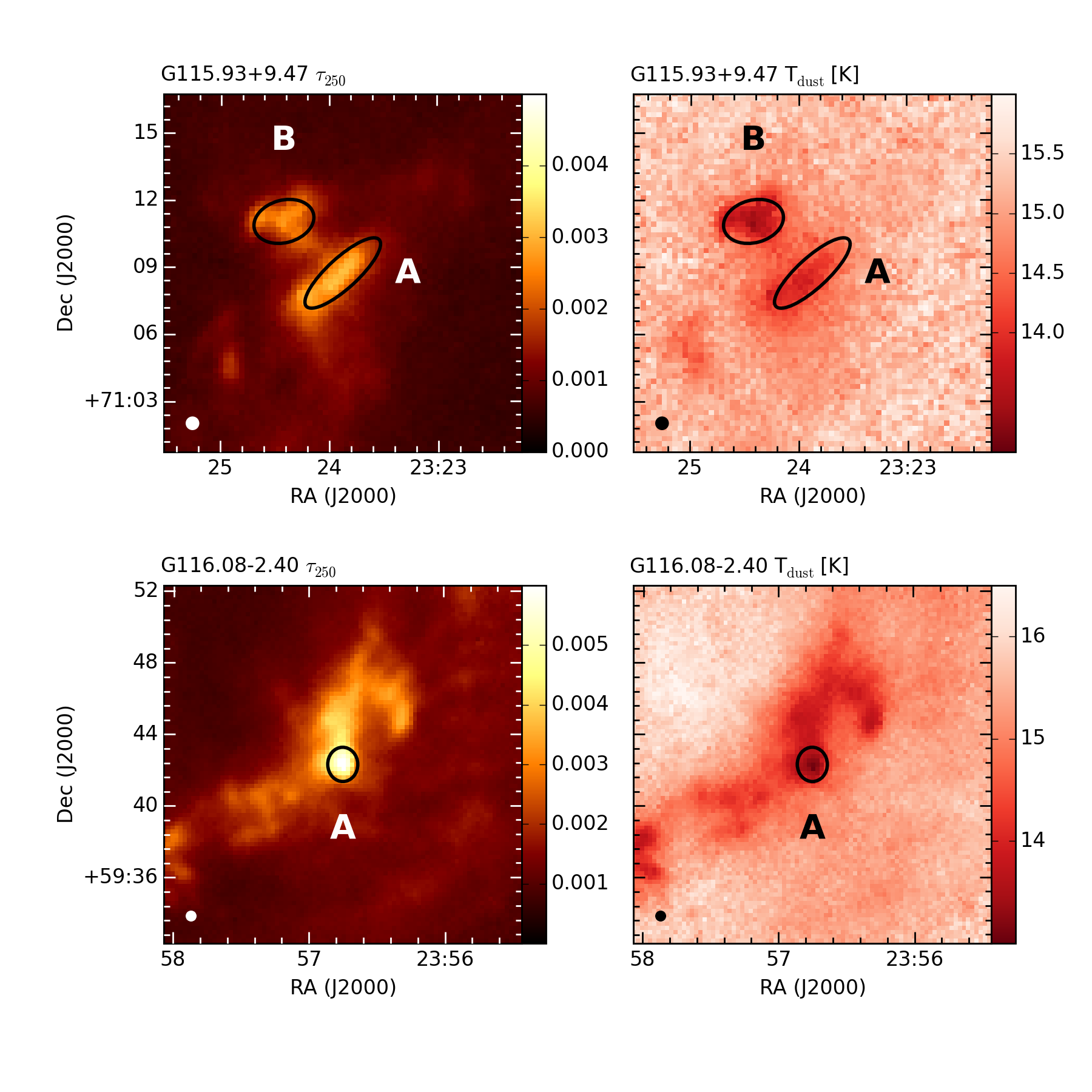}
       \caption{$Herschel$-based $\tau_{250}$ and $T_{\rm dust}$ maps of the clumps. Black ellipses mark the 2D Gaussians fitted to the clumps and they are marked with the letter ID of the respective clumps. A white ellipse in the lower left corner shows the beam size of the $Herschel$ map (37$\arcsec$).}
 	\label{herschel1}
 \end{figure*}
 
 \begin{figure*}[th]
 \centering
        \includegraphics[width=.45\linewidth, trim=0 2cm 0 0]{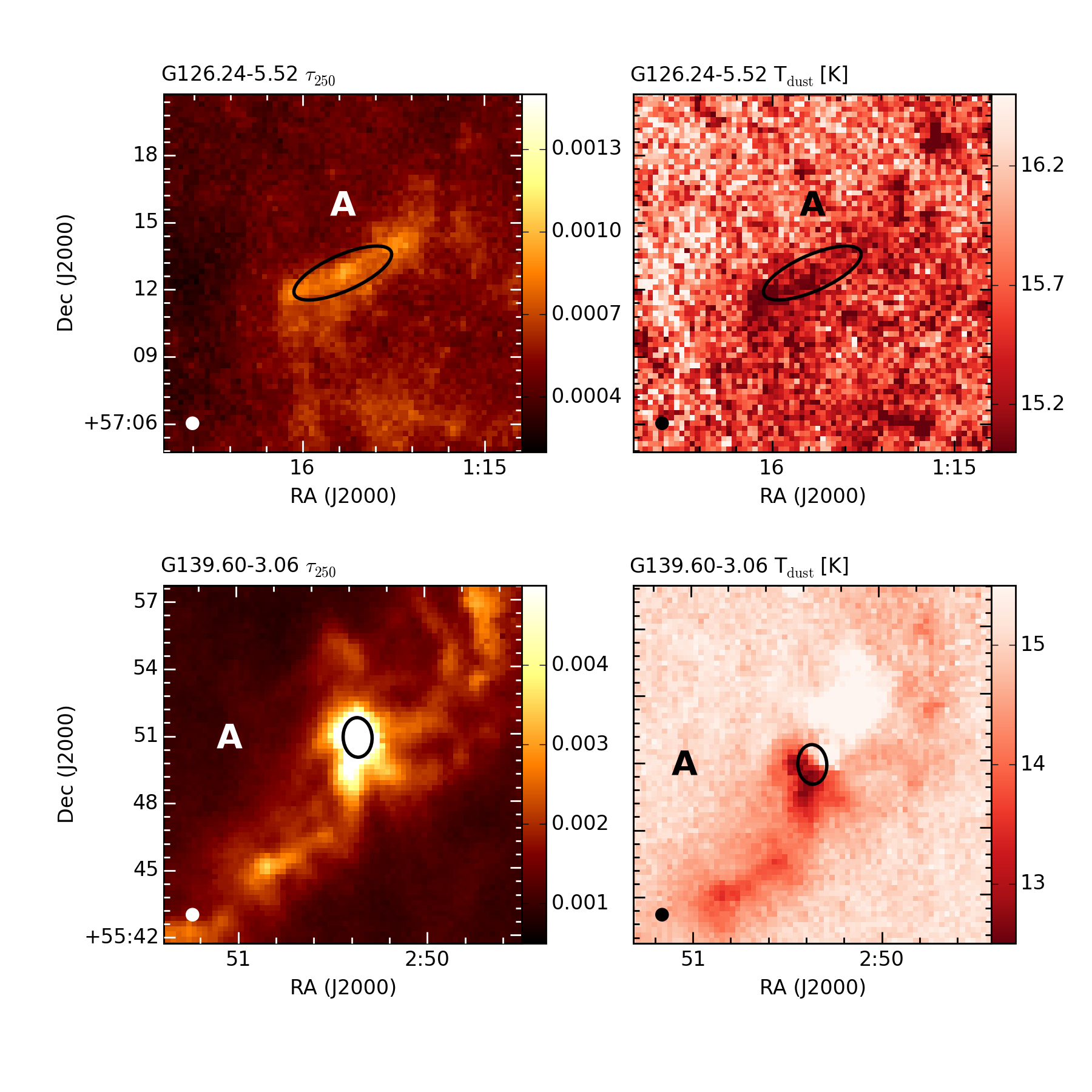}  
        \includegraphics[width=.45\linewidth, trim=0 2cm 0 0]{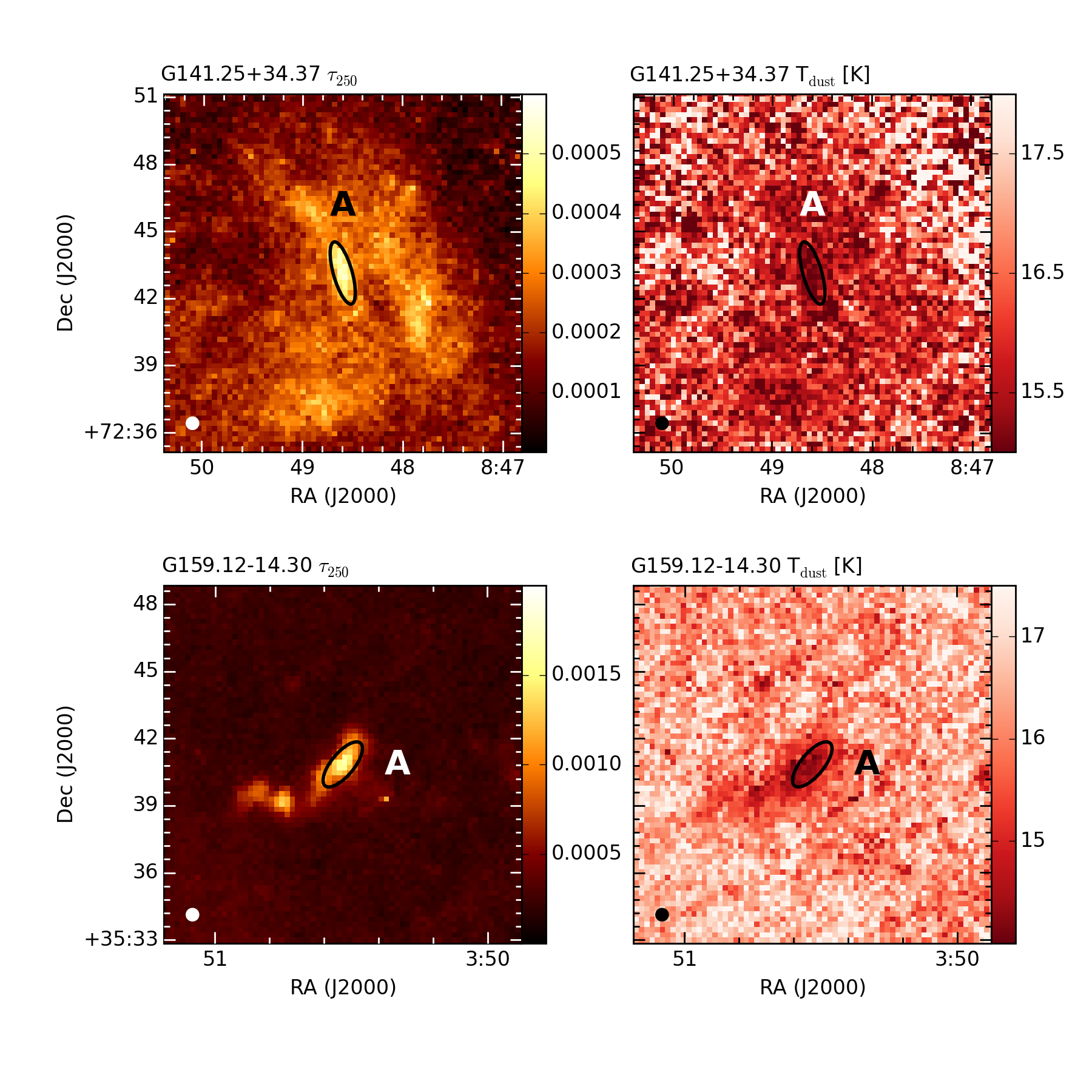}  
        \includegraphics[width=.45\linewidth, trim=0 2cm 0 0]{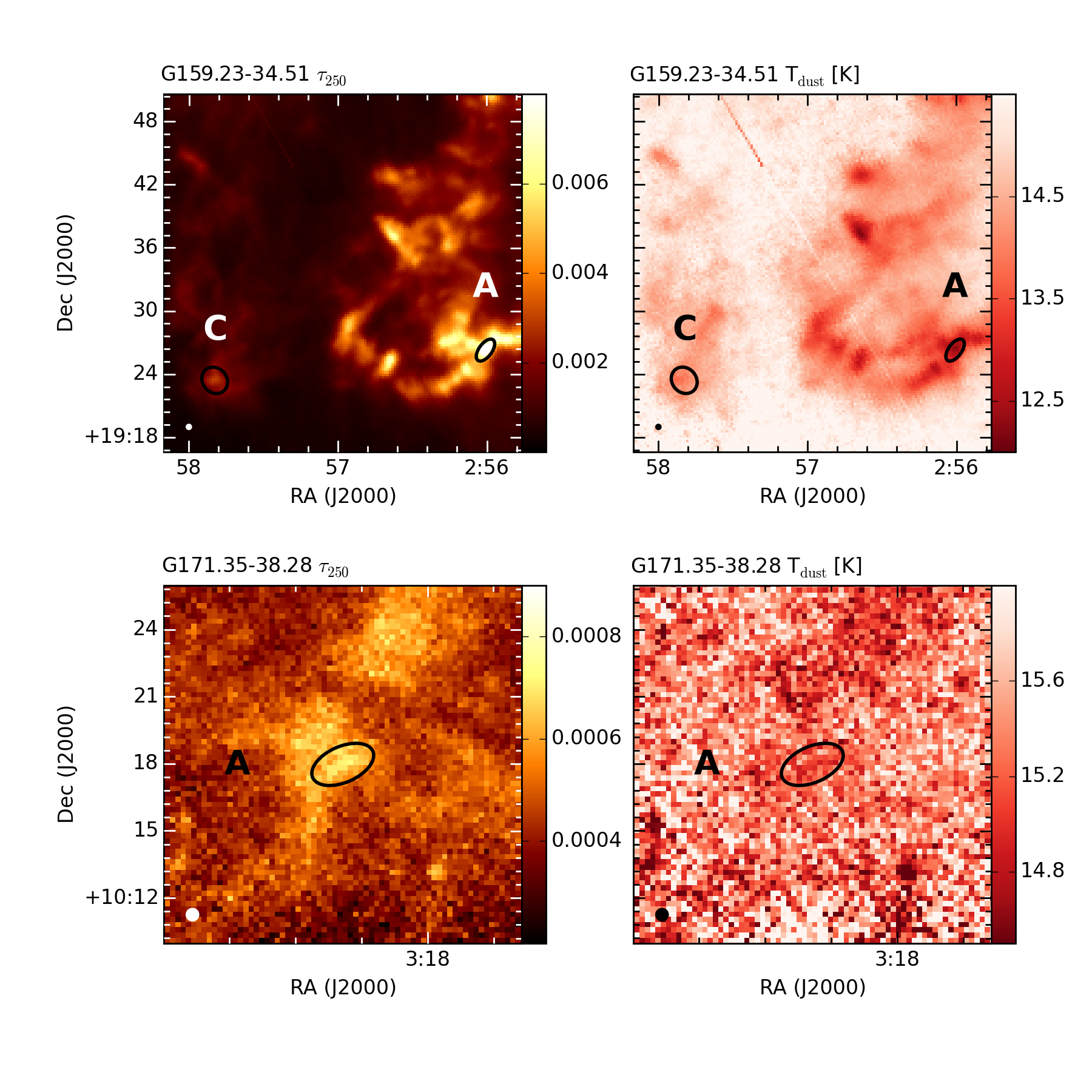}  
        \includegraphics[width=.45\linewidth, trim=0 2cm 0 0]{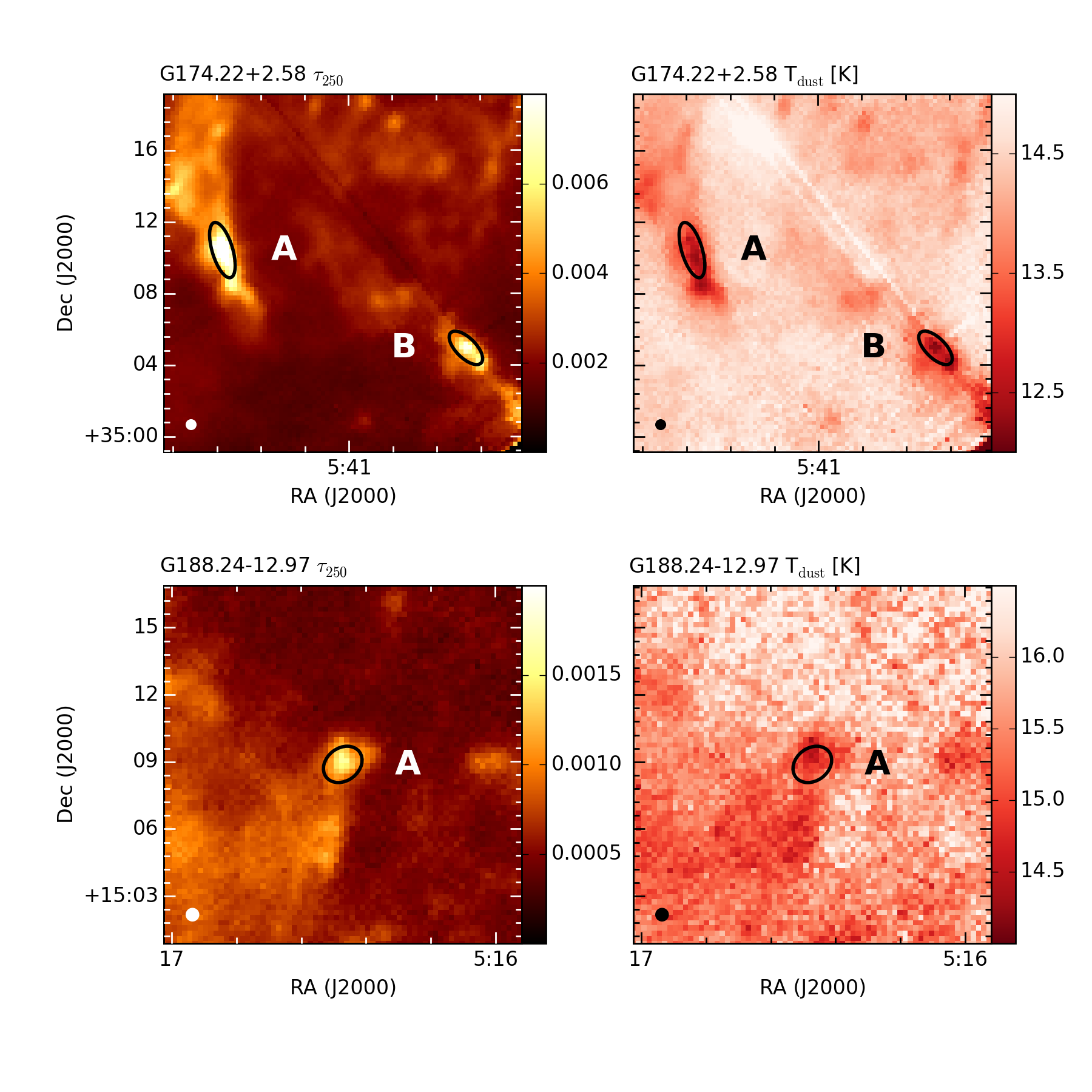}  
        \includegraphics[width=.45\linewidth, trim=0 2cm 0 0]{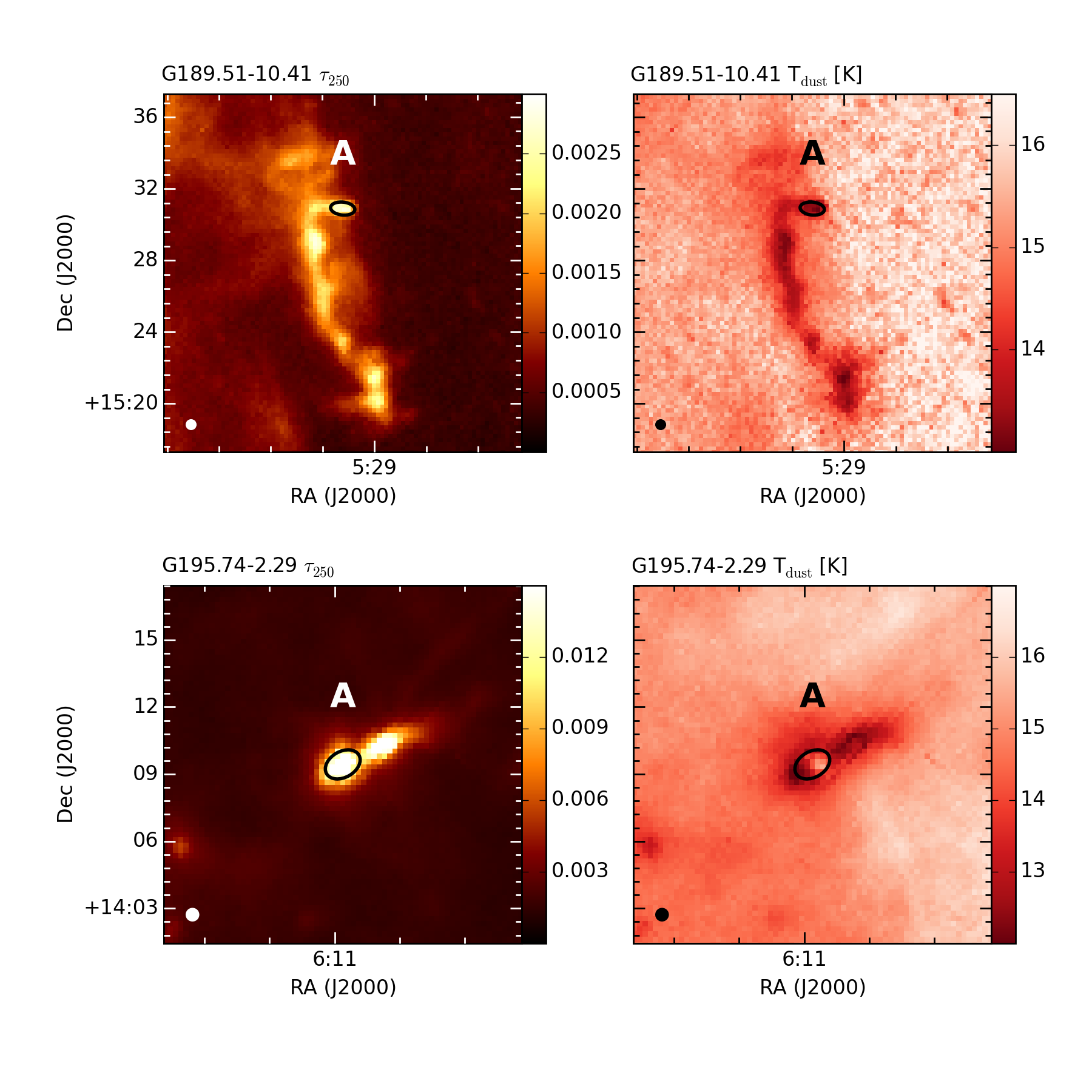}  
        \includegraphics[width=.45\linewidth, trim=0 2cm 0 0]{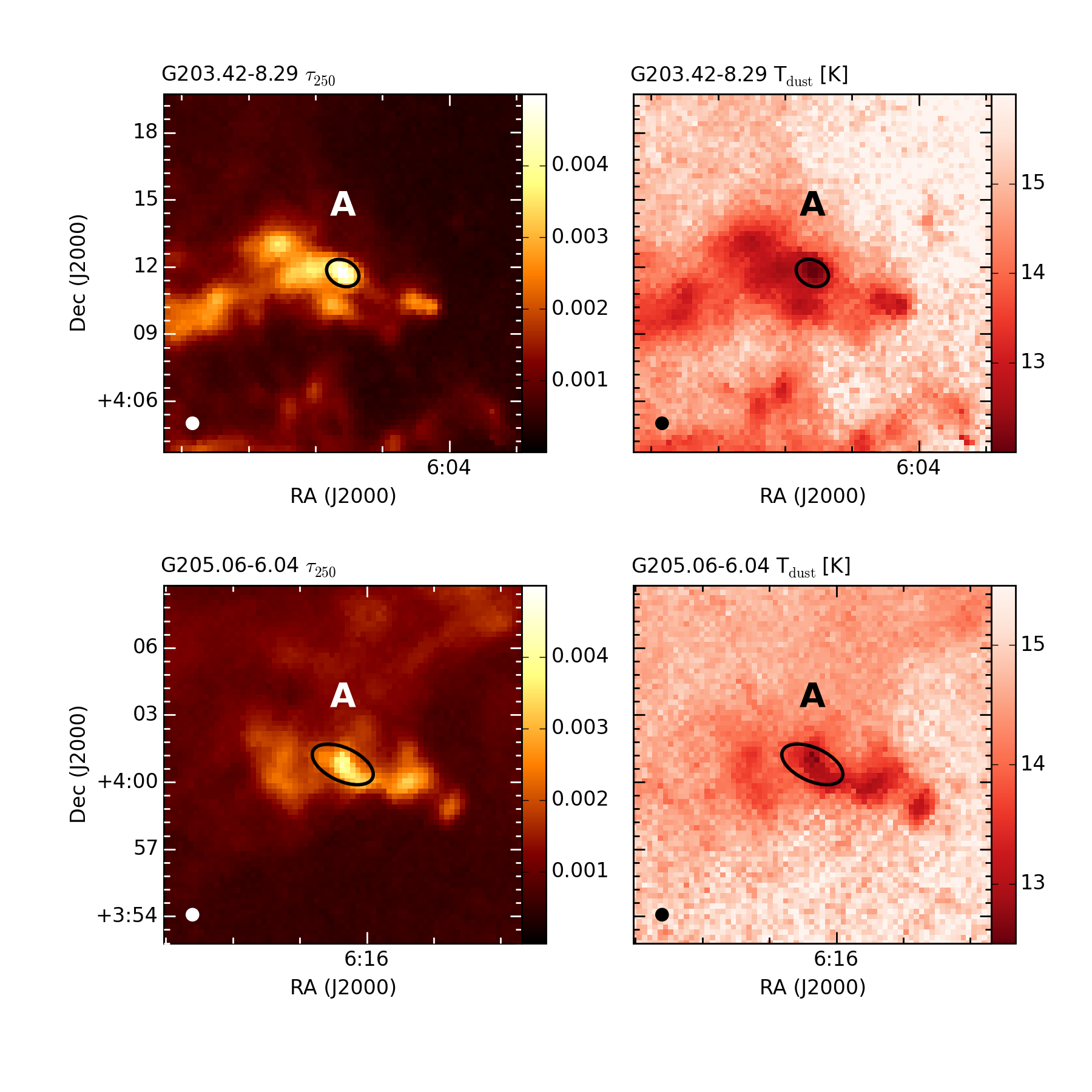}
      \caption{Cont. $Herschel$-based $\tau_{250}$ and $T_{\rm dust}$ maps of the clumps. Black ellipses mark the 2D Gaussians fitted to the clumps and they are marked with the letter ID of the respective clumps. A white ellipse in the lower left corner shows the beam size of the $Herschel$ map (37$\arcsec$).}
 	\label{herschel2}
 \end{figure*}

\section{Spectra}

Fig. \ref{spec1}-\ref{spec5} show the $^{12}$CO(1$-$0) and $^{13}$CO(1$-$0) spectra observed at the central position of each clump. Red line shows the Gaussian profile fit(s) to the lines. Where only $^{13}$CO(1$-$0) measurements were made, only one box appears. 

\begin{figure*}[th]
	\centering		
		\includegraphics[angle=-90,width=.45\linewidth]{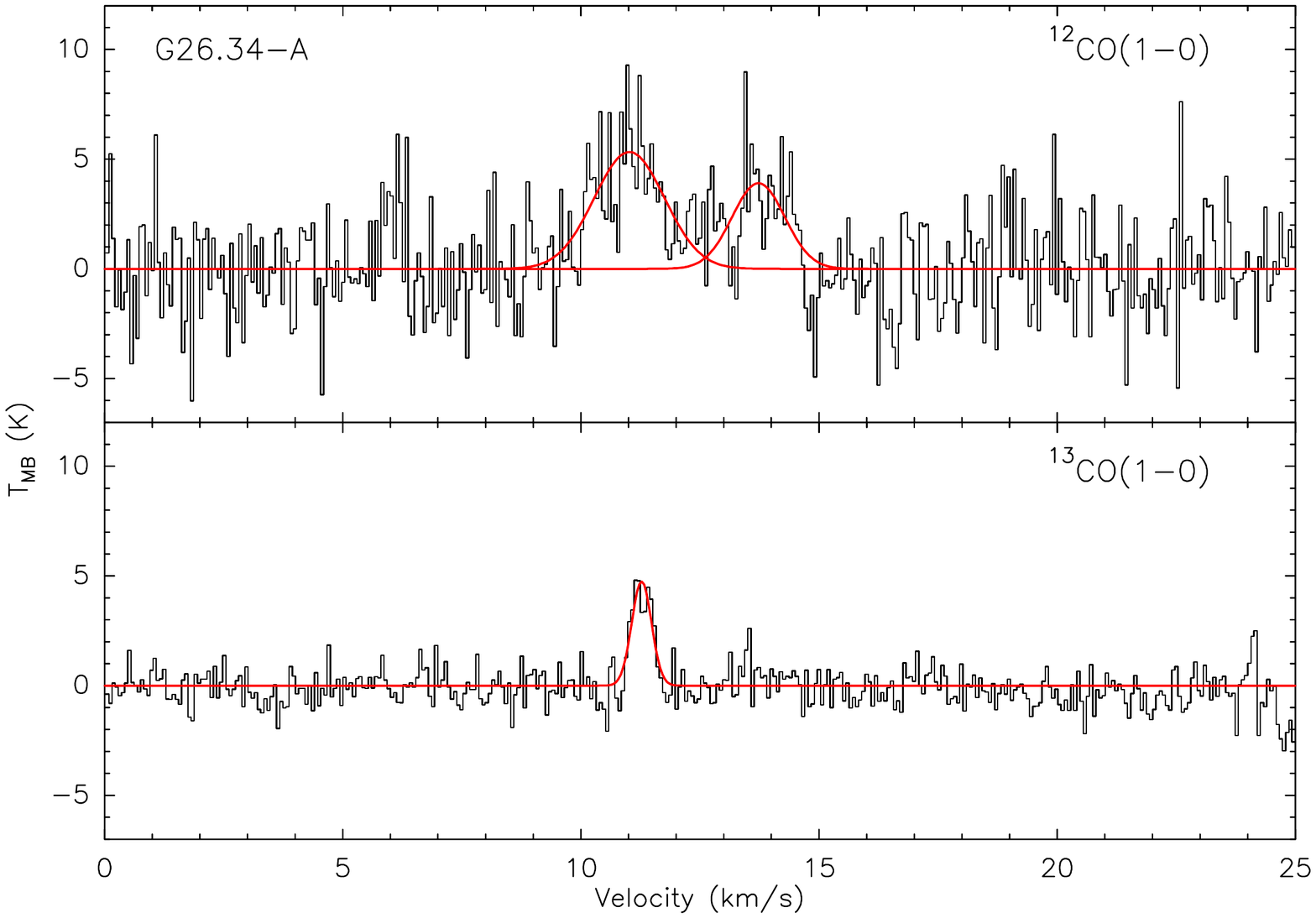}
		\includegraphics[angle=-90,width=.45\linewidth]{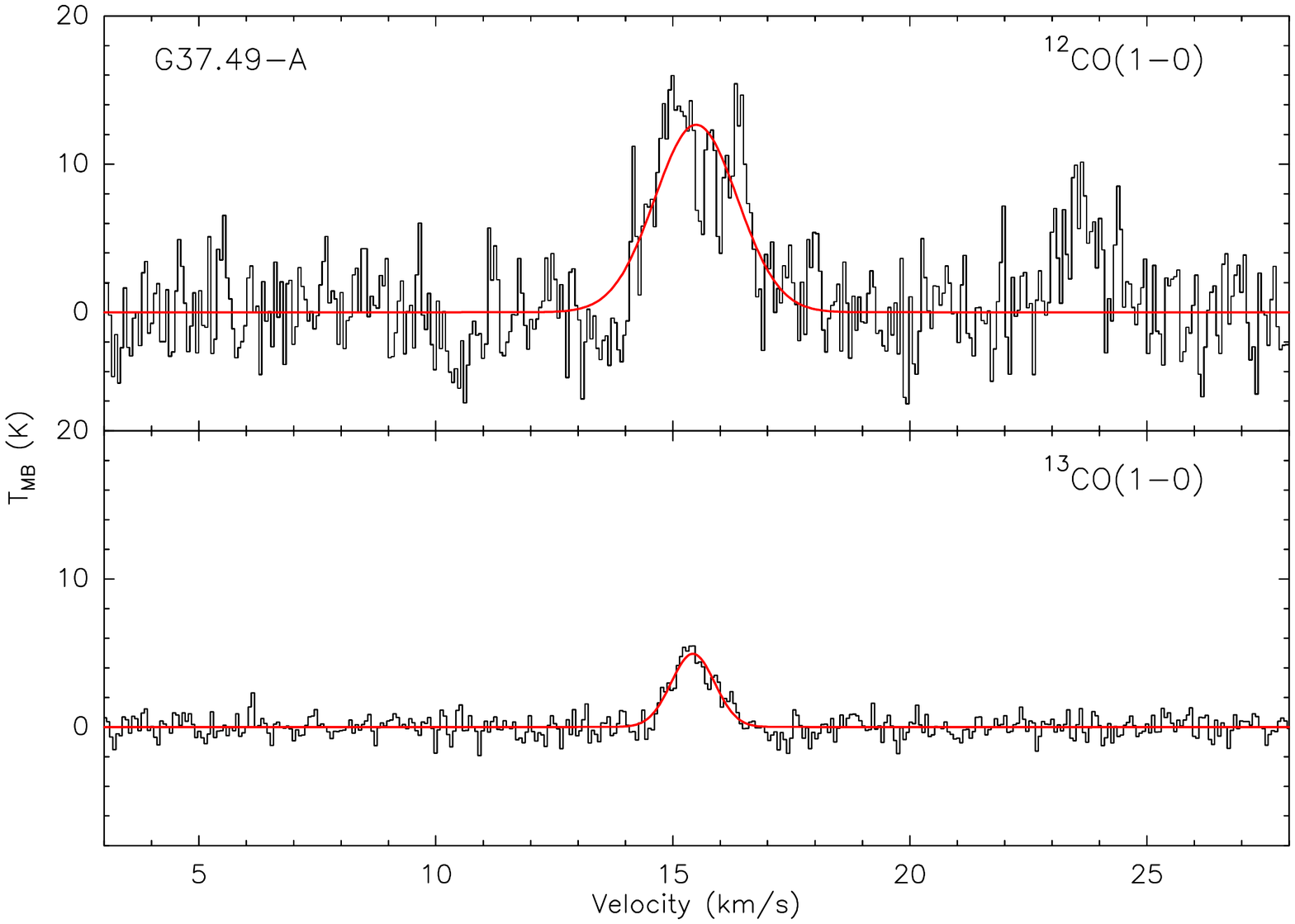}
		\includegraphics[angle=-90,width=.45\linewidth]{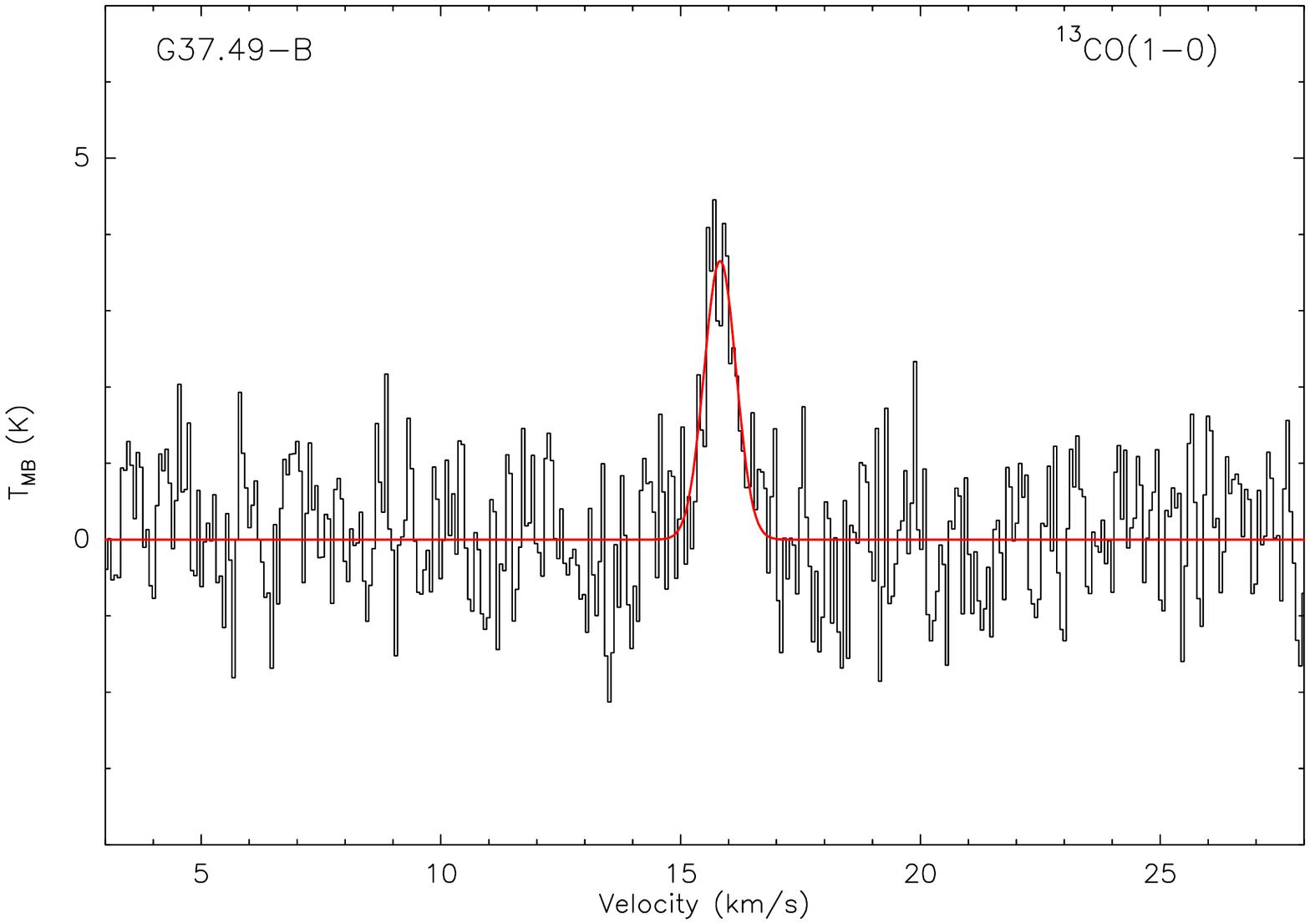}
		\includegraphics[angle=-90,width=.45\linewidth]{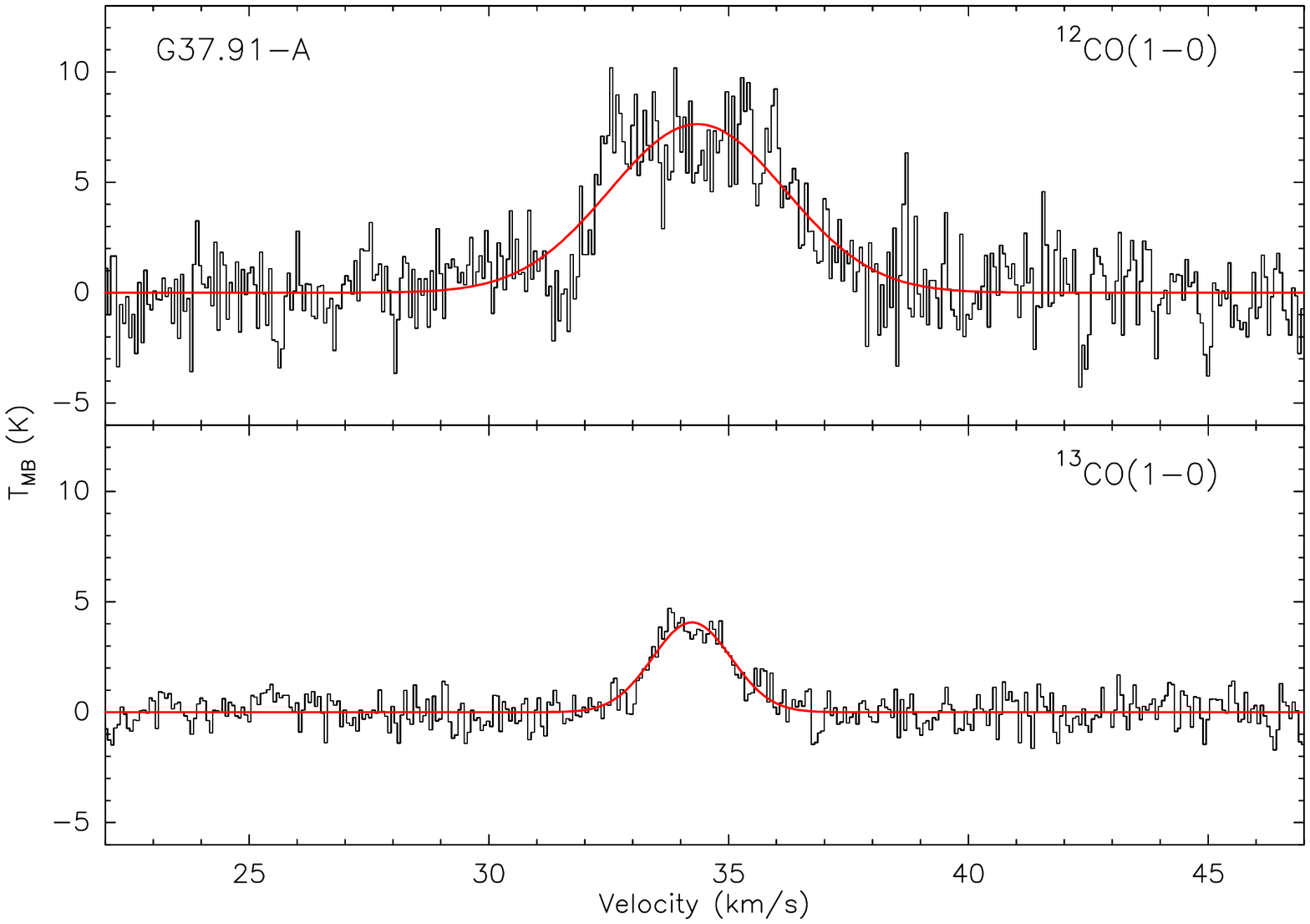}
		\includegraphics[angle=-90,width=.45\linewidth]{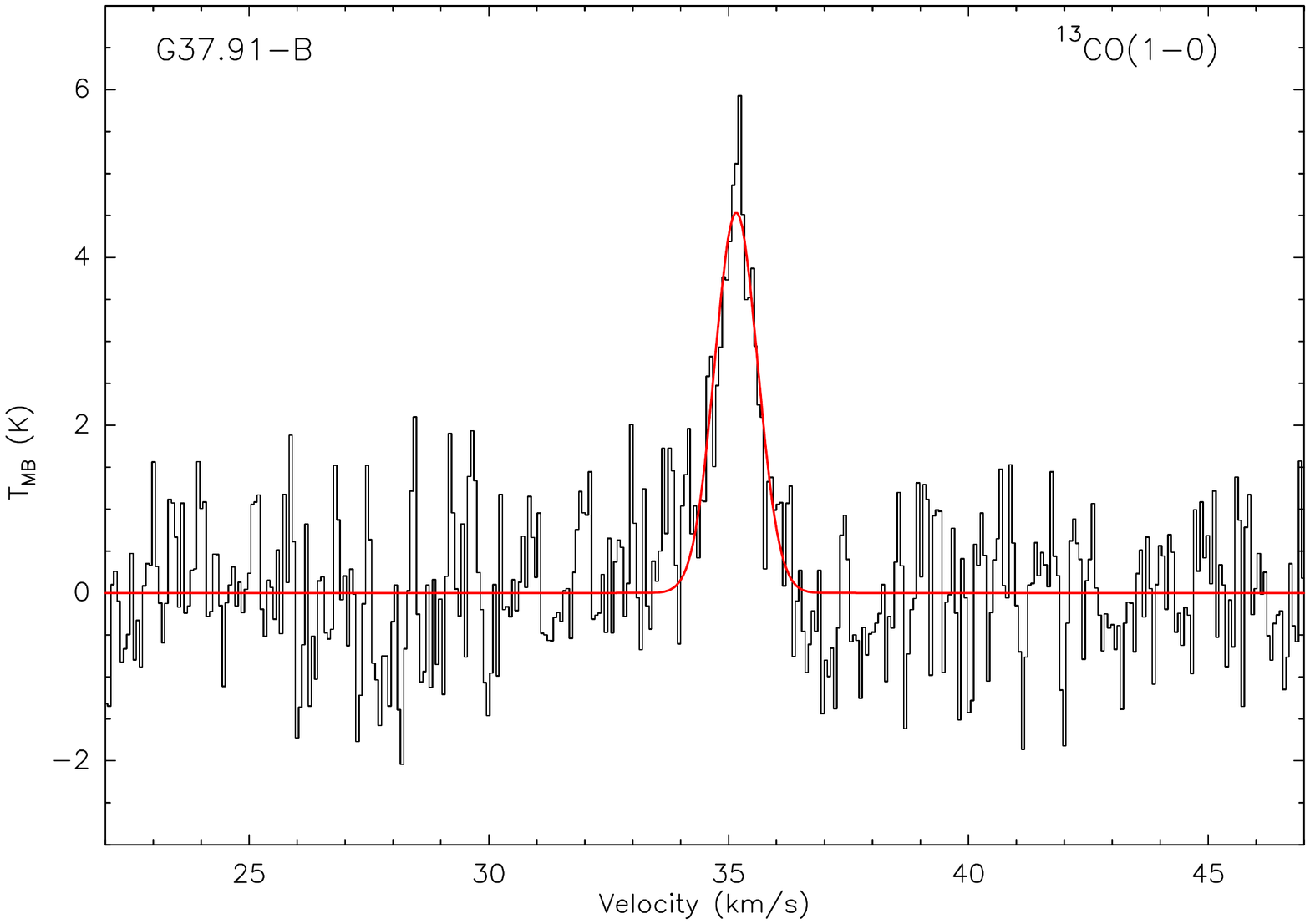}
		\includegraphics[angle=-90,width=.45\linewidth]{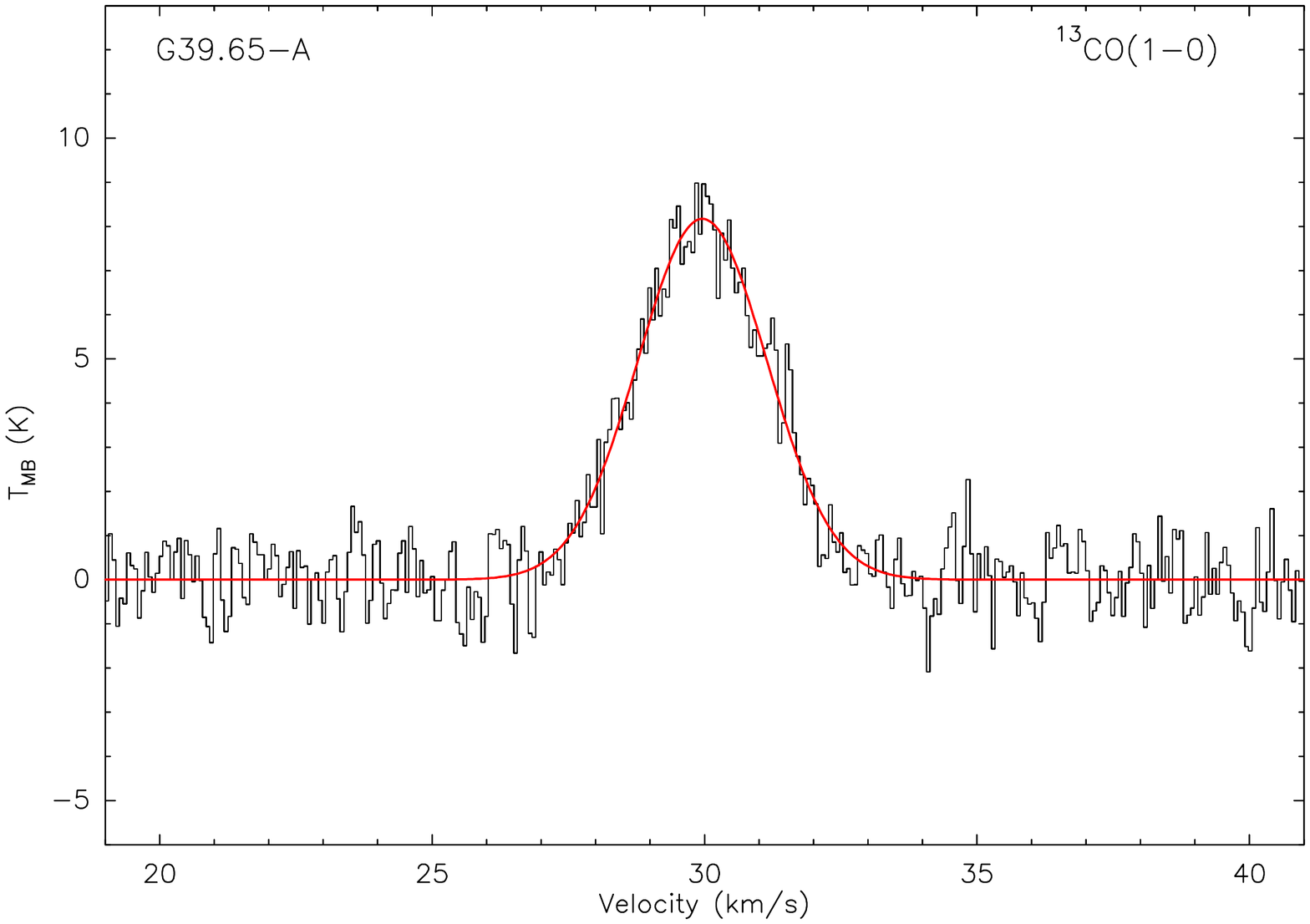}
		\includegraphics[angle=-90,width=.45\linewidth]{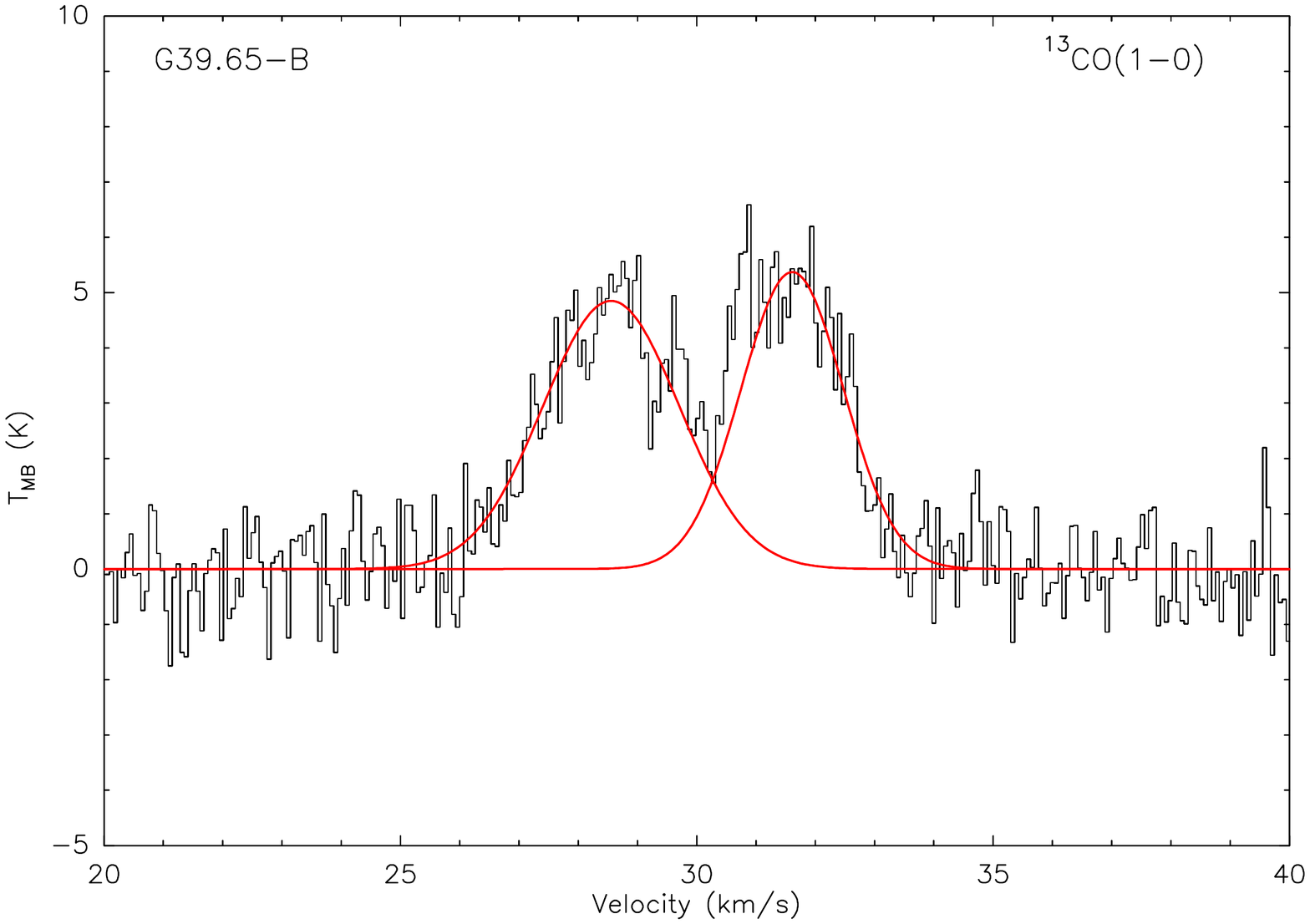}
		\includegraphics[angle=-90,width=.45\linewidth]{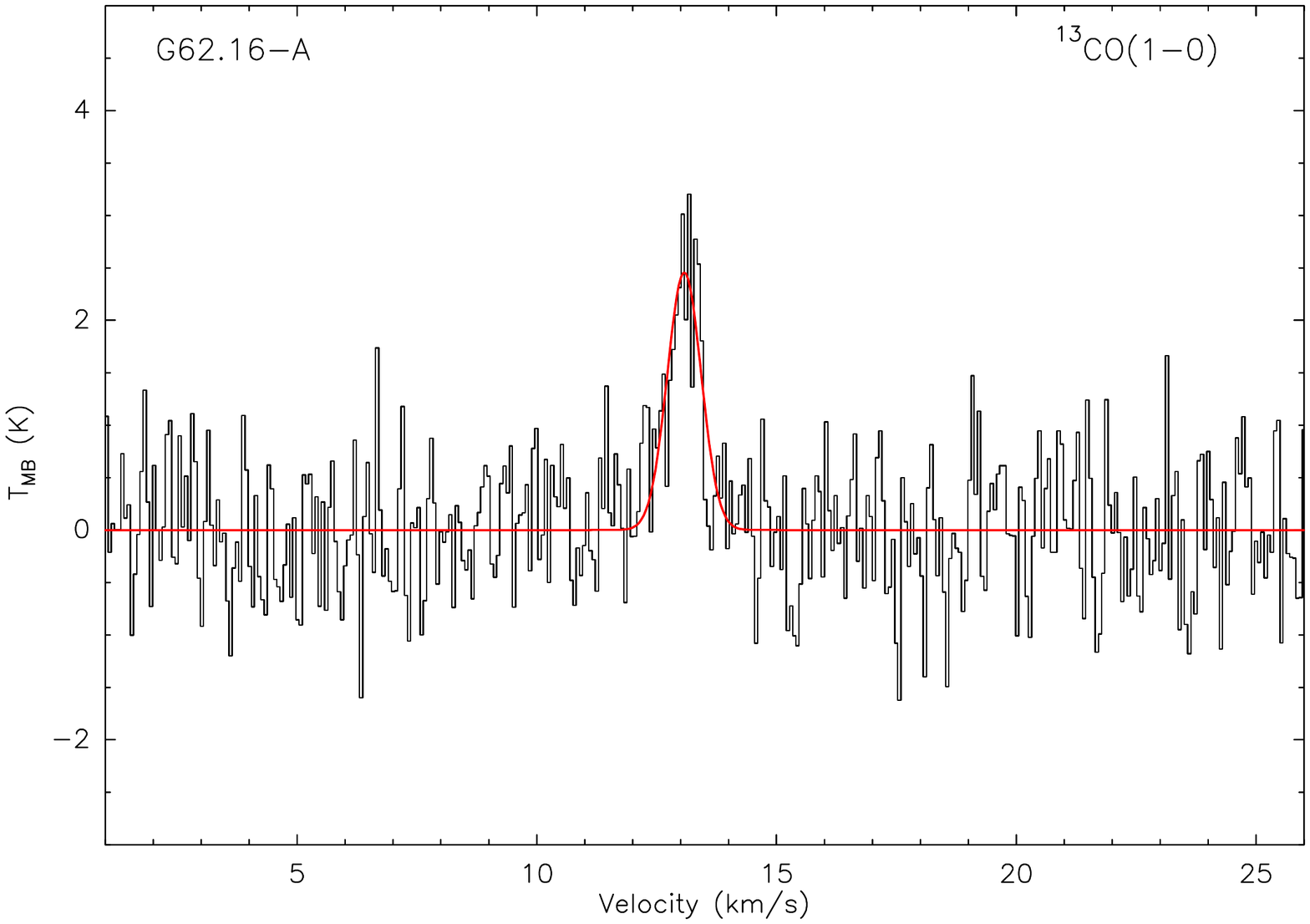}
     \caption{$^{12}$CO(1$-$0) and $^{13}$CO(1$-$0) spectra at the central position of each observed clump. Red line shows the Gaussian profile fit(s) to the lines.}
	\label{spec1}
\end{figure*}
\begin{figure*}[th]
	\centering		
    	\includegraphics[angle=-90,width=.45\linewidth]{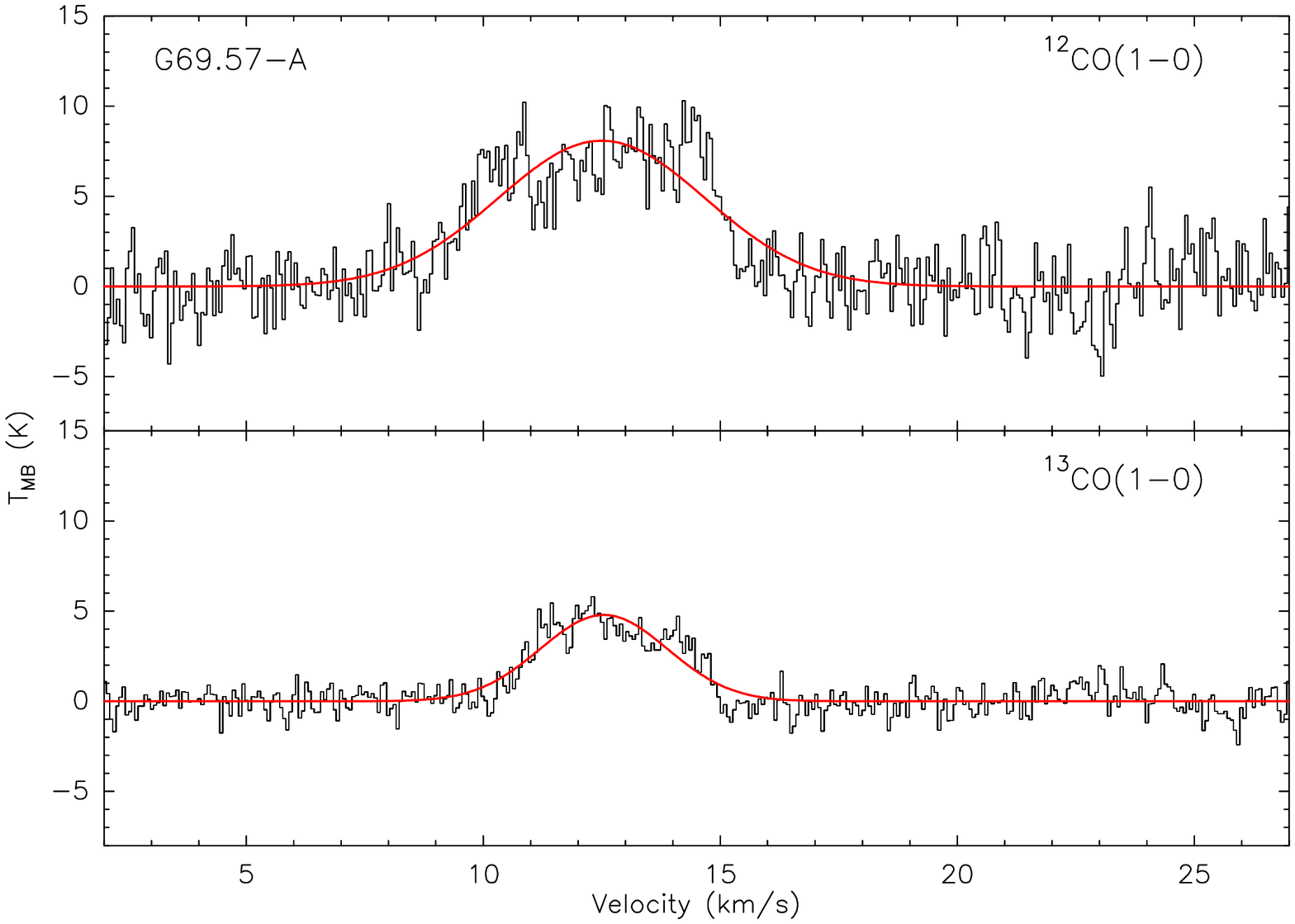}
		\includegraphics[angle=-90,width=.45\linewidth]{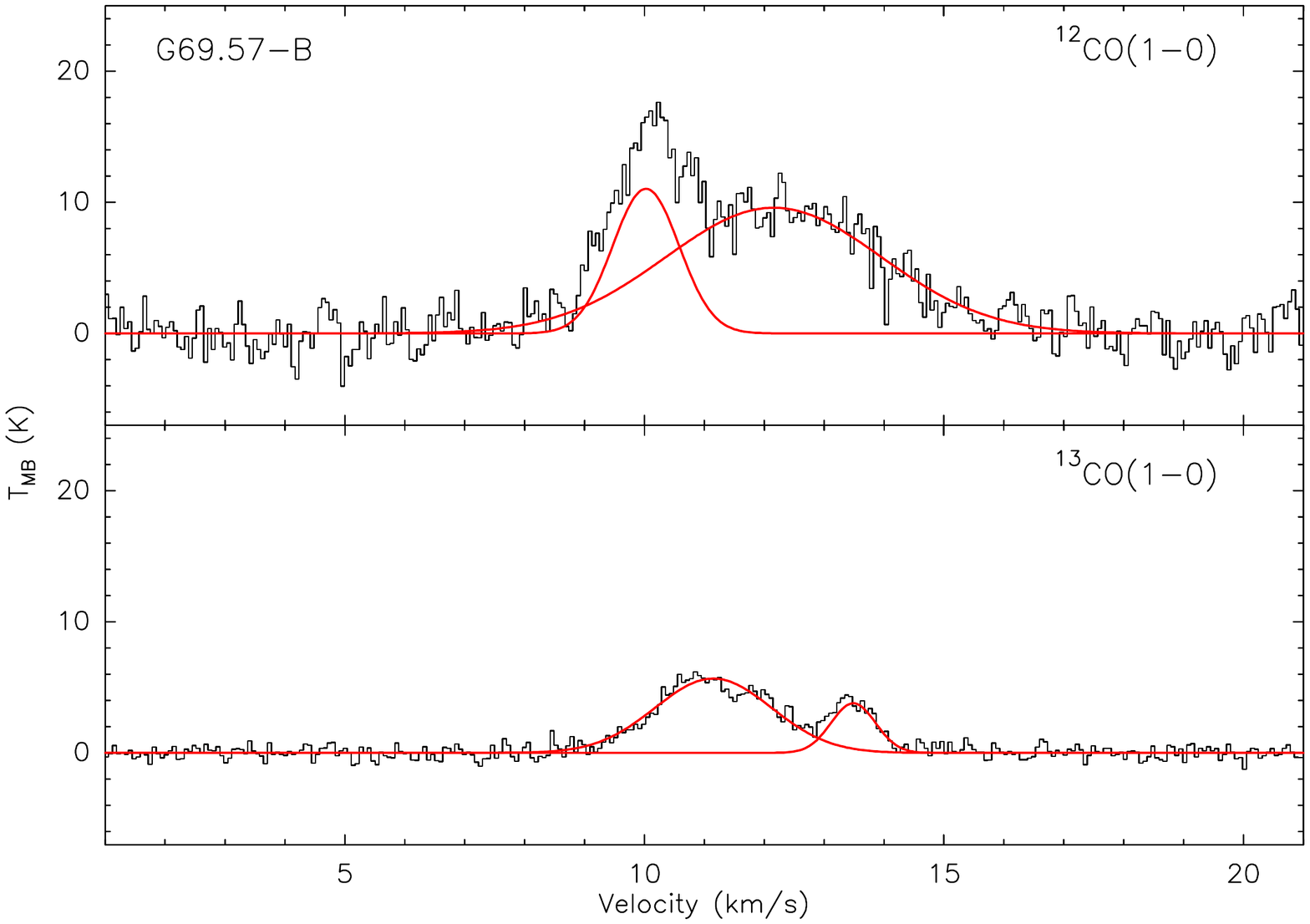}
		\includegraphics[angle=-90,width=.45\linewidth]{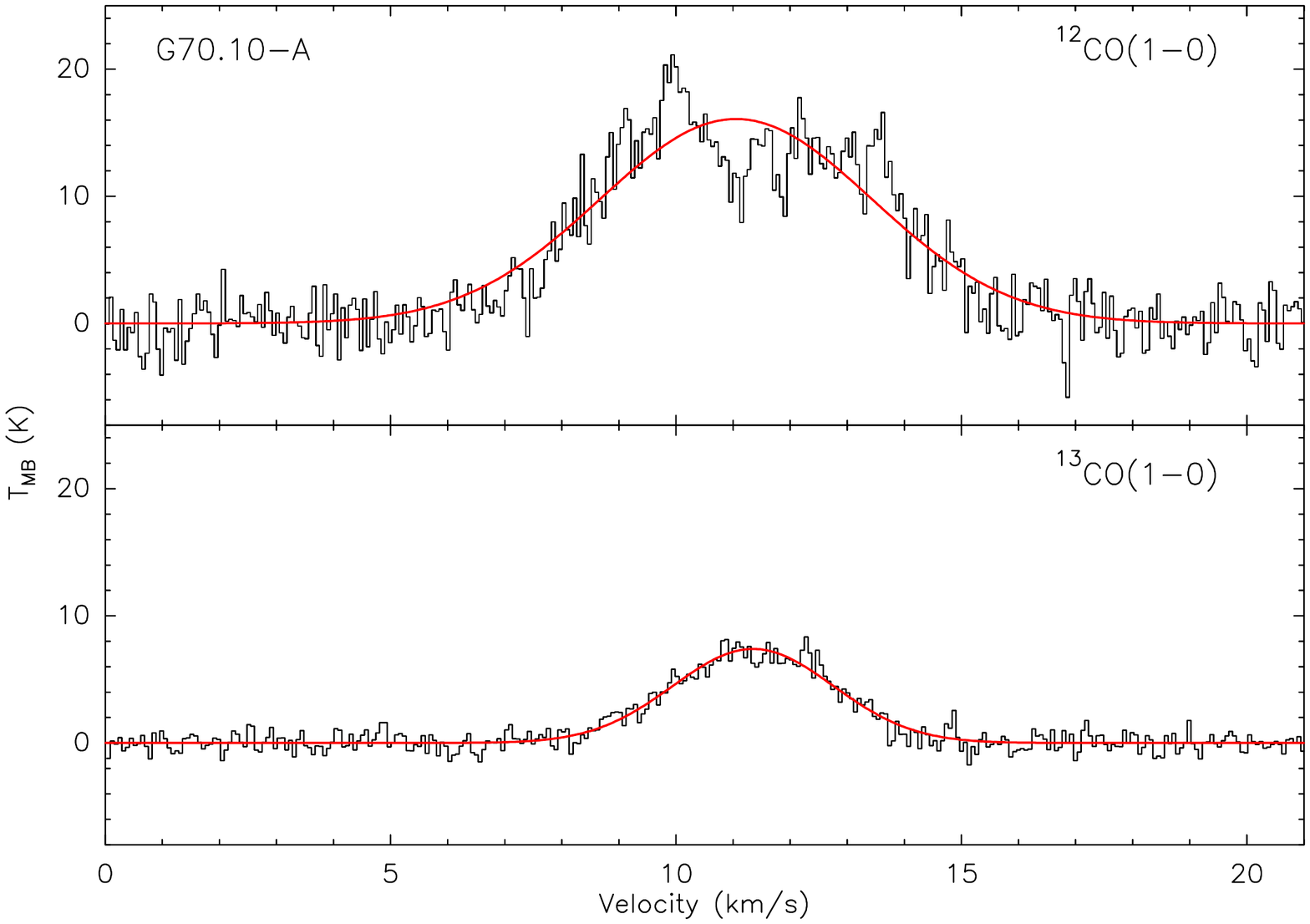}
		\includegraphics[angle=-90,width=.45\linewidth]{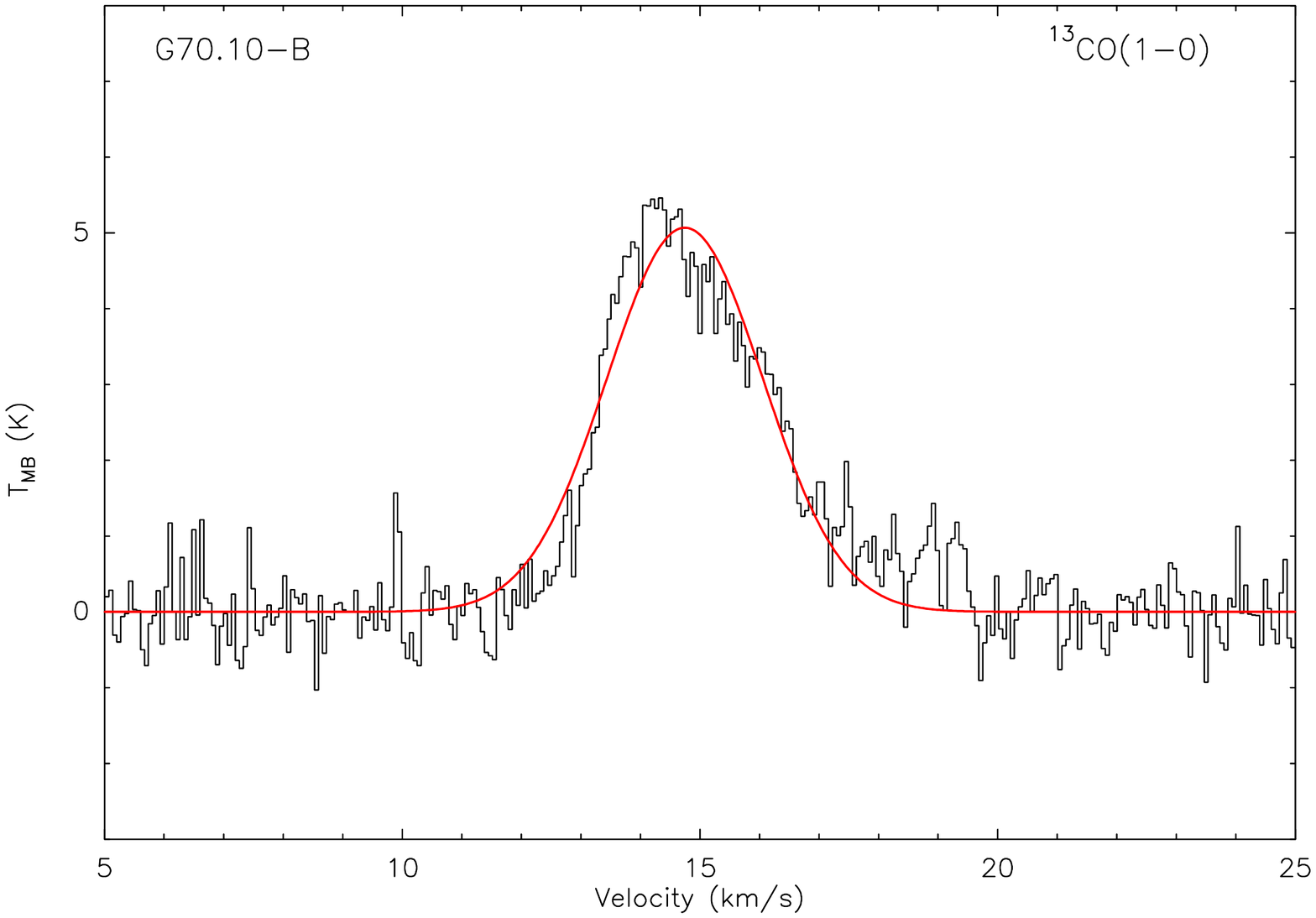}
		\includegraphics[angle=-90,width=.45\linewidth]{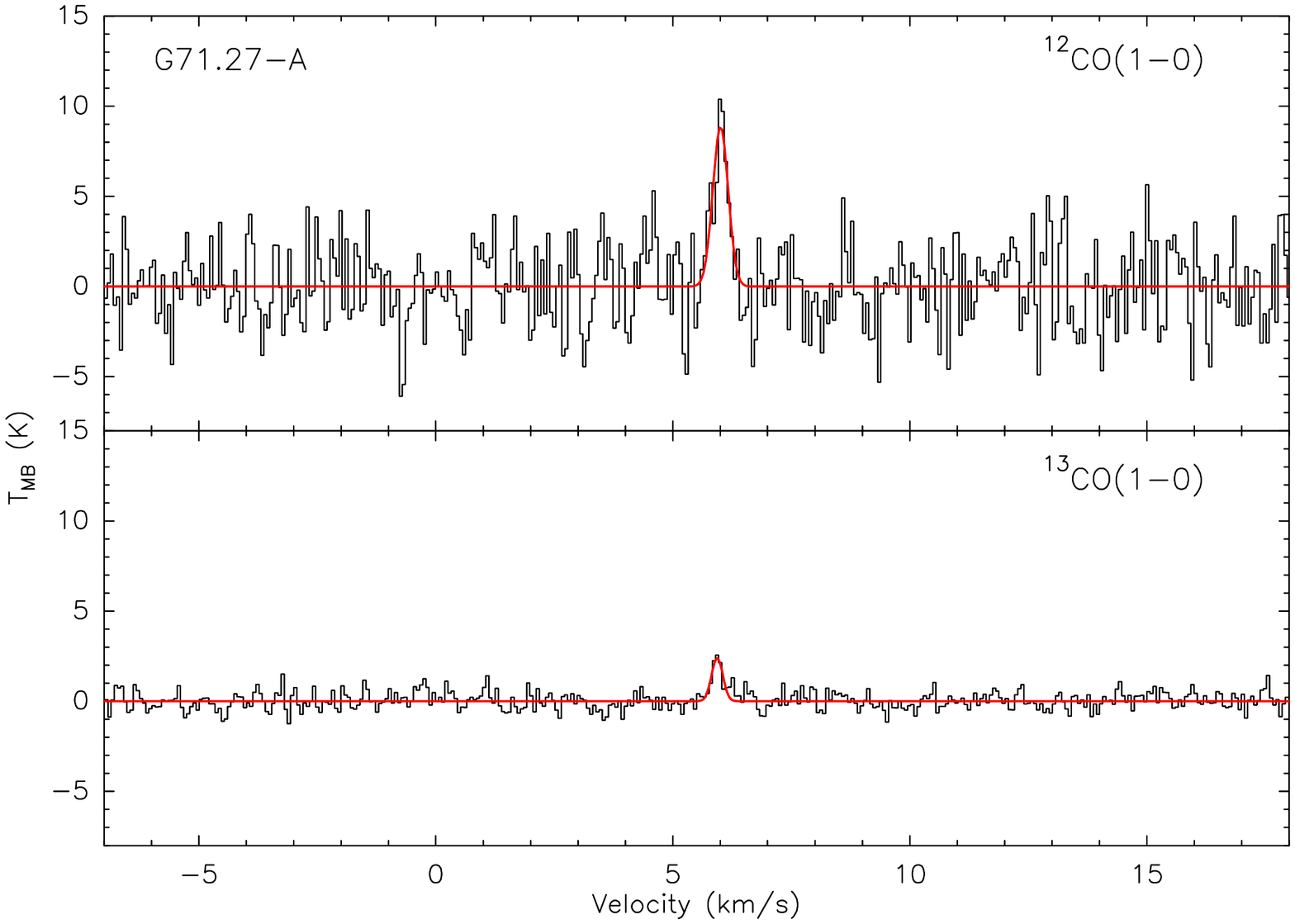}
		\includegraphics[angle=-90,width=.45\linewidth]{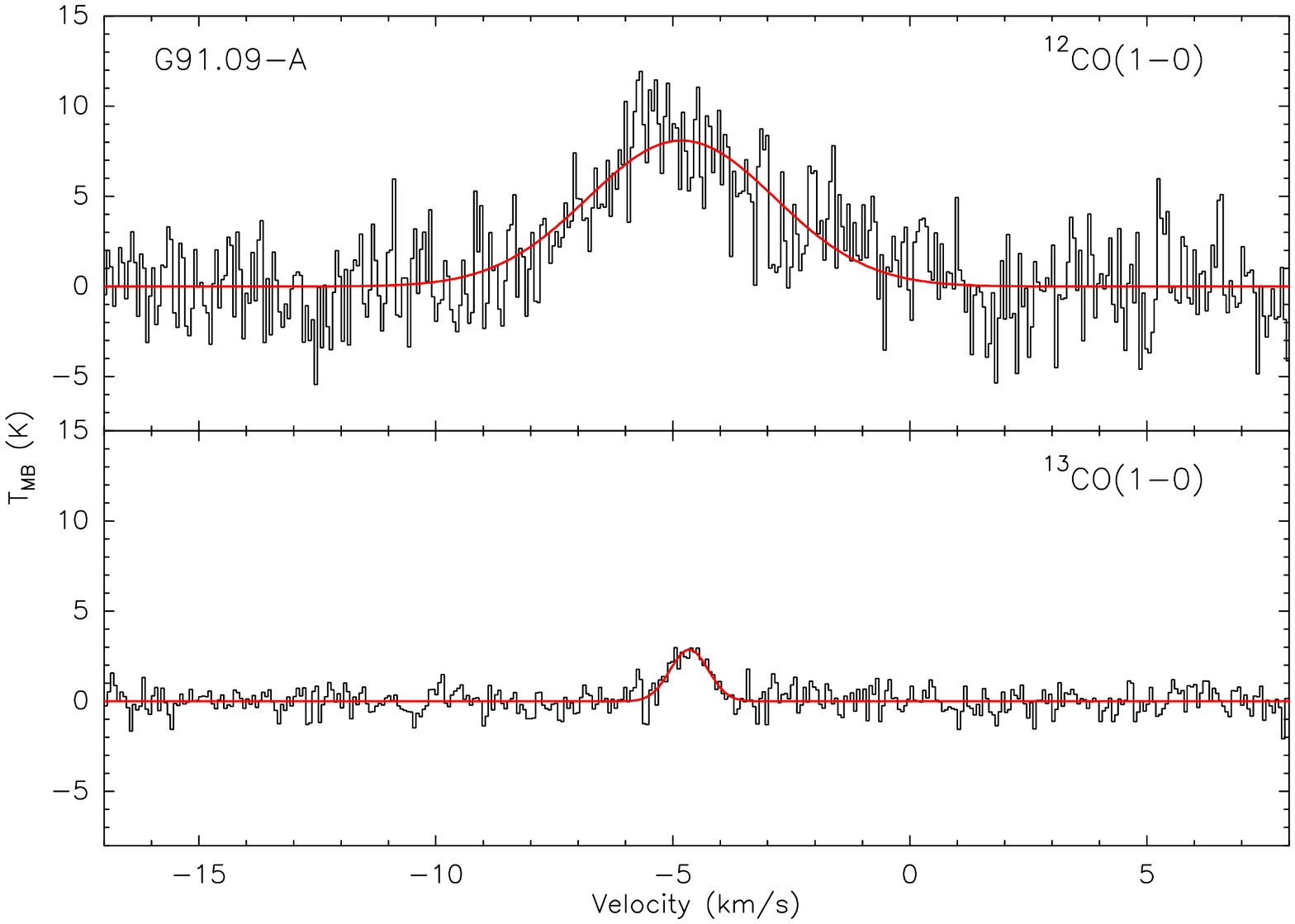}
		\includegraphics[angle=-90,width=.45\linewidth]{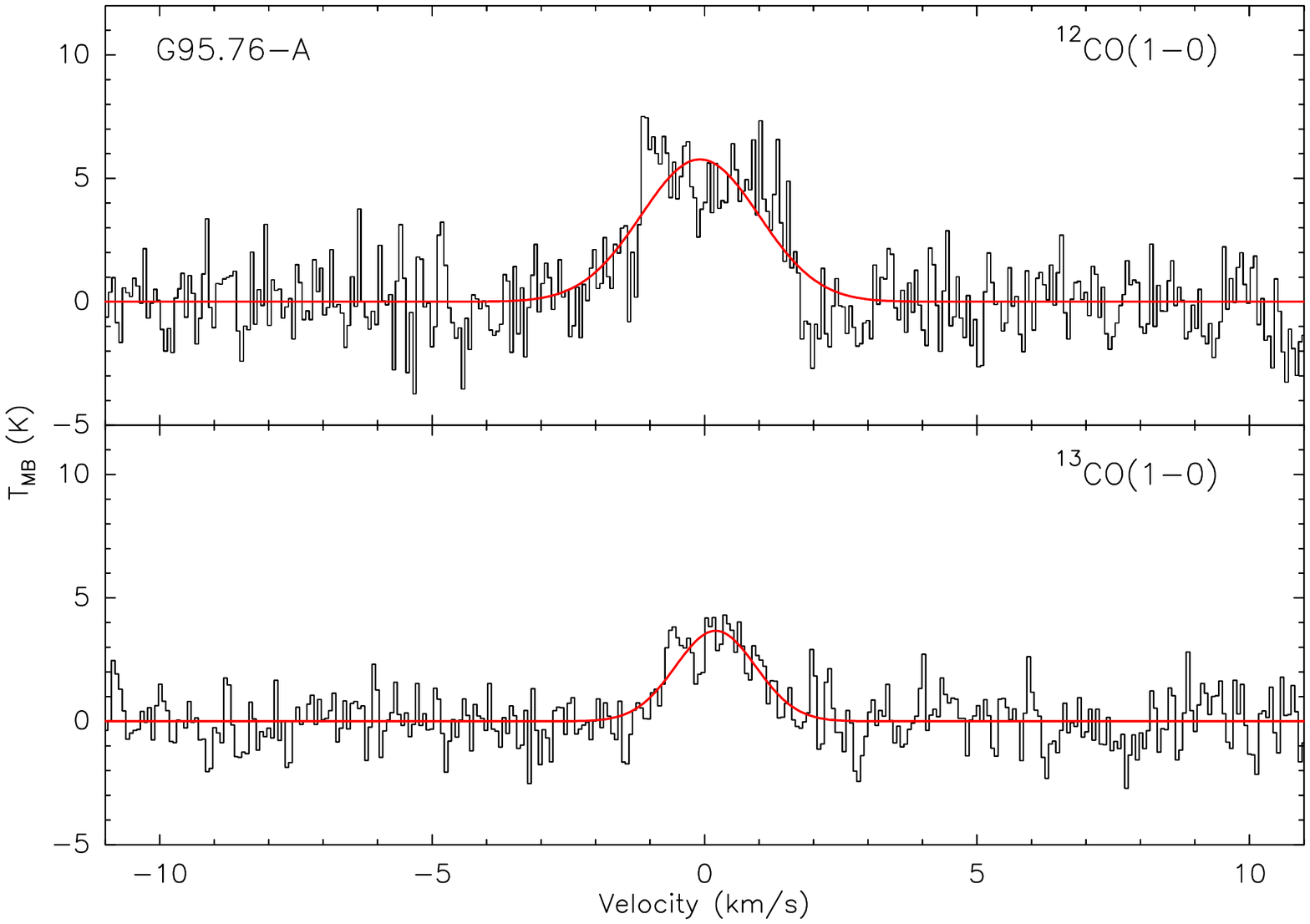}
		\includegraphics[angle=-90,width=.45\linewidth]{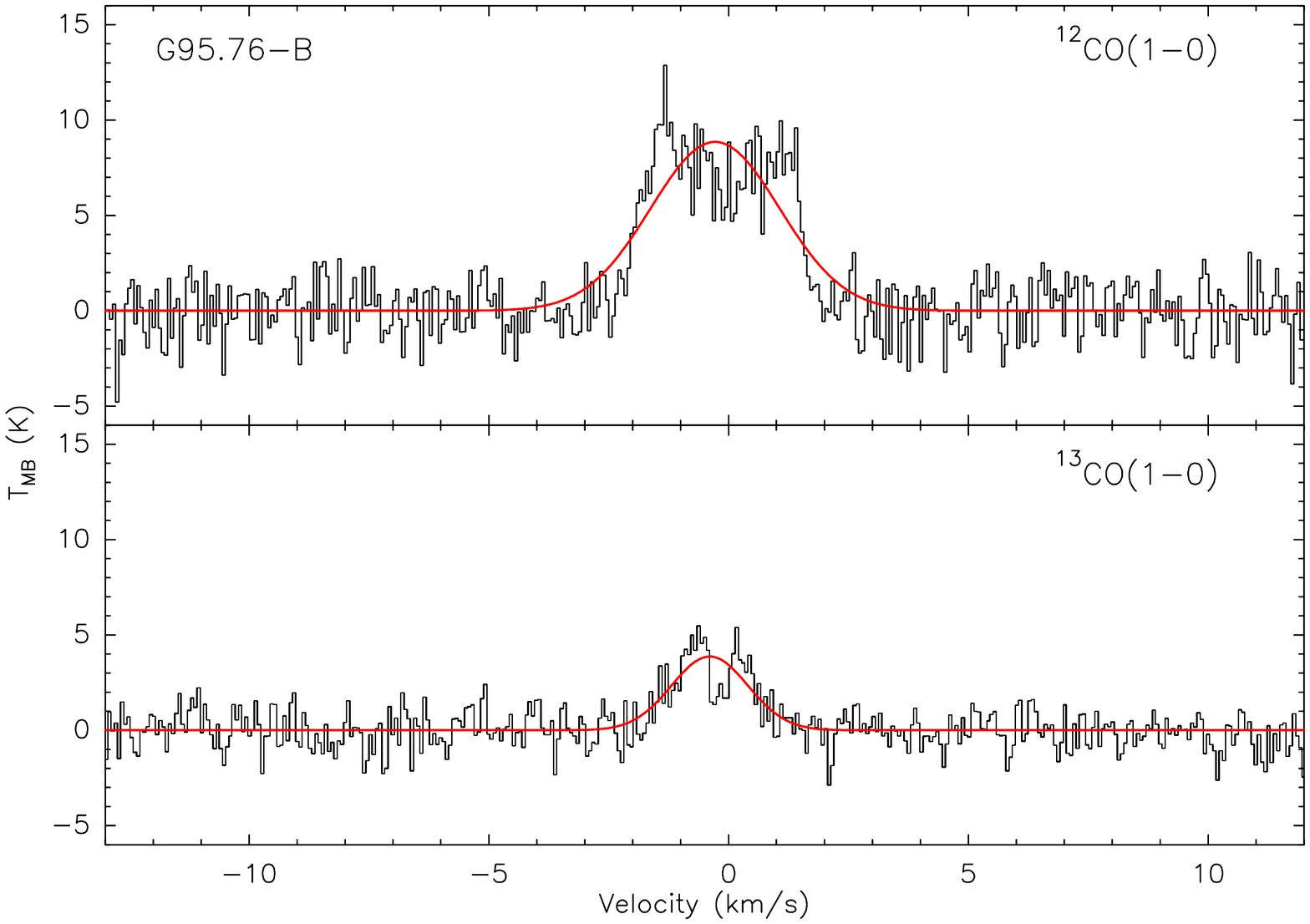}
    \caption{Cont. $^{12}$CO(1$-$0) and $^{13}$CO(1$-$0) spectra at the central position of each observed clump.}
	\label{spec2}
\end{figure*}
\begin{figure*}[th]
	\centering	
		\includegraphics[angle=-90,width=.45\linewidth]{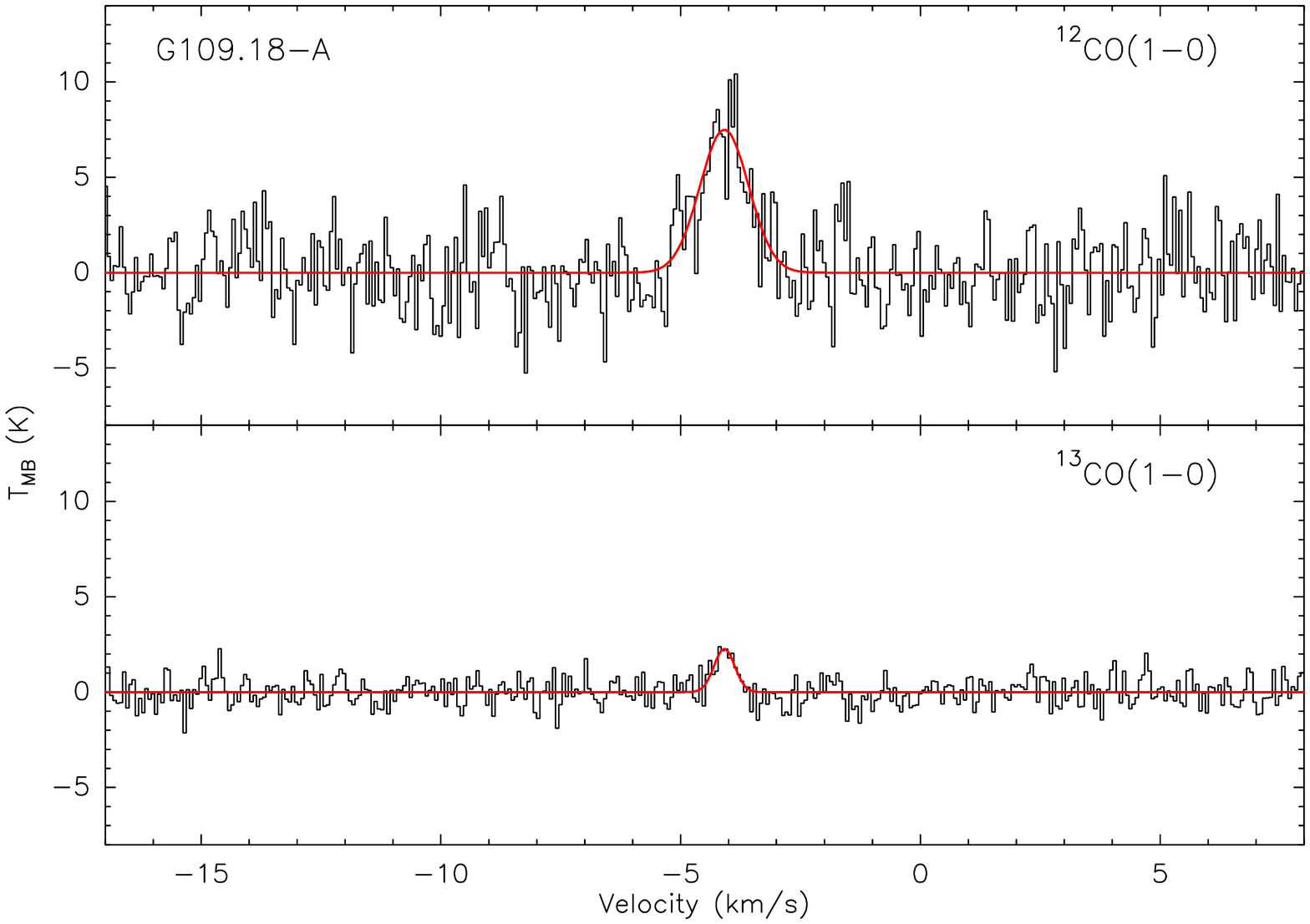}
        \includegraphics[angle=-90,width=.45\linewidth]{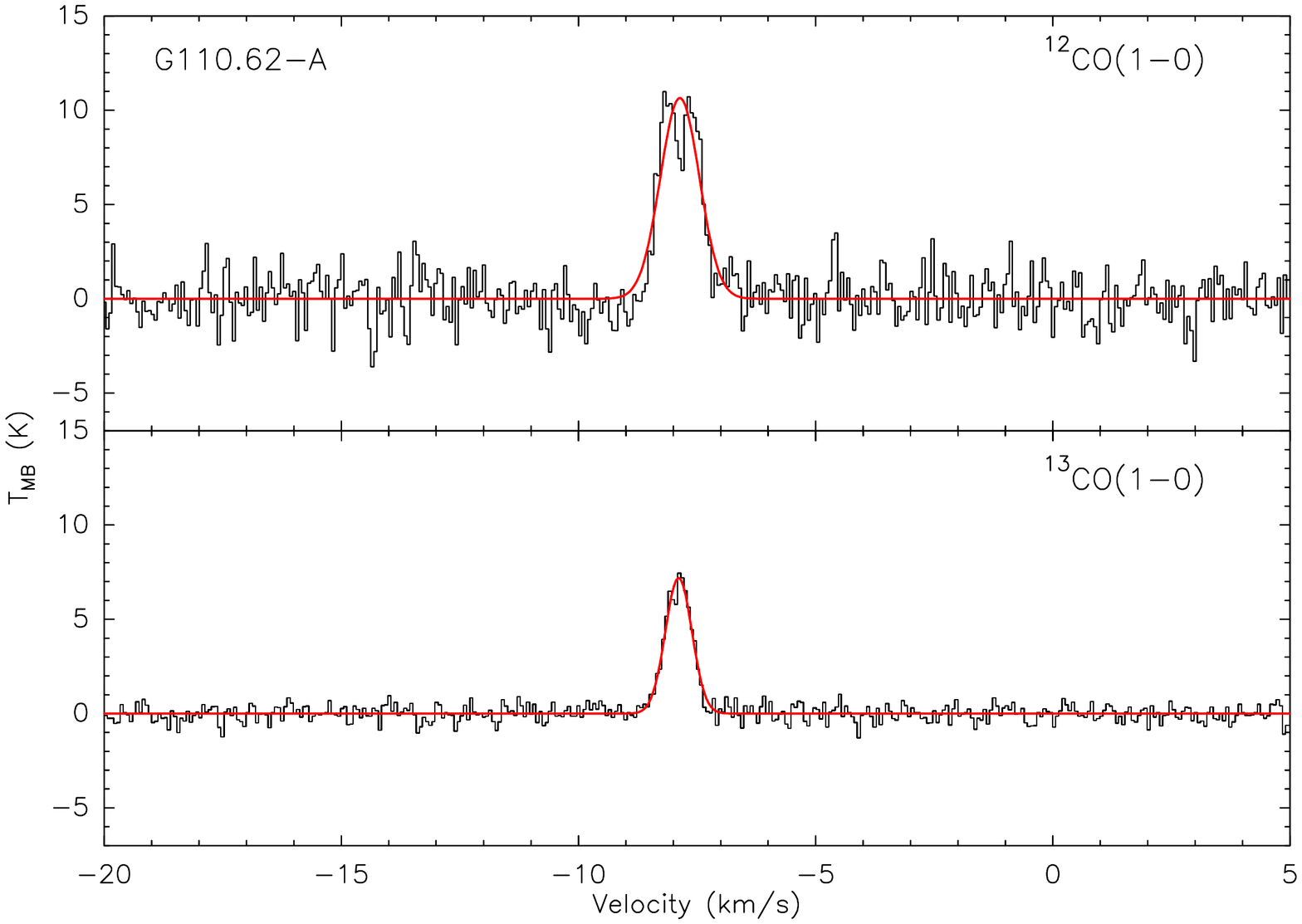}
		\includegraphics[angle=-90,width=.45\linewidth]{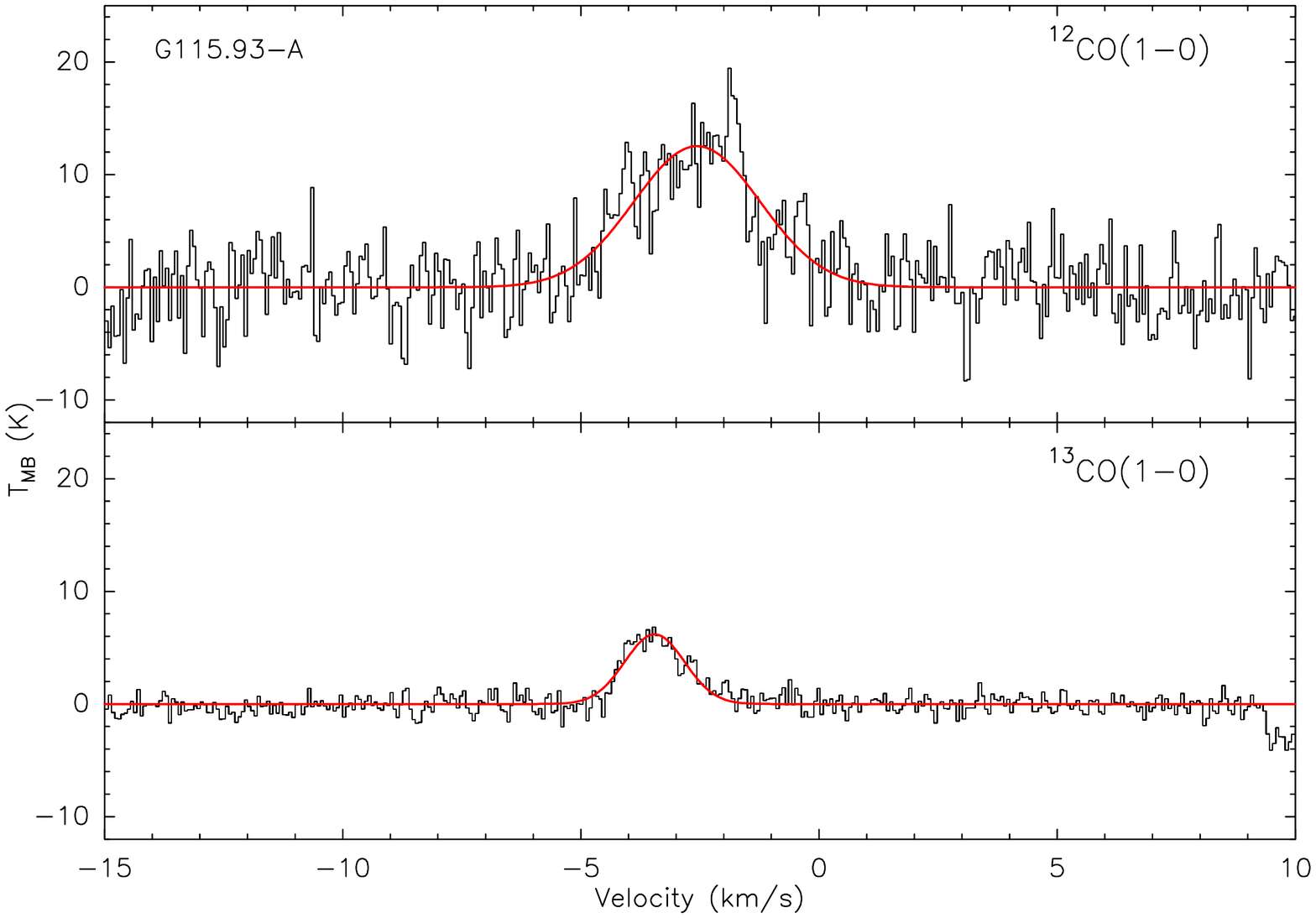}
		\includegraphics[angle=-90,width=.45\linewidth]{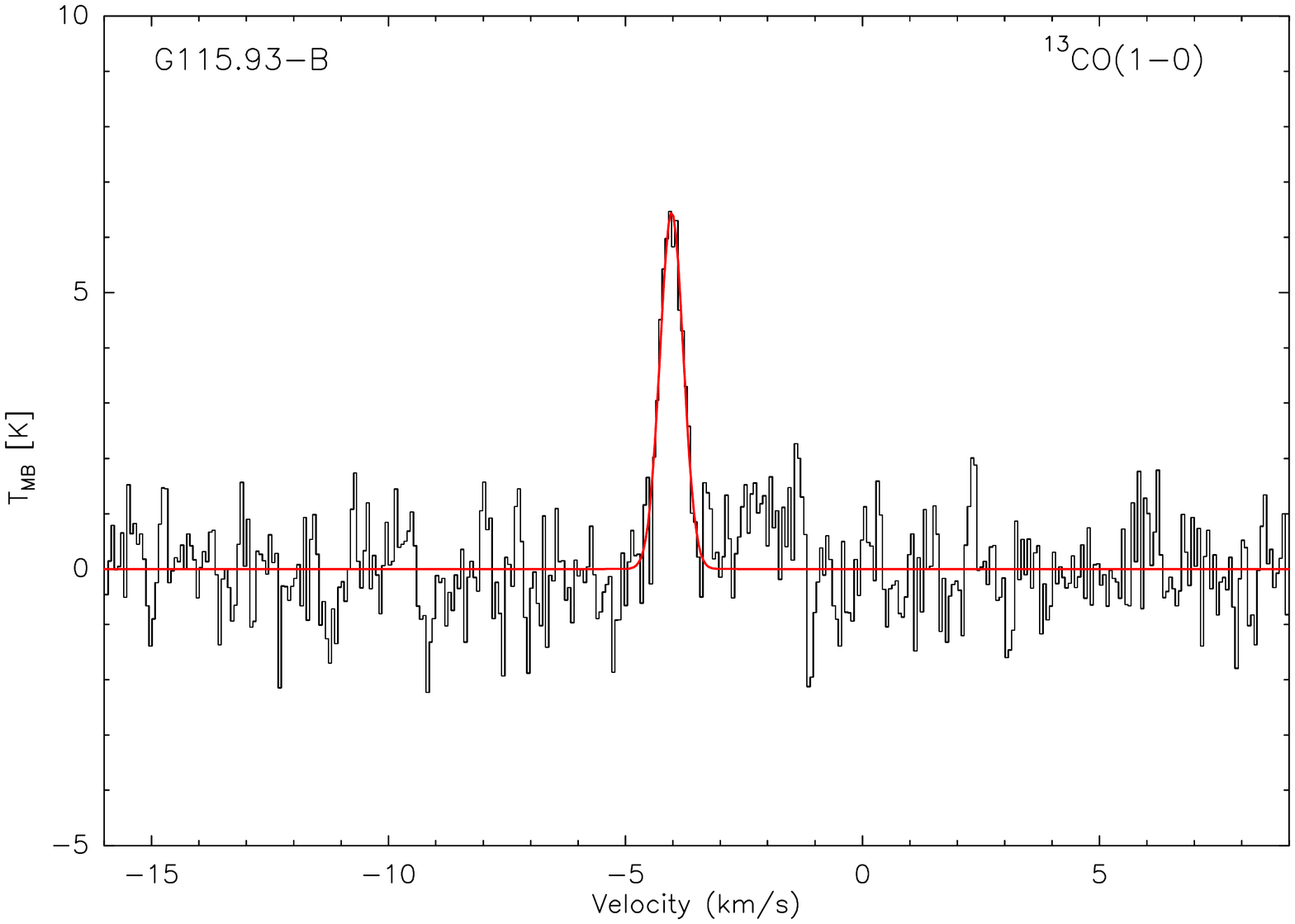}
		\includegraphics[angle=-90,width=.45\linewidth]{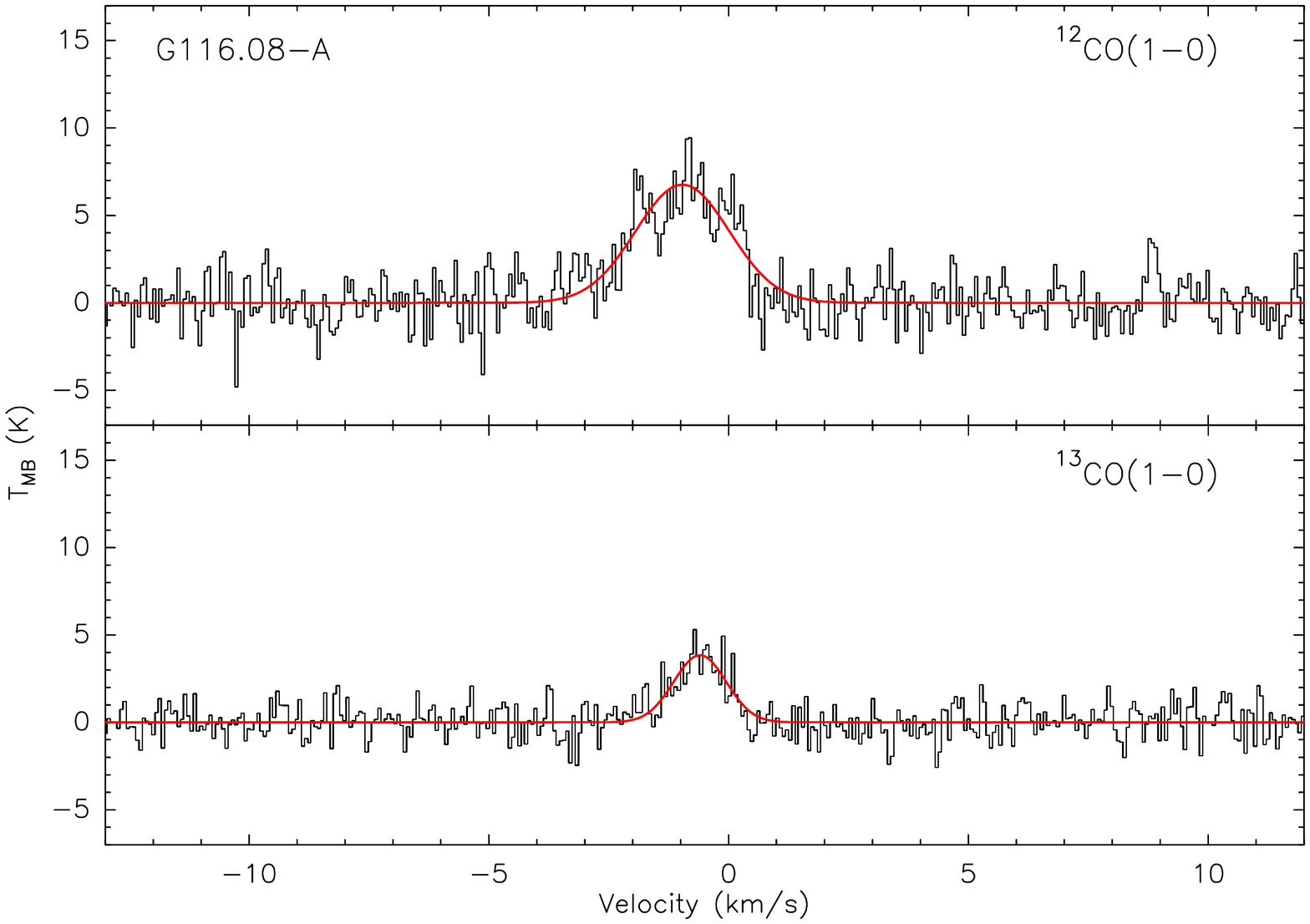}
		\includegraphics[angle=-90,width=.45\linewidth]{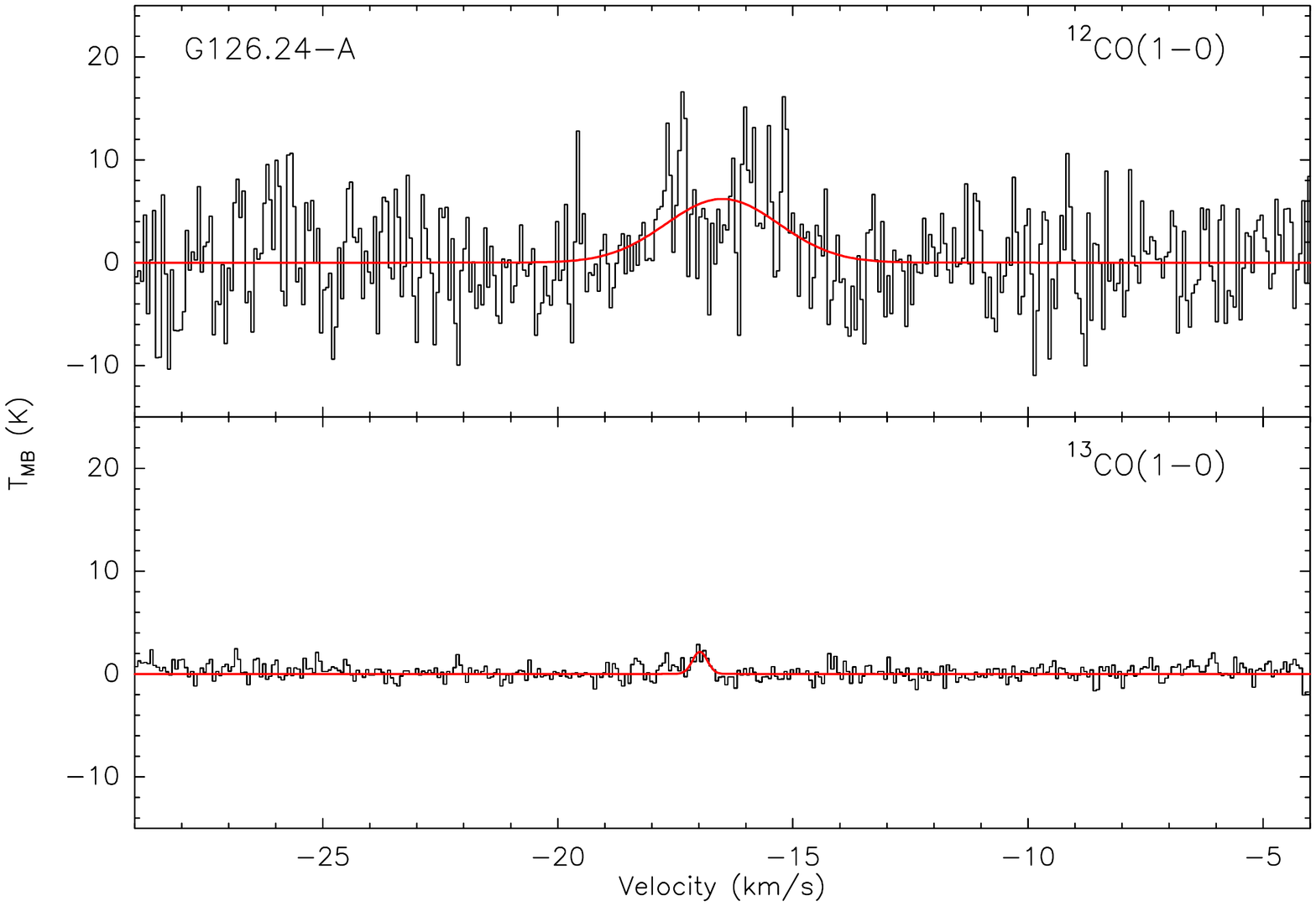}
		\includegraphics[angle=-90,width=.45\linewidth]{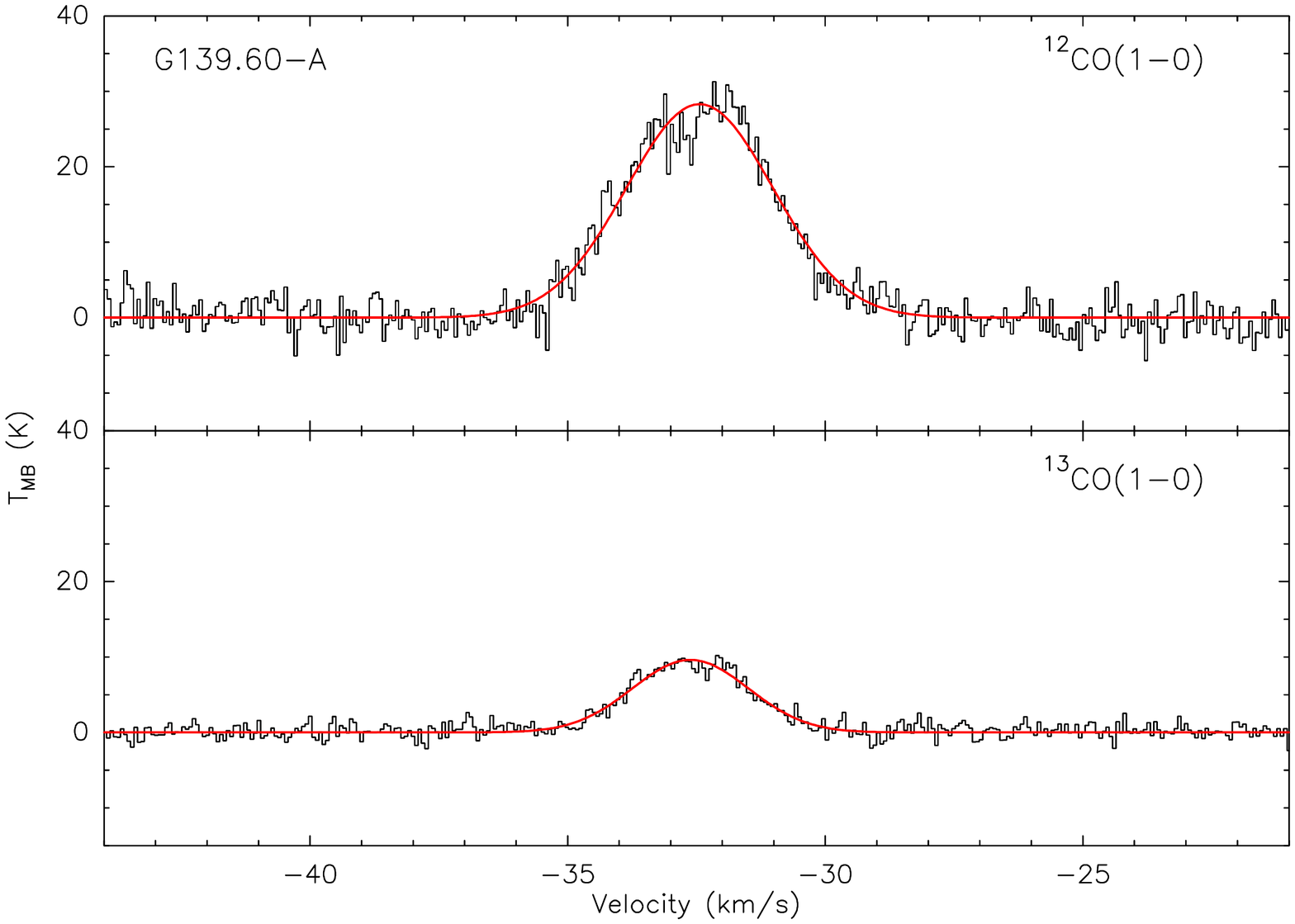}
		\includegraphics[angle=-90,width=.45\linewidth]{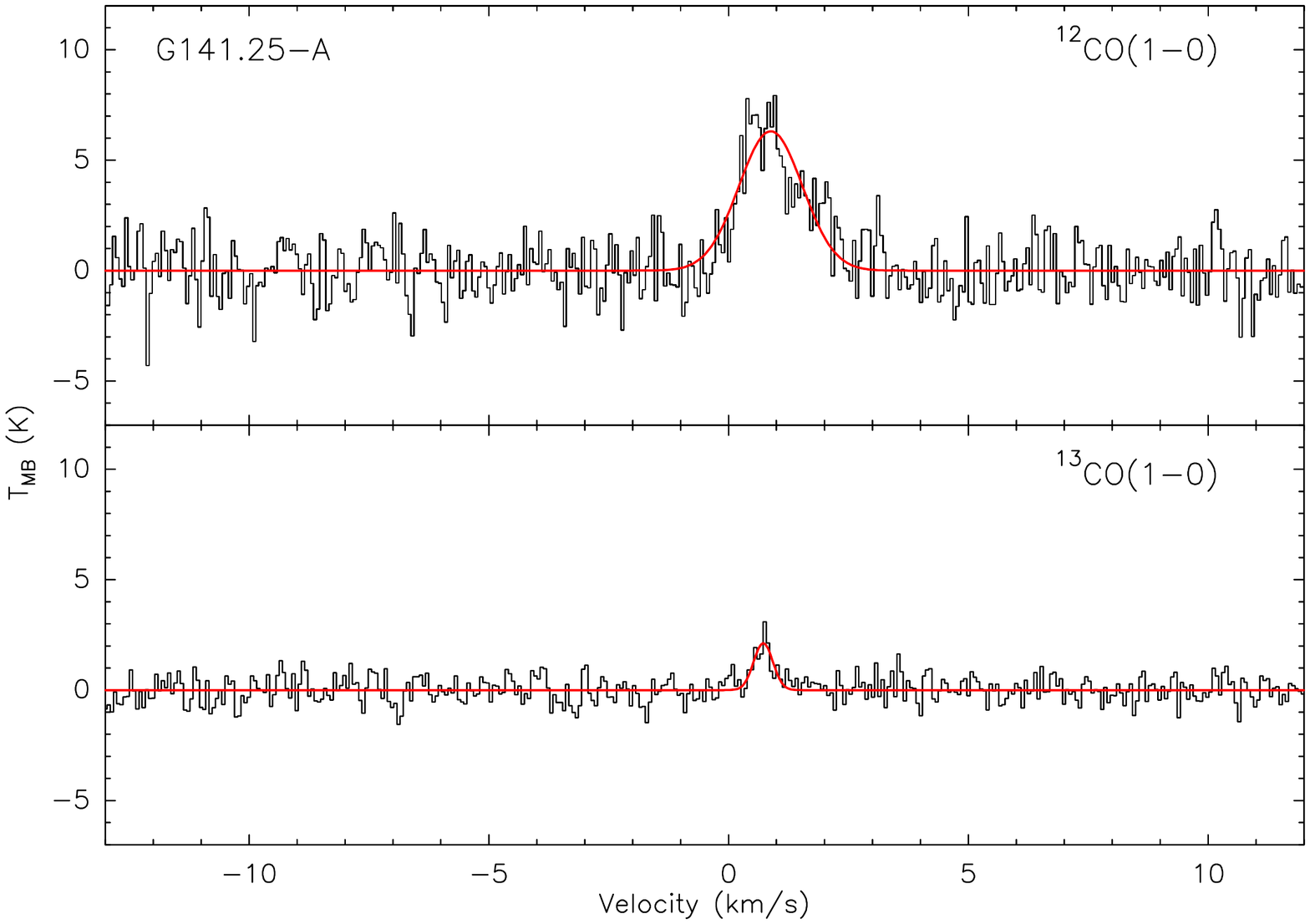}
    \caption{Cont. $^{12}$CO(1$-$0) and $^{13}$CO(1$-$0) spectra at the central position of each observed clump. Red line shows the Gaussian profile fit(s) to the lines.}
	\label{spec3}
\end{figure*}
\begin{figure*}[th]
	\centering		
		\includegraphics[angle=-90,width=.45\linewidth]{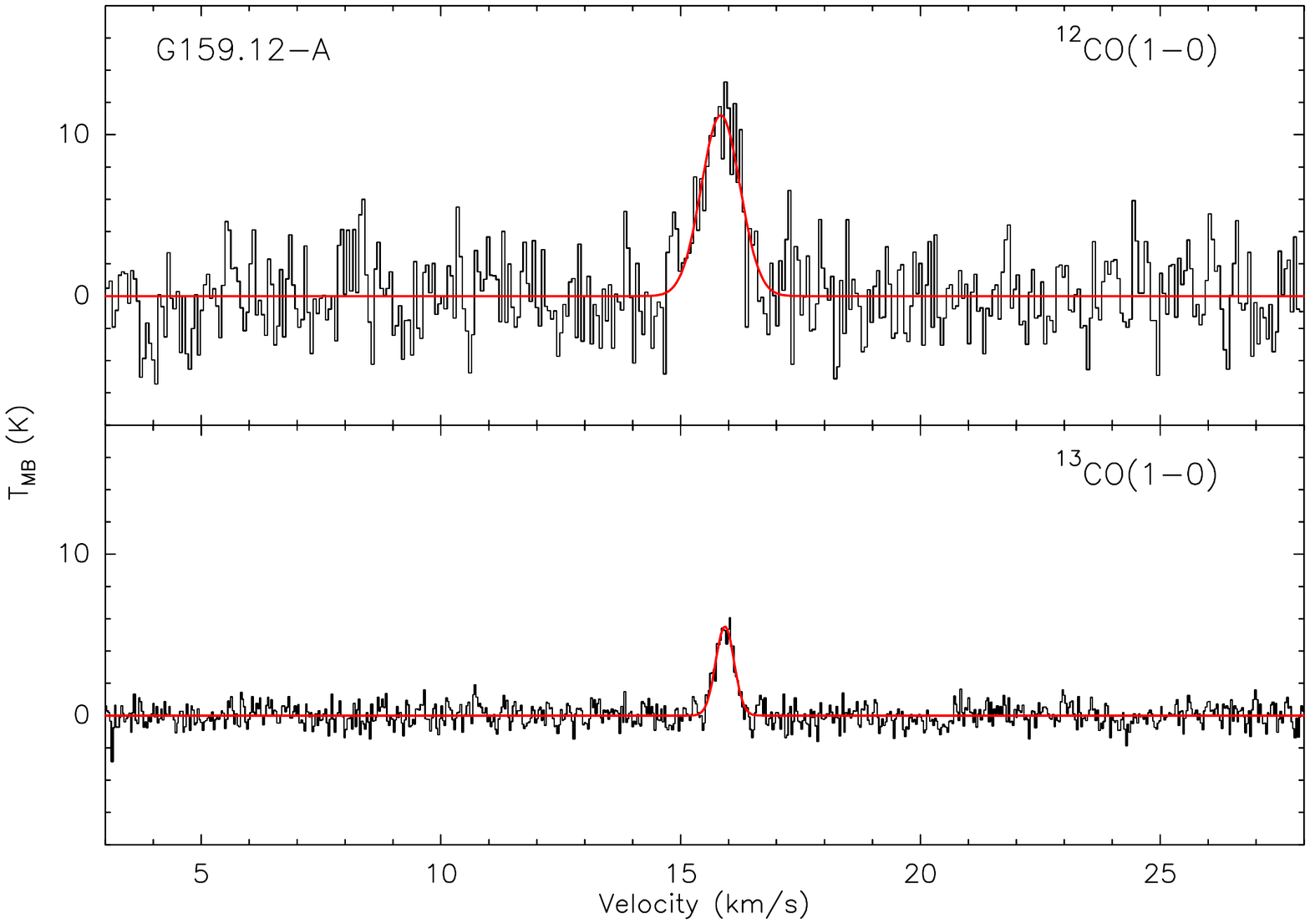}
        \includegraphics[angle=-90,width=.45\linewidth]{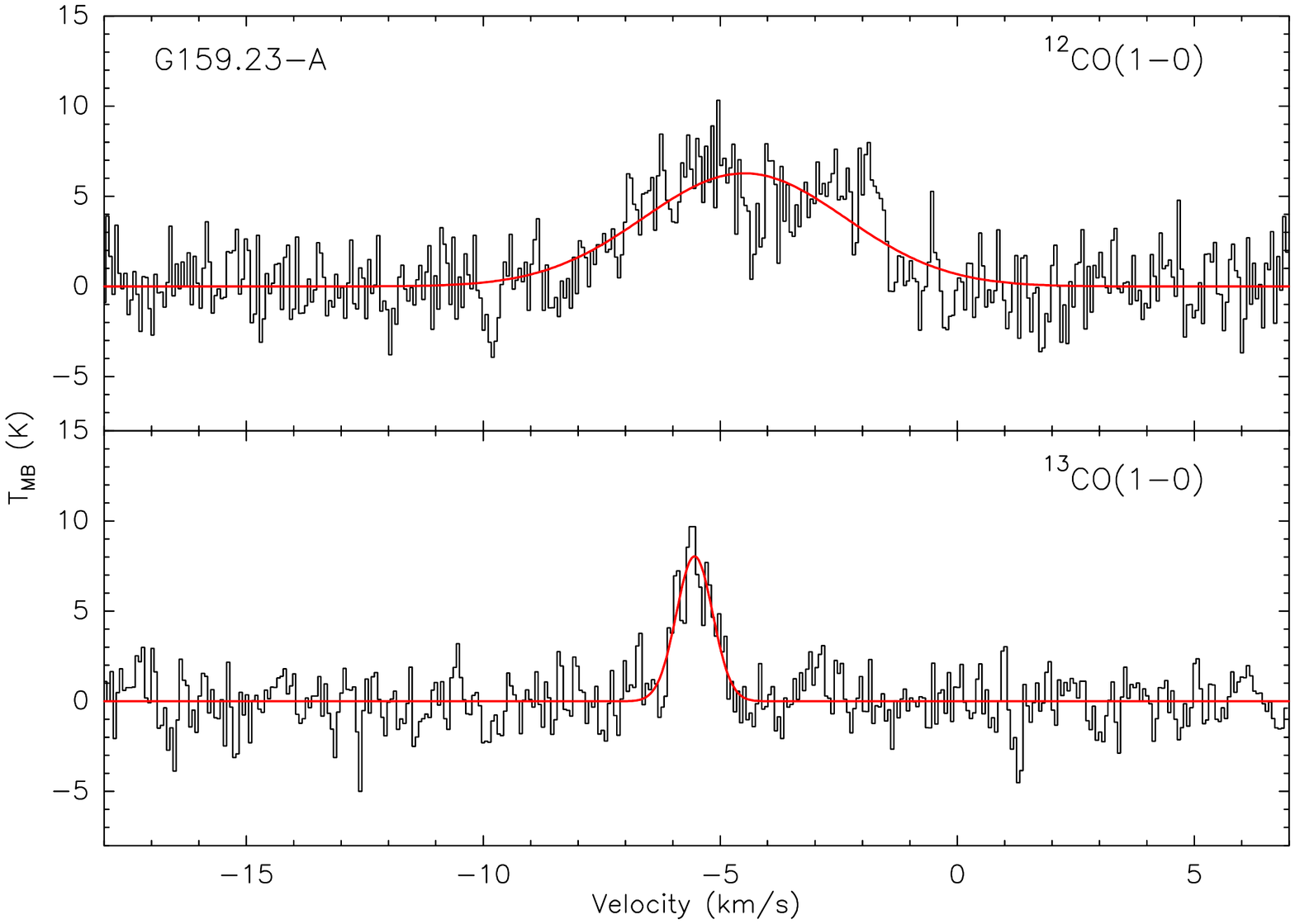}
		\includegraphics[angle=-90,width=.45\linewidth]{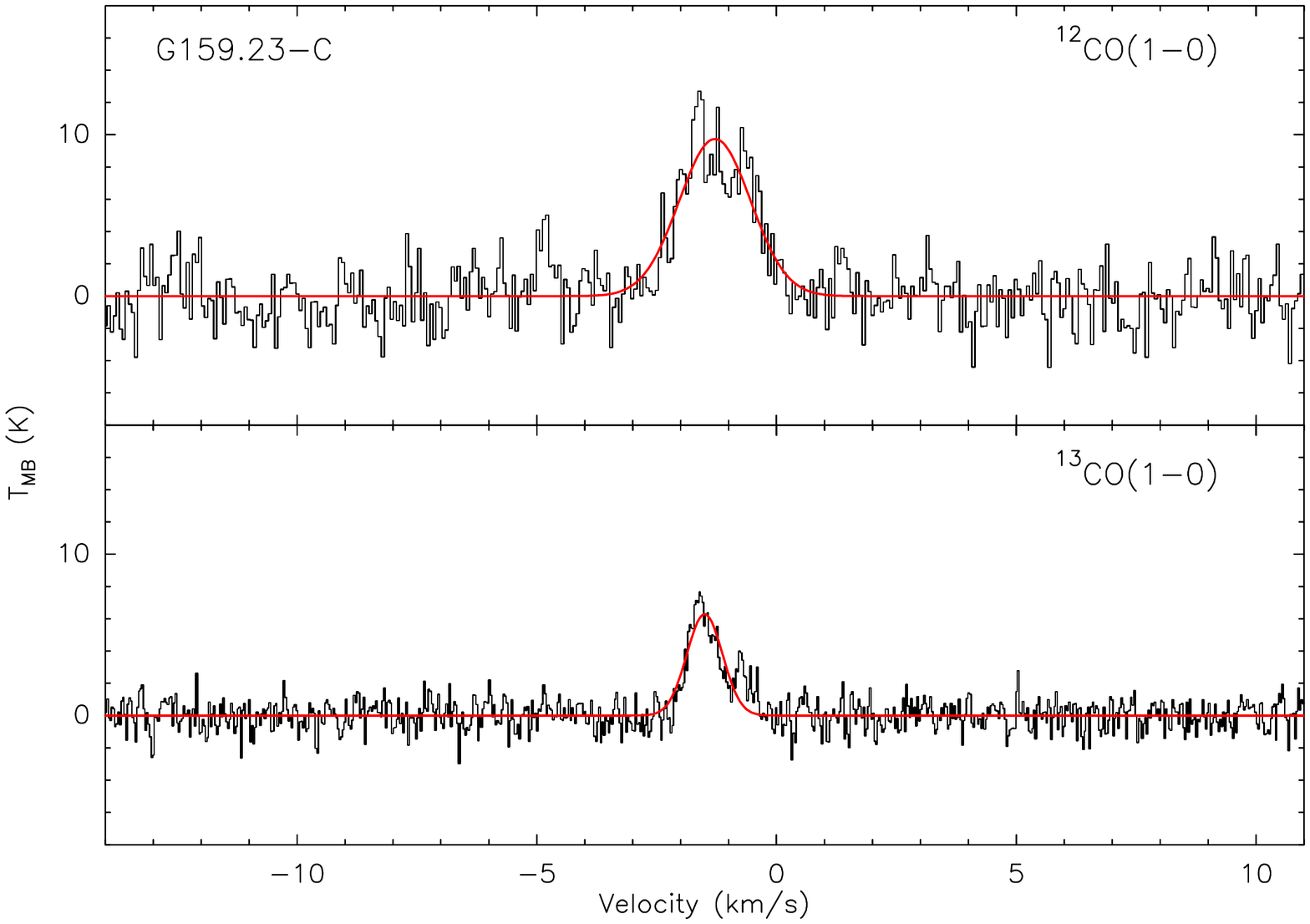}
		\includegraphics[angle=-90,width=.45\linewidth]{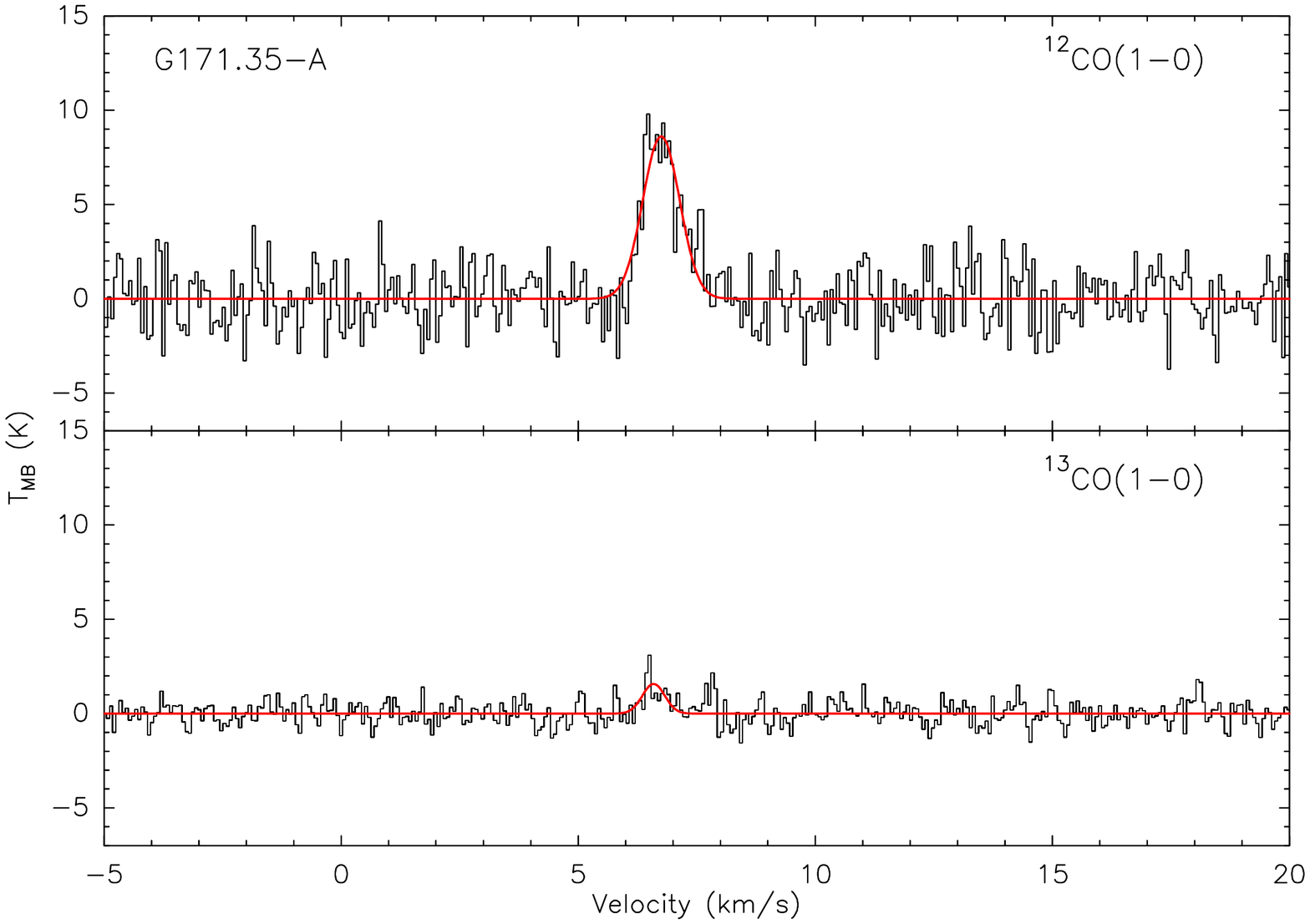}
		\includegraphics[angle=-90,width=.45\linewidth]{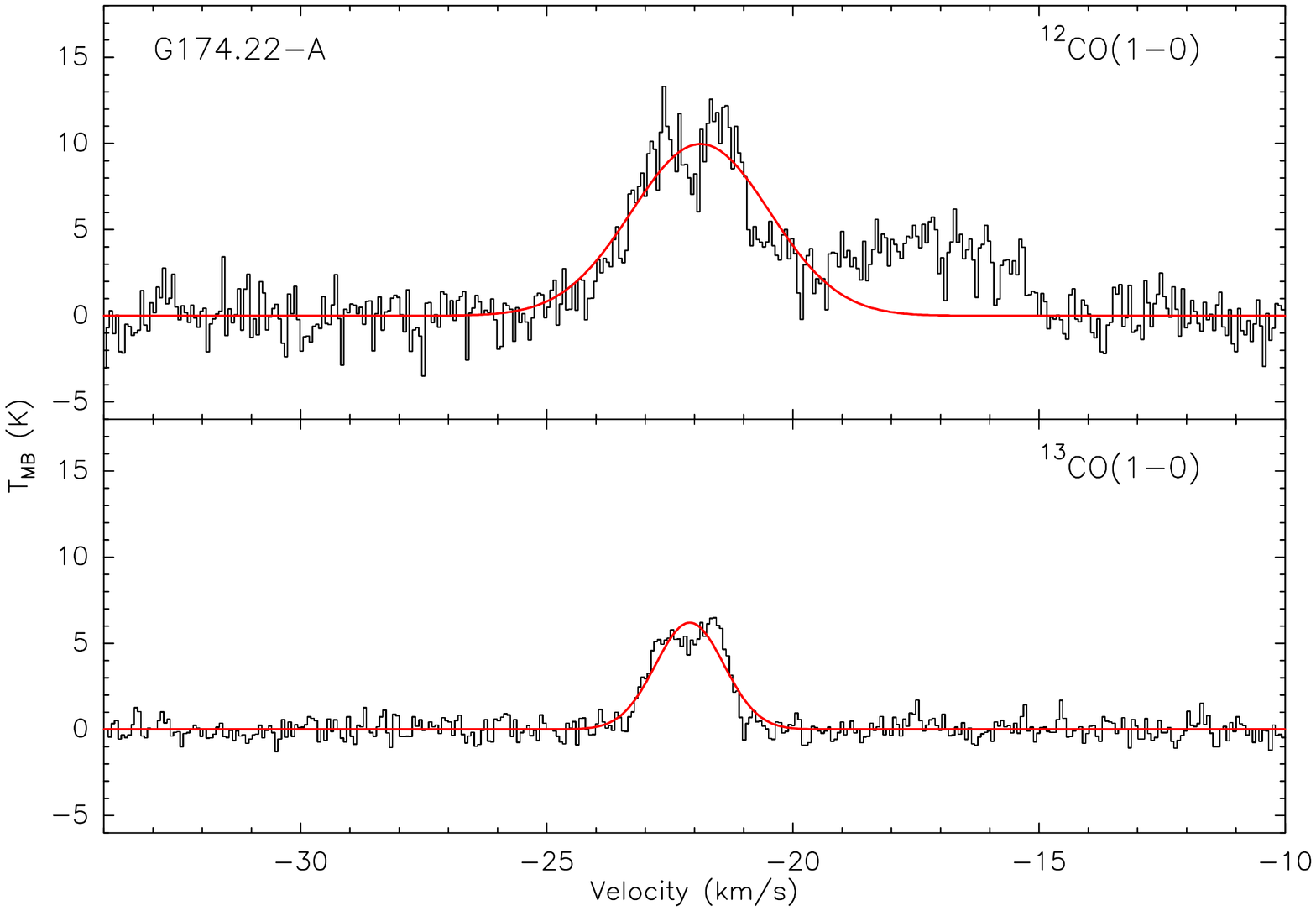}
		\includegraphics[angle=-90,width=.45\linewidth]{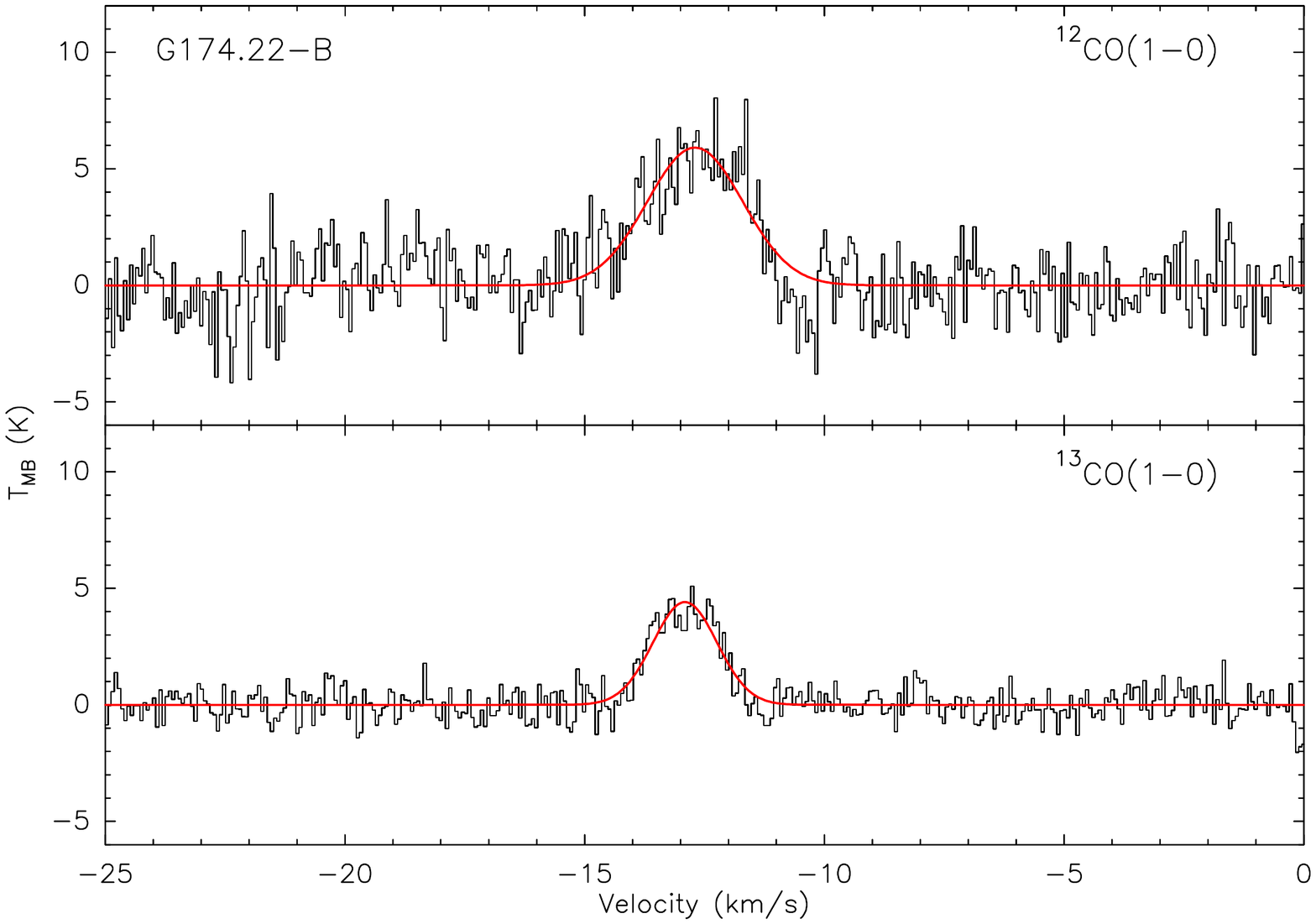}
		\includegraphics[angle=-90,width=.45\linewidth]{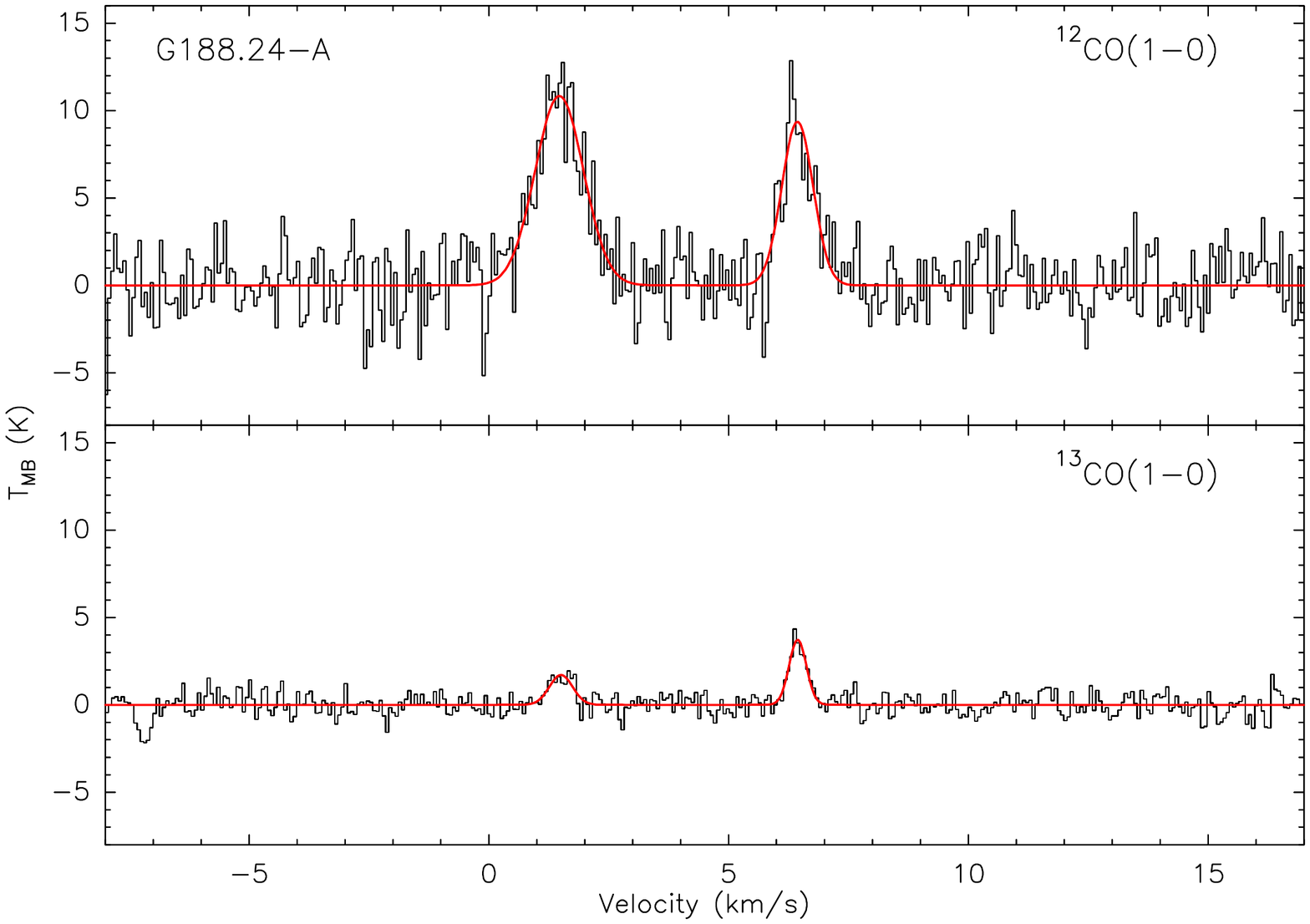}
		\includegraphics[angle=-90,width=.45\linewidth]{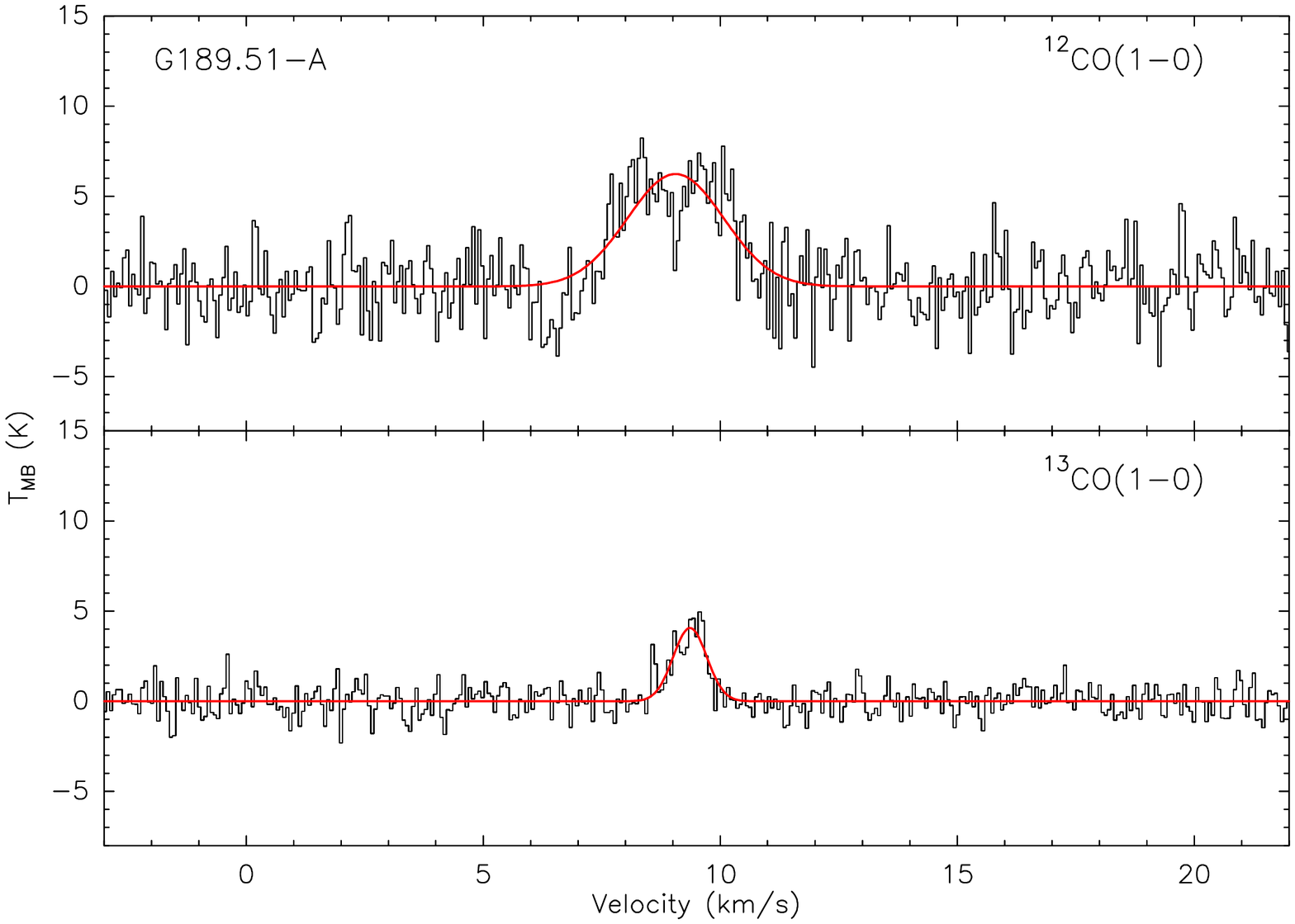}
    \caption{Cont. $^{12}$CO(1$-$0) and $^{13}$CO(1$-$0) spectra at the central position of each observed clump. Red line shows the Gaussian profile fit(s) to the lines.}
	\label{spec4}
\end{figure*}
\begin{figure*}[th]
	\centering		
		\includegraphics[angle=-90,width=.45\linewidth]{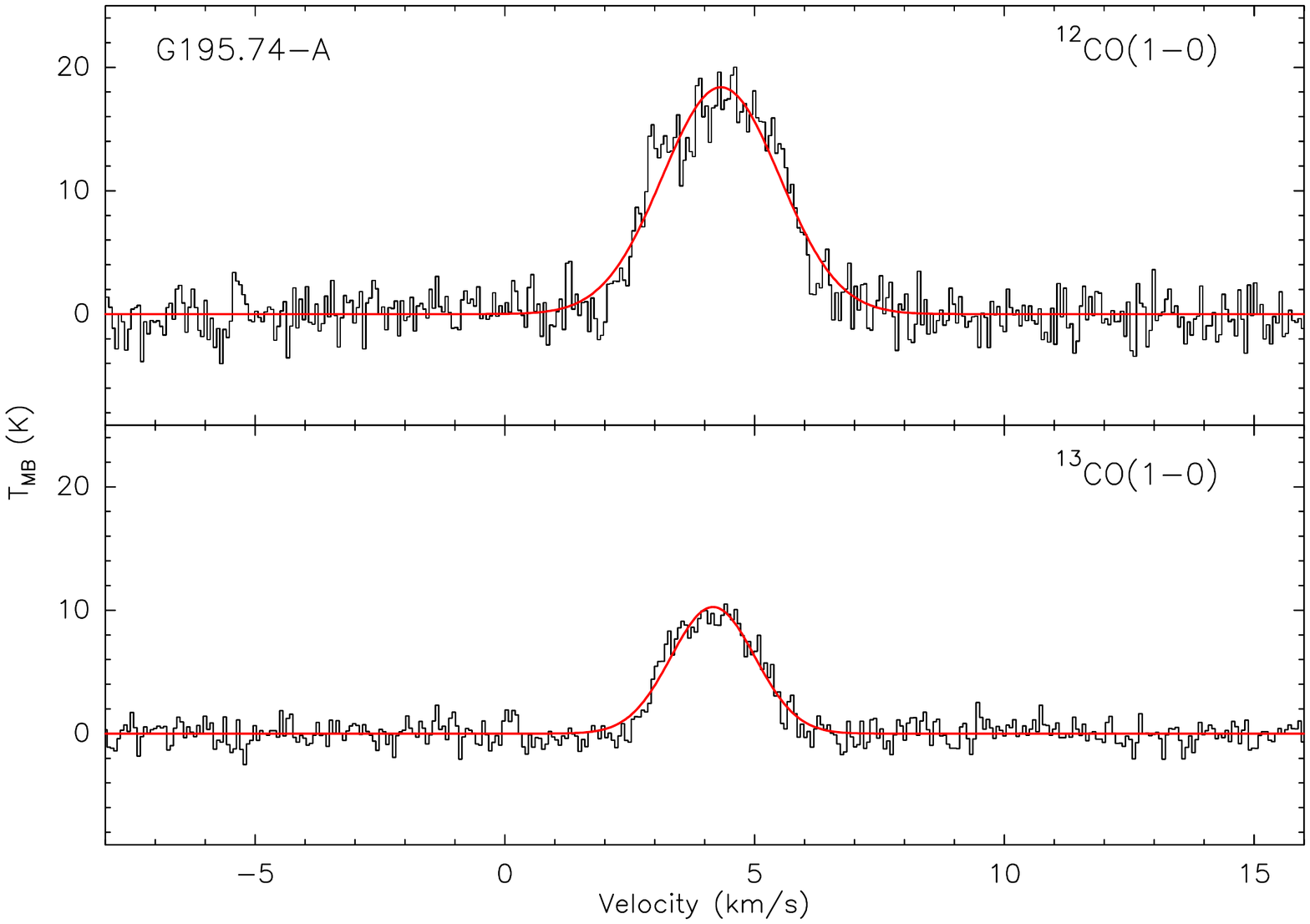}
    	\includegraphics[angle=-90,width=.45\linewidth]{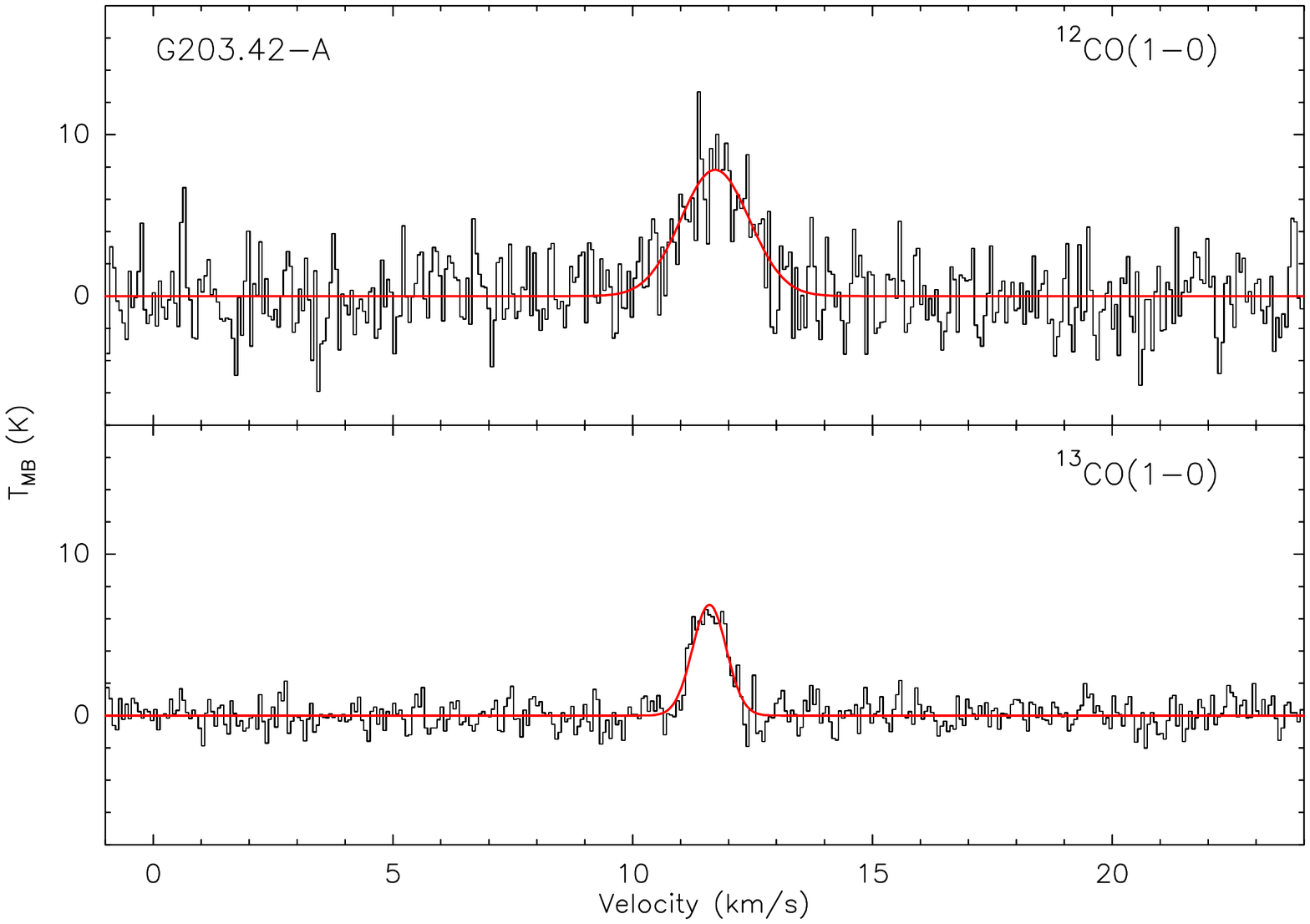}
		\includegraphics[angle=-90,width=.45\linewidth]{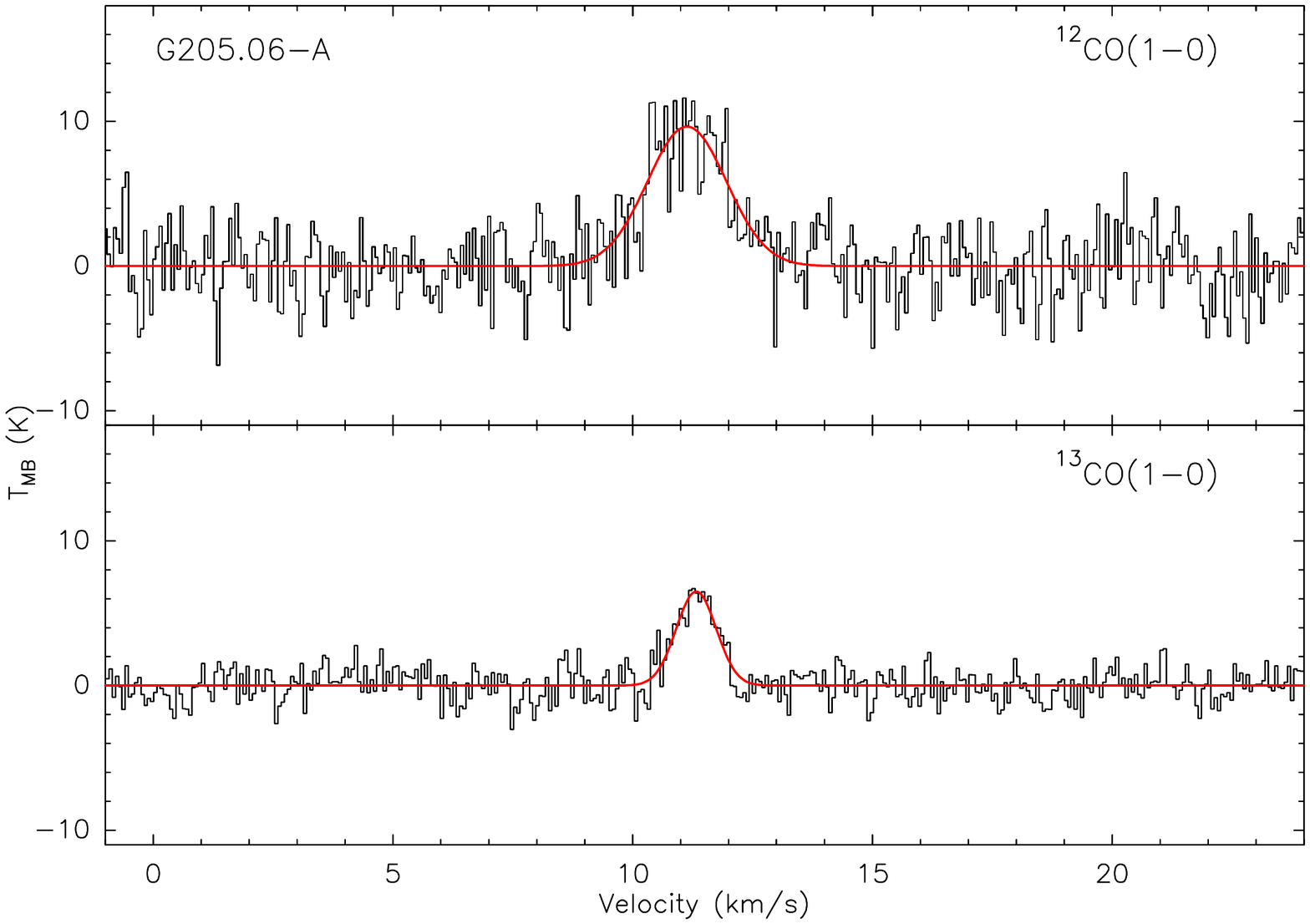}
    \caption{Cont. $^{12}$CO(1$-$0) and $^{13}$CO(1$-$0) spectra at the central position of each observed clump. Red line shows the Gaussian profile fit(s) to the lines.}
	\label{spec5}
\end{figure*}

\end{appendix}
 
\end{document}